\documentclass[11pt]{article}
\usepackage[a4paper, margin=1in]{geometry}
\usepackage[sorting = anyt, maxbibnames=9,maxcitenames=2, backend = bibtex, style=authoryear, doi=false, url=false,dashed=false,firstinits=true]{biblatex}

\AtEveryBibitem{\clearfield{issn}}
\AtEveryCitekey{\clearfield{issn}}

\AtEveryBibitem{\clearlist{language}}
\AtEveryCitekey{\clearlist{language}}

\AtEveryBibitem{\clearfield{month}}
\AtEveryCitekey{\clearfield{month}}
\usepackage{float}

\renewbibmacro{in:}{}
\usepackage[titletoc,page]{appendix}

\usepackage{soul}
\usepackage{setspace}
\usepackage{gensymb}
\usepackage{enumerate}
\usepackage{kotex}
\usepackage{abstract,lipsum}
\usepackage{color}
\usepackage{xcolor}
\soulregister\ref{7}
\soulregister\cite{7}
\soulregister\citeyearpar{7}
\usepackage{amssymb}
\usepackage{amsmath, nccmath}
\usepackage{amsthm}
\usepackage{amsfonts}
\usepackage{upgreek}
\usepackage{mathtools}
\usepackage{mathrsfs}
\usepackage{caption}
\usepackage{bm}
\usepackage{indentfirst}

\usepackage[symbol]{footmisc}

\DeclareMathAlphabet\mathrsfso{U}{rsfso}{m}{n}
\DeclareMathAlphabet\mathbfcal{OMS}{cmsy}{b}{n}
\usepackage[colorlinks=true,linkcolor=blue,citecolor=blue,urlcolor=blue]{hyperref}
\usepackage{graphicx}
\usepackage{array}
\usepackage{enumitem}
\SetLabelAlign{Center}{\hfil#1\hfil}
\SetLabelAlign{CenterWithParen}{\hfil(\makebox[1.0em]{#1})\hfil}
\usepackage{multirow}
\usepackage{makecell}

\begin{document}
\begin{center}
       \fontsize{19pt}{19pt}\selectfont Bayesian inference and uncertainty quantification for modeling of body-centered-cubic single crystals
       \vspace*{0.3in}

       \fontsize{10pt}{10pt}\selectfont Seunghyeon Lee$^{1}$, Thao Nguyen$^{3}$, Darby J. Luscher$^{4}$, Saryu J. Fensin$^{5}$, John S. Carpenter$^{6}$,\\ Hansohl Cho$^{2}$\\
       \vspace*{0.1in}
       \fontsize{9pt}{9pt}\selectfont $^{1}$Department of Aerospace Engineering, $^{2}$Department of Mechanical Engineering, Korea Advanced Institute of Science and Technology, Daejeon 34141, Republic of Korea \\ 
        $^{3}$Theoretical Division, $^{4}$X Computational Physics Division, $^{5}$Materials Physics and Applications Division, $^{6}$Sigma Division, Los Alamos National Laboratory, Los Alamos, NM 87545, US\\
\end{center}
\vspace*{0.1in}
\fontsize{8pt}{8pt}\selectfont 
\begin{tabular}{@{}l@{ }l}
E-mails: & sh1ee@kaist.ac.kr (S. Lee), thaonguyen@lanl.gov (T. Nguyen), djl@lanl.gov (D.J. Luscher), \\
         & hansohl@kaist.ac.kr (H. Cho)
\end{tabular}
\renewenvironment{abstract}
{\small
\noindent \rule{\linewidth}{.5pt}\par{\noindent \bfseries \abstractname.}}
{\medskip\noindent \rule{\linewidth}{.5pt}
}

\vspace*{0.1in}
\onehalfspacing
\begin{abstract}
\fontsize{11pt}{11pt}\selectfont
\onehalfspacing

Uncertainties in the high-dimensional space of material parameters pose challenges for the predictive modeling of bcc single crystals, especially under extreme loading conditions. In this work, we identify the key physical assumptions and associated uncertainties in constitutive models that describe the deformation behavior of bcc single crystal molybdenum subjected to quasi-static to shock loading conditions. We employ two representative physics-based bcc single crystal plasticity models taken from our previous work (\cite{nguyen2021dynamic,LEE2023103529}), each prioritizing different key deformation mechanisms. The Bayesian model calibration (BMC) is used for probabilistic estimates of material parameters in both bcc crystal plasticity models. In conjunction with the BMC procedure, the global sensitivity analysis is conducted to quantify the impact of uncertainties in the material parameters on the key simulation results of quasi-static to shock responses. The sensitivity indices at various loading conditions clearly illustrate the physical basis underlying the predictive capabilities of the two distinct bcc crystal plasticity models at low to high strain rates. Both of the calibrated bcc models are then further validated beyond the calibration regime, by which we further identify critical physical mechanisms that govern the transient elastic-plastic responses of single crystal molybdenum under shock loading. The statistical inference framework demonstrated here facilitates the further development of continuum crystal plasticity models that account for a broad range of deformation mechanisms.

\end{abstract}

\onehalfspacing
\section{Introduction}
\label{sec:intro}

The plastic deformation of body-centered-cubic (bcc) refractory metals and their alloys has been widely investigated for a broad variety of applications under thermo-chemo-mechanical extremes. However, predictive modeling of bcc materials remains challenging due to the complex nature of slip activity, patterns and interactions (\cite{Bulatov2006,Queyreau2009,MADEC2017166}). Furthermore, the deformation features in these materials are highly rate- and temperature-dependent, which further complicates the modeling of their responses (\cite{Seeger2001,Seeger2002AnomalousMetals,Weygand2015}). Additionally, the breakdown of the standard Schmid law throughout the major slip systems poses a challenge to the further development of continuum constitutive models especially at the single crystal level (\cite{GROGER20085401,dezerald_plastic_2016,groger_symmetry-adapted_2021}). In this context, the microstructure-informed and the physics-based bcc crystal plasticity models that account for the key deformation mechanisms have been widely proposed (\cite{Cereceda2016,lhadi_micromechanical_2018,cho_anomalous_2018,bronkhorst_local_2021,nguyen2021dynamic,LEE2023103529,tsekpuia_microstructure-based_2023}).

Recently, guided by the dislocation dynamics simulations (\cite{Madec2004,Queyreau2009,MADEC2017166}) and the atomistic simulations (\cite{GROGER20085401,Weygand2015,dezerald_plastic_2016}), the physics-based bcc crystal plasticity models have been further advanced. This approach has shown success in capturing the key features in temperature dependence of yield strength (\cite{monnet_dislocation-dynamics_2013,patra__2023,lim_multi-scale_2015,Cereceda2016}), slip instability and segregation (\cite{LEE2023103529,dequiedt_slip_2023}), slip mode transition (\cite{lim_investigating_2020,bertin_crystal_2023})  and texture evolution (\cite{Weinberger2012IncorporatingModels,Cereceda2016,LEE2023103529}) in single- and polycrystal bcc materials. In parallel, physics-based crystal plasticity models that account for the kinetics of slip (\cite{KOCKSUF1975,regazzoni_dislocation_1987}) and the coupled mobile$-$immobile dislocation density evolution laws (\cite{estrin_local_1986}) have been proposed to describe the responses of single crystals in the weak shock regime (\cite{austin2011dislocation}). This approach for weak shock applications has been further extended by \cite{lloyd_plane_2014} and \cite{LUSCHER201763} to better represent the orientation-dependent shock responses at the single crystal level. Despite extensive modeling efforts over the past decade, a comprehensive mechanistic understanding of the relationships and interactions between the distinct deformation mechanisms in bcc materials remains largely lacking, and our ability to capture the material behavior across a wide range of loading conditions and crystallographic orientations is still limited. Moreover, incorporating multiple deformation mechanisms into the crystal plasticity modeling framework inevitably leads to an increase in the number of model parameters. More specifically, the various dislocation evolution mechanisms and their coupling with the hardening laws and kinetic relations, essential for capturing the key deformation features, often result in strong correlations between the model parameters. Furthermore, they lead to substantial uncertainties in the model parameters, thus undermining the reliability of model predictions for conditions beyond the calibration regime.

In order to overcome the challenges associated with high-dimensional parameter identification procedures for the complex bcc crystal plasticity models that incorporate the multiple, distinct deformation mechanisms, systematic statistical approaches have been proposed (e.g., \cite{rousselier_macroscopic_2012,sedighiani_efficient_2020,NGUYEN2021104284,depriester_crystal_2023,dindarlou_optimization_2024}). More specifically, the Bayesian model calibration (BMC), introduced by \cite{kennedy2001bayesian} and \cite{doi:10.1198/016214507000000888}, has been widely employed for quantifying modeling uncertainties in the Bayesian statistical setting. The BMC framework identifies the model parameters that best match experimental data and quantifies the uncertainties in these parameters as probability distributions. Then, the resulting calibrated parameters can be used to model the material responses under unexplored experimental conditions. This BMC framework has proven very useful in calibrating and validating constitutive models; e.g., the Johnson-Cook strength model (\cite{10.1063/1.5051442}), the Preston–Tonks–Wallace strength model (\cite{nguyen2024calibration}), the single crystal plasticity models (\cite{nguyen2021dynamic,NGUYEN2021104284,venkatraman_bayesian_2022}), and the phase-field fracture models (\cite{noii_bayesian_2021,noii_bayesian_2022}). Furthermore, as noted in \cite{10.1063/1.5051442}, the posterior distribution obtained from the BMC procedure can guide experimental designs by analyzing the posterior sensitivity to data selection. However, it should be noted that, when the crystal plasticity models with specific parameter sets fail to reproduce the experimental data, the BMC procedure provides limited insight into the sources of these discrepancies. The variance-based global sensitivity analysis proposed by \cite{sobol1993sensitivity} and \cite{sobol_global_2001} offers a promising means of complementing the BMC procedure. It decomposes the total simulated output variance into the additive partial variances attributable to the main effects of each parameter on the model prediction, as well as to their interactions across the entire input parameter space. This has been shown to facilitate dimensionality reduction in high-dimensional inverse parameter estimation problems (\cite{sargsyan2014dimensionality,schill_simultaneous_2021}) and to provide insight into the underlying physical mechanisms of physics-based models (\cite{robbe_global_2023, nelms_uncertainty_2024}). Furthermore, conducting the global sensitivity analysis provides guidance for diagnosing the sources of discrepancies between experiments and modeling results by identifying the parameters that play key roles in the model predictions (\cite{huan2018global}).

In this work, we aim to identify key physical assumptions and the associated uncertainties for the predictive modeling of bcc molybdenum under quasi-static to shock loading conditions. To this end, we employ two representative physics-based bcc single crystal plasticity models taken from our previous work (\cite{nguyen2021dynamic,LEE2023103529}), each prioritizing different key deformation mechanisms. First, the BMC procedure is used for probabilistic estimates of parameters in both models using the same experimental data spanning diverse strain rates, temperatures, and crystallographic orientations. In conjunction with the BMC procedure, the global sensitivity analysis is conducted to quantify the impact of uncertainties in the model parameters on the key simulation results of quasi-static to shock responses. The global sensitivity analysis of uniaxial stress responses at various strain rates and temperatures shows that the influence of individual parameters is strongly correlated with loading conditions. The sensitivity indices at various loading conditions clearly illustrate the physical basis underlying the predictive capabilities of the two distinct bcc models for the rate- and temperature-dependence. Then, both of the physics-based bcc crystal plasticity models are further validated at extreme shock loading conditions beyond the calibration regime. By assessing the predictive capabilities of the two calibrated bcc models for the recent plate impact experiments (\cite{10.1063/5.0082267}), we further identify critical physical mechanisms that govern the elastic-plastic transition in single crystal molybdenum under shock loading.

The paper is organized as follows. We summarize the two physics-based bcc crystal plasticity models in Section \ref{section:sc model}. In Section \ref{Sec:Bayesian}, the material parameters in the two bcc models are inferred using the BMC procedure; we then conduct the global sensitivity analysis on the model predictions of uniaxial stress-strain responses across various temperatures (293 to 413 K), strain rates (10$^{-4}$ to 10$^{5}$ s$^{-1}$) and crystallographic orientations. The calibrated bcc crystal plasticity models are further validated against the plate impact responses in Section \ref{sec:plate_impact}. Here, the key physical assumptions to capture the transient behavior during the plate impact tests are identified by the additional global sensitivity analysis on the plate impact simulations. Finally, the implications of our results on uncertainty quantification via the Bayesian model calibration and global sensitivity analysis are discussed Sections \ref{sec:discussion}. We close the paper with a brief summary and future directions in Section \ref{sec:conclusion}.

\section{Single crystal plasticity model}
\label{section:sc model}
\subsection{Kinematics}
A motion $\boldsymbol{\varphi}$ is defined as a one-to-one mapping $\mathbf{x} = \boldsymbol{\varphi}(\mathbf{X},t)$ with a material point $\mathbf{X}$ in a reference configuration and $\mathbf{x}$ in a deformed spatial configuration. The deformation gradient can be multiplicatively decomposed into, 
\begin{equation}
\mathbf{F} \stackrel{\text{def}}{=} \frac{\partial \boldsymbol{\varphi}}{\partial \mathbf{X}}= \mathbf{F}^{\textrm{e}} \mathbf{F}^{\textrm{p}},
\end{equation}
where $\mathbf{F}^{\textrm{e}}$ and $\mathbf{F}^{\textrm{p}}$ represent the elastic and plastic parts of deformation gradients, respectively. The velocity gradient ($\mathbf{L}$) is additively decomposed into the elastic distortion ($\mathbf{L}^{\textrm{e}}$) and plastic distortion ($\mathbf{L}^{\textrm{p}}$) rate tensors by, 
\begin{equation}
\label{eqn:RSS}
\mathbf{L} \stackrel{\text{def}}{=} \frac{\partial \mathbf{v}}{\partial \mathbf{x}} = \dot{\mathbf{F}}\mathbf{F}^{-1} =  \mathbf{L}^{\textrm{e}} + \mathbf{F}^{\textrm{e}} \mathbf{L}^{\textrm{p}} \mathbf{F}^{\textrm{e} -1},
\end{equation}
where $\mathbf{v}$ denotes the spatial velocity field, and $\mathbf{L}^{\textrm{e}}=\dot{\mathbf{F}}^\textrm{e}\mathbf{F}^{\textrm{e}-1}$ and $\mathbf{L}^{\textrm{p}}=\dot{\mathbf{F}}^\textrm{p}\mathbf{F}^{\textrm{p}-1}$. Here, the magnitude of the plastic distortion tensor is related to the dislocation slip rate ($\dot{\gamma_\textrm{p}}^{\alpha}$) on prescribed slip systems, i.e.,
\begin{equation}
\label{eq:plastic_distortion}
\mathbf{L}^{\textrm{p}} = \sum_{\alpha=1}^{n_{\textrm{slip}}} \dot{\gamma_\textrm{p}}^{\alpha} \mathbf{\mathbb{S}}^{\alpha}_0 \quad \textrm{with} \quad  \mathbf{\mathbb{S}}^{\alpha}_0=\mathbf{s}^{\alpha}_0 \otimes \mathbf{m}^{\alpha}_0,
\end{equation}
where $\mathbf{\mathbb{S}}^{\alpha}_0$ is the Schmid tensor, $\mathbf{s}^{\alpha}_0$ is the slip direction vector, and $\mathbf{m}^{\alpha}_0$ is the slip plane normal vector defined in the intermediate space. $n_{\textrm{slip}}$ represents the number of slip systems. Moreover, the elastic strain tensor is defined as,
\begin{equation}
\mathbf{E}^{\textrm{e}}=\frac{1}{2}(\mathbf{C}^{\textrm{e}} - \mathbf{I}),
\end{equation}
where $\mathbf{C}^{\textrm{e}} = \mathbf{F}^{\textrm{e T} }\mathbf{F}^{\textrm{e}}$ is the elastic right Cauchy-Green tensor defined in the intermediate space.

\subsection{Elasticity}

In this study, the stress response to the shear and volumetric components of the strain tensor is computed separately. The deviatoric part of the Cauchy stress tensor is modeled with anisotropic crystal elasticity. For the deviatoric part, the second Piola-Kirchhoff stress defined in the intermediate space is calculated by,
\begin{equation}
\label{elastic}
\mathbf{T}^\textrm{e}=\bm{\mathcal{C}} \mathbf{E}^{\textrm{e}},
\end{equation}
where $\bm{\mathcal{C}}$ is the fourth-order elasticity tensor with three temperature- and pressure-dependent elastic constants $C_{11}(P,\theta)$, $C_{12}(P,\theta)$, and $C_{44}(P,\theta)$. We assume that the elastic constants are linearly dependent on the temperature and pressure, i.e.,
\begin{equation}
C_{ij}(P,\theta) = C_{ij,0} + C_{ij,\theta} \theta + C_{ij,P}P.
\end{equation}
Here, $C_{ij,0}$ is the elastic constants at 0 K, and $C_{ij,T}$ and $C_{ij,P}$ represent their temperature and pressure dependences. Then, the Cauchy stress is calculated through the relation,
\begin{equation}
\textbf{T}_{\textrm{cpl}} = J^{-1} \mathbf{F}^{\textrm{e}} \mathbf{T}^{\textrm{e}} \mathbf{F}^{\textrm{e} \, \textrm{T}} \quad \textrm{with} \quad J=\textrm{det}\mathbf{F}.
\end{equation}
The resolved shear stress that drives slip on the $\alpha$-th slip system is defined as the projection of the stress via the Schmid tensor, i.e.,
\begin{equation}
\label{eqn:RSS}
\tau^{\alpha} = \mathbf{C}^{\textrm{e}}\mathbf{T}^{\textrm{e}}:\mathbf{\mathbb{S}}^{\alpha}_0 = \boldsymbol{\tau}:\mathbf{\mathbb{S}}^{\alpha},
\end{equation}
where $\boldsymbol{\tau} = J\textbf{T}_{\textrm{cpl}} $ is the Kirchhoff stress and $\mathbf{\mathbb{S}}^{\alpha}=\mathbf{F}^{\textrm{e}} \mathbf{\mathbb{S}}^{\alpha}_0 \mathbf{F}^{\textrm{e -1}}$ is the Schmid tensor defined in the deformed spatial configuration.

The hydrostatic part of the stress response is modeled with the Mie-Gr{\"u}niesen equation of state (\cite{mie_zur_1903,gruneisen_theorie_1912}), i.e.,
\begin{equation}
\label{eqn:eos}
P_{\textrm{eos}} = \frac{\rho_0 C_0^2 \chi \left[ 1-\frac{\Gamma_0}{2} \chi \right]}{(1-s \chi)^2} + \rho_0 \Gamma_0 e.
\end{equation}
Here, $\chi=1-\rho_0 / \rho$ is the compression ratio. $\rho_0$ and $\rho$ are the density at reference state and deformed state, respectively. $C_0$ is the bulk speed of sound, $\Gamma_0$ is the Gr{\"u}niesen gamma, $s = dU_s/dU_p$ is the Hugoniot slope coefficient, and $e$ is the specific internal energy. Then, the Cauchy stress is corrected via, 
\begin{equation}
\mathbf{T}=\mathbf{T}_{\textrm{cpl}} - \left( P_{\textrm{eos}} - P_{\textrm{cpl}}\right)\mathbf{I},
\end{equation}
where $P_{\textrm{cpl}}=-\frac{1}{3} \textrm{tr} \mathbf{T}_{\textrm{cpl}}$ is the mean pressure calculated from the crystal plasticity models. Since the Schmid tensors in the intermediate space and the deformed spatial configuration are traceless,
\begin{equation}
\mathbf{I}:\mathbf{\mathbb{S}}^{\alpha}= \textrm{tr}(\mathbf{F}^{\textrm{e}} \mathbf{\mathbb{S}}^{\alpha}_0 \mathbf{F}^{\textrm{e -1}})= \textrm{tr}\mathbf{\mathbb{S}}^{\alpha}_0 = \mathbf{s}^{\alpha}_0 \cdot \mathbf{m}^{\alpha}_0=0,
\end{equation}
the pressure correction does not influence the slip rate ($J\mathbf{T}:\mathbf{\mathbb{S}}^{\alpha}=J\mathbf{T}_{\textrm{cpl}}:\mathbf{\mathbb{S}}^{\alpha}$). Hence, the pressure is updated explicitly at the end of the increment.

\subsection{Dislocation density evolution and flow rule}

As noted in Section \ref{sec:intro}, we here employ the physics-based bcc single crystal plasticity models to identify the model-specific, key deformation mechanisms through the uncertainty quantification (UQ)-based calibration and validation. Thermally activated flow rules, widely used with a fixed pre-exponential factor, have found success in capturing the rate- and temperature dependent yield and hardening behavior, often without explicit coupling to the total or mobile dislocation density (\cite{kothari_elasto-viscoplastic_1998, monnet_dislocation-dynamics_2013}) via the Orowan relation (\cite{EOrowan_1940}). Moreover, the bcc single crystal plasticity models have been further extended to account for the relationship between the macroscopic strain hardening behavior and the evolution of dislocation density (\cite{ma_dislocation_2006,Cereceda2016,LEE2023103529}). Especially, the dislocation-dynamics-informed crystal plasticity models have showed that a detailed description of dislocation interaction strengths across the slip systems is crucial for capturing slip system activation and slip patterns in the bcc single crystals loaded upon high symmetry directions (\cite{stainier_micromechanical_2002,LEE2023103529,dequiedt_slip_2023}).

Depending upon the stress level, the plastic slip rate has been found to be significantly influenced by the thermal activation of pinned dislocations as well as by the velocity of dislocations traveling between obstacles (\cite{KOCKSUF1975,regazzoni_dislocation_1987,austin2011dislocation}). In particular, at extreme strain rates where phonon drag effects on dislocation motion between obstacles become significant, the coupling between the flow rule and the evolution of mobile dislocation density via the Orowan relation has been found to play a critical role in capturing the competition between the thermally activated and (phonon) drag-dominated glide mechanisms. Although these models account for the complex phenomena associated with mobile dislocations especially at high velocities, identifying reliable parameters remains challenging due to the coupled influence of multiple dislocation glide and evolution mechanisms within these models. This leads to significant inconsistencies in parameter values across the literature (\cite{austin2011dislocation, LUSCHER201763,NGUYEN2021104284,nguyen2021dynamic}). Furthermore, accurately determining the ratio of mobile to immobile dislocations is essential for applying these crystal plasticity models that make use of the Orowan relation with the mobile dislocation evolution; however, theoretical studies for the identification procedures of these parameters remain limited (\cite{sills_dislocation_2018,denoual2024dislocation}).

In this work, two distinct physics-based bcc single crystal plasticity models are considered: (1) a model that accounts for the mobile dislocation kinetics including the mobile–immobile transition (\cite{nguyen2021dynamic}), and (2) a model that incorporates dislocation-dynamics-informed hardening laws and dislocation density evolution (\cite{LEE2023103529}). Through the UQ-based validation of two distinct models, we aim to identify the model-specific deformation mechanisms and key physical assumptions required to capture the deformation behavior of bcc molybdenum at quasi-static to shock loading conditions. Next, we summarize the two bcc crystal plasticity models.

\subsubsection{Model 1}
\label{section: vumat model}
The overall framework of Model 1 is based on the crystal plasticity models that incorporate a transition from the thermally-activated dislocation glide to the drag-dominated regime (\cite{austin2011dislocation,lloyd_plane_2014}). The modeling framework has been extended to include the characteristics of both edge and screw dislocations (\cite{LUSCHER201763,NGUYEN2021104284}). More recently, this framework was further extended to the bcc-specific crystal plasticity model (\cite{nguyen2021dynamic}), labeled Model 1 throughout the main body of this work, which also includes the twinning/anti-twinning asymmetry of glides and has been validated for bcc tantalum.

Model 1 considers the simple geometric relations to account for dislocation interactions (\cite{ma_dislocation_2006,lloyd_plane_2014,LUSCHER201763}) associated with the strain hardening and dislocation density evolution throughout the slip systems. The total dislocation density is projected onto the slip plane normal to obtain the forest dislocation densities  ($\rho^{\alpha}_{\textrm{for}}$) and onto the in-plane direction to obtain the coplanar dislocation densities ($\rho^{\alpha}_{\textrm{cop}}$), i.e.,
\begin{equation}
\label{forest disloc.}
\rho^{\alpha}_{\textrm{for}} = \sum^{n_{\textrm{slip}}}_{\beta = 1} A^{\alpha \beta}_{\textrm{for}} \rho^{\beta} \quad \textrm{with} \quad A^{\alpha \beta}_{\textrm{for}}= \frac{1}{2} | \mathbf{n}^{\alpha} \cdot \mathbf{s}^{\beta} | + \frac{1}{2} | \mathbf{n}^{\alpha} \cdot (\mathbf{n}^{\beta} \times \mathbf{s}^{\beta} ) |,
\end{equation}
\begin{equation}
\label{coplanar disloc.}
\rho^{\alpha}_{\textrm{cop}} = \sum^{n_{\textrm{slip}}}_{\beta = 1} A^{\alpha \beta}_{\textrm{cop}} \rho^{\beta} \quad \textrm{with} \quad \quad A^{\alpha \beta}_{\textrm{cop}}= \frac{1}{2} | \mathbf{n}^{\alpha} \times \mathbf{s}^{\beta} | + \frac{1}{2} | \mathbf{n}^{\alpha} \times (\mathbf{n}^{\beta} \times \mathbf{s}^{\beta} ) |.
\end{equation}
Here, the $| \mathbf{n}^{\alpha} \cdot \mathbf{s}^{\beta} |$ and $| \mathbf{n}^{\alpha} \times \mathbf{s}^{\beta} |$ represent the projection coefficients for screw dislocations, and $| \mathbf{n}^{\alpha} \cdot (\mathbf{n}^{\beta} \times \mathbf{s}^{\beta} ) |$ and $| \mathbf{n}^{\alpha} \times (\mathbf{n}^{\beta} \times \mathbf{s}^{\beta} ) |$ represent the projection coefficients for edge dislocations. In Equations \ref{forest disloc.} and \ref{coplanar disloc.}, an equal mixture of screw and edge dislocation characteristics is assumed. Then, the athermal resistance due to the forest dislocation is expressed by,
\begin{equation}
\label{eq:forest_resistance}
\tau^{\alpha}_{\textrm{a}} = \tau_{\textrm{a,0}} + C_{\textrm{for}} b \mu_{\star} \sqrt{\rho^{\alpha}_{\textrm{for}} },
\end{equation}
where $\tau_{\textrm{a,0}}$ is the reference athermal resistance, $b$ is the Burgers vector, $\mu_{\star}$ is the effective shear modulus, and $C_{\textrm{for}}$ is the forest hardening coefficient. In addition, the hardening due to parallel dislocations is expressed by,
\begin{equation}
\label{eq:lattice_resis}
\tau^{\alpha}_{\textrm{r}} = \tau^{\alpha}_{\textrm{r,0}} + C_{\textrm{cop}} b \mu_{\star} \sqrt{\rho^{\alpha}_{\textrm{cop}} },
\end{equation}
where $\tau^{\alpha}_{\textrm{r,0}}$ is the lattice resistance or Peierls stress, and $C_{\textrm{cop}}$ is the hardening coefficient due to the parallel dislocations.

In Model 1, it is further assumed that the total dislocation density in each slip system $\alpha$ is decomposed into the mobile ($\rho_M^\alpha$) and the immobile ($\rho_I^\alpha$) parts, i.e.,
\begin{equation}
\rho^{\alpha} = \rho^{\alpha}_M + \rho^{\alpha}_I.
\end{equation}
The plastic strain rate on the $\alpha$-th slip system is directly connected to the mobility of dislocations via the Orowan relation (\cite{EOrowan_1940}),
\begin{equation}
\dot{\gamma_\textrm{p}}^{\alpha} = b \rho^{\alpha}_M v^{\alpha},
\label{eq_model1:orowan}
\end{equation}
and the mean velocity of dislocations ($v^{\alpha}$) is defined as (\cite{KOCKSUF1975,austin2011dislocation}),
\begin{equation}
\label{eqn:mean_velo}
v^{\alpha} = \frac{\bar{L}^{\alpha}}{t^{\alpha}_w + t^{\alpha}_r} \textrm{sign} (\tau^{\alpha}),
\end{equation}
where $\bar{L}^{\alpha}$ is the mean spacing between obstacles, $t^{\alpha}_w$ is the waiting time for thermal activation, and $t^{\alpha}_r$ is the running time for the damped glide between the obstacles. The waiting time for thermal activation is defined as,
\begin{equation}
\label{eq:thermal_act}
t^{\alpha}_w= \frac{1}{\textit{f}_D} \left( \textrm{exp}\left[ \frac{\Delta G}{k_B \theta}\left\langle 1 - \left\langle \frac{|\tau^{\alpha}| - \tau^{\alpha}_{\textrm{a}} }{\tau^{\alpha}_{\textrm{r}}} \right\rangle^p \right\rangle^q \right ]-1 \right),
\end{equation}
where $\textit{f}_D$ is the attempt frequency of waiting dislocations, $\Delta G$ is the activation energy, $k_B$ is the Boltzmann's constant, and $p$ and $q$ are the parameters associated with the shape of stress-dependent kink-pair formation energy. $\langle \, \cdot \, \rangle = \frac{1}{2} \big( \, \lvert \,\cdot\,\rvert + (\,\cdot\,)\,\big)$ is the Macaulay bracket. In addition, the running velocity associated with the damped glide between the obstacles is expressed by,
\begin{equation}
\label{eqn:running}
v^{\alpha}_r =\frac{\bar{L}^{\alpha}}{t^{\alpha}_r} = \frac{b}{B_{\star}^{\alpha} } (| \tau^{\alpha} | -\tau^{\alpha}_{\textrm{a}})\quad \textrm{with} \quad B_{\star}^{\alpha} = \frac{B_0}{1-(v^{\alpha}_r/c_s)^2}, \\
\end{equation}
where $B_0$ is the drag coefficient at $v^{\alpha}_r =0$, $B_{\star}^{\alpha}$ is the dislocation velocity dependent drag coefficient considering a relativistic effect, and $c_s$ is the elastic shear wave speed. The relativistic relation of the drag coefficient limits the running velocity ($v^{\alpha}_r$) at high stresses to the elastic shear wave speed ($c_s$) as noted in \cite{austin2011dislocation}.

The dislocation density evolutions are described by multiplication ($\dot{\rho}_{\textrm{mult}}^{\alpha}$), annihilation ($\dot{\rho}_{\textrm{ann}}^{\alpha}$), trapping ($\dot{\rho}_{\textrm{trap}}^{\alpha}$) of mobile dislocations and dynamic recovery ($\dot{\rho}_{\textrm{rec}}^{\alpha}$) of immobile dislocations. The evolution equations for mobile and immobile dislocation densities are then written as,
\begin{equation}
\label{eq:mobile_evo}
\dot{\rho}_M^{\alpha}=\dot{\rho}_{\textrm{mult}}^{\alpha} - \dot{\rho}_{\textrm{ann}}^{\alpha} - \dot{\rho}_{\textrm{trap}}^{\alpha},
\end{equation}
\begin{equation}
\label{eq:immobile_evo}
\dot{\rho}_I^{\alpha}=\dot{\rho}_{\textrm{trap}}^{\alpha} -\dot{\rho}_{\textrm{rec}}^{\alpha} \approx \textit{f}_{\textrm{rec}}^{\alpha} \dot{\rho}_{\textrm{trap}}^{\alpha},
\end{equation}
where $\textit{f}_{\textrm{rec}}^{\alpha}$ is the recovered fraction for approximating the effect of dynamic recovery. Each term of the mobile dislocation density evolution equation is expressed by,
\begin{equation}
\label{eq:multi}
\dot{\rho}_{\textrm{mult}}^{\alpha} = C_M \sqrt{\rho^{\alpha}_{\textrm{for}}} \rho_M^{\alpha} |v^{\alpha} |,
\end{equation}
\begin{equation}
\label{eq:anni}
\dot{\rho}_{\textrm{ann}}^{\alpha} = \frac{1}{4} C_A d^{\alpha}_{A} (\rho_M^{\alpha} )^2 |  v^{\alpha} |,
\end{equation}
\begin{equation} 
\label{eq:trapp}
\dot{\rho}_{\textrm{trap}}^{\alpha} = C_T \sqrt{\rho^{\alpha}_{\textrm{for}}} \rho_M^{\alpha} |v^{\alpha} |,
\end{equation}
where $C_M$, $C_A$, and $C_T$ are the coefficients for dislocation multiplication, annihilation, and trapping, respectively. $d^{\alpha}_{A}$ is the capture radius for mobile dislocation annihilation. Considering asymptotic state ($\dot{\rho}_M^{\alpha}=0$ when $\rho_M^{\alpha}=\rho_{M, sat}^{\alpha}$) in Equation \ref{eq:mobile_evo}, we impose the following relations,
\begin{equation}
C_A = 4 ( C_M - C_T),
\end{equation}
\begin{equation}
d^{\alpha}_{A} = \frac{\sqrt{\rho^{\alpha}_{\textrm{for}}}}{\rho_{M, sat}^{\alpha}},
\end{equation}
where $\rho_{M, sat}^{\alpha}$ is the slip system specific saturation value of the mobile dislocation density. In Equation \ref{eq:immobile_evo}, the recovery fraction is defined as (\cite{nguyen2021dynamic}),
\begin{equation}
\textit{f}_{\textrm{rec}}^{\alpha}=1-\left(\frac{\rho_I^\alpha}{\rho_{I, sat}^{\alpha}}\right)^{1 / n_{\textrm{rec}}},
\end{equation}
where $\rho_{I, sat}^{\alpha}$ is the slip system specific saturation value of the immobile dislocation density, and $n_{\mathrm{rec}}$ is the inverse saturation exponent. In addition, the rate- and temperature-dependent saturation value of the immobile dislocation density is expressed by,
\begin{equation}
\label{eq:sat_dd}
\rho_{I, sat}^{\alpha} = \rho_{0, \, sat} \left( \frac{\dot{\gamma}^{\alpha}}{\dot{\gamma}_0} \right)^{k_B \theta/A} \geq \rho_{I, annealed}^{\alpha},
\end{equation}
where $\rho_{0, sat}$ is the reference saturation dislocation density, $\dot{\gamma}_0$ is the reference strain rate, and $A$ is the reference thermal energy. Moreover, $\rho_{M, sat}^{\alpha}$ is assumed to be a 10$\%$ of $\rho_{I, sat}^{\alpha}$ (i.e., $\rho_{M, sat}^{\alpha} = 0.1 \rho_{I, sat}^{\alpha}$).

\subsubsection{Model 2}
\label{section: umat model}

The overall framework of Model 2 closely follows those of \cite{asaro_overview_1985}, \cite{kalidindi1992crystallographic} and \cite{kothari_elasto-viscoplastic_1998}. These models have been validated primarily against stress–strain responses and texture evolutions at both single- and polycrystalline levels in diverse loading conditions. More recently, the bcc-specific extensions incorporating dislocation-density-based hardening laws have been proposed and validated for bcc tantalum  (\cite{cho_anomalous_2018,bronkhorst_local_2021}). In these continuum crystal plasticity models, the dislocation density evolution and Taylor hardening laws have been further refined based on the dislocation dynamics simulation results (\cite{madec_role_2003,devincre2008,Queyreau2009,MADEC2017166}). More specifically in \cite{LEE2023103529}, labeled Model 2 in this work, the flow rule and dislocation-density-based hardening law were constructed based upon the recent dislocation dynamic simulation results especially for bcc crystals (\cite{MADEC2017166}). 

The plastic strain rate on the $\alpha$-th slip system is described by the thermal activation of dislocation motion,
\begin{equation}
\label{eqn:slip}
\begin{aligned}
\dot{\gamma_\textrm{p}}^{\alpha}=\dot{\gamma_0} \, \textrm{exp} \left(-\frac{\Delta G}{k_B \theta}\left< 1- \left( \frac{\lvert \tau^{\alpha} \rvert - \tau^{\alpha}_{\textrm{a}}}{\widetilde{\tau^{\alpha}_{\textrm{r}}}} \right)^p \right>^q \right) \quad \textrm{for} \quad \lvert \tau^{\alpha} \rvert > \tau^{\alpha}_{\textrm{a}}\, , &\\ \textrm{otherwise} \quad \dot{\gamma_\textrm{p}}^{\alpha}=0. &\\
\end{aligned}
\end{equation}
Here, the resolved shear stress is approximated to $\tau^{\alpha} = \mathbf{T}^{\textrm{e}}:\mathbf{\mathbb{S}}^{\alpha}_0$\footnote{Note that, in \cite{LEE2023103529}, the non-Schmid behavior (e.g., tension-compression asymmetry) in bcc materials was taken into account especially at low temperature. Accordingly, the resolved stresses (e.g., Equations \ref{eqn:slip} in Model 2) were modified to incorporate the additional non-Schmid stresses on the \{110\} planes (\cite{GROGER20085401,groger_multiscale_2008b,cho_anomalous_2018}). However, in this work, the three-term formulation is not considered.} from Equation \ref{eqn:RSS}. Furthermore, the temperature-dependent lattice resistance, $\widetilde{\tau^{\alpha}_{\textrm{r}}}$ is expressed by,
\begin{equation}
\widetilde{\tau^{\alpha}_{\textrm{r}}} = \tau^{\alpha}_{\textrm{r,0}} \frac{\mu}{\mu_0},
\end{equation}
where $\mu_0=\sqrt{\mathcal{C}_{44,0} \Big( \frac{\mathcal{C}_{11,0}-\mathcal{C}_{12,0}}{2} \Big)}$ and $\mu=\sqrt{\mathcal{C}_{44} \Big( \frac{\mathcal{C}_{11}-\mathcal{C}_{12}}{2} \Big)}$ are the effective shear moduli at 0K and at the current temperature, respectively. Moreover, the lattice resistance is assumed to be independent of the dislocation density in this model.

In Model 2, the modified Taylor hardening law is employed to represent interaction-type specific dislocation interaction coefficients. The evolution of the athermal resistance in the $\alpha$-th slip system is expressed by,
\begin{equation}
\tau^{\alpha}_{\textrm{a}} = \tau_{\textrm{a,0}} + \mu b \sqrt{\sum_{\beta=1}^{n_{\textrm{slip}}} a^{\alpha\beta} \rho^{\beta}},
\label{eqn:taylor-hardening}
\end{equation}
where $a^{\alpha\beta}$ is the dislocation interaction matrix whose elements are informed from the dislocation dynamics calculations of bcc crystals in \cite{MADEC2017166}. The dislocation density in each of the slip systems in Equation \ref{eqn:taylor-hardening} is taken to evolve according to a multiplication-annihilation type model of \cite{dequiedt_heterogeneous_2015},
\begin{equation}
\dot{\rho}^{\; \alpha} = \frac{1}{b} \left(\; \frac{1}{\mathcal{L}^{\alpha}} - 2y_c^{\alpha} \,\rho^{\,\alpha} \;\right) \left\lvert\dot{\gamma_\textrm{p}}^{\alpha}\right\rvert,
\end{equation}
where $\mathcal{L}^{\alpha}$ is the mean free path of dislocations, and $y_c^{\alpha}$ is the annihilation capture radius. The mean free path is inversely proportional to the forest dislocation density, i.e.,
\begin{equation}
\label{mean_free}
\frac{1}{\mathcal{L}^{\alpha}} = \sqrt{\sum_{\beta=1}^{n_{\textrm{slip}}} d^{\,\alpha\beta} \rho^{\,\beta}}\\,
\end{equation}
with $d^{\,\alpha\beta} = \frac{a^{\,\alpha\beta}}{k_1^2}$ for self interaction or coplanar interaction, and $d^{\,\alpha\beta} = \frac{a^{\,\alpha\beta}}{k_2^2}$ for other interactions, where $k_1$ and $k_2$ are the mean free path coefficients (\cite{devincre2008,dequiedt_heterogeneous_2015}). Moreover, the temperature- and rate-dependent annihilation capture radius is expressed by,
\begin{equation}
y_c^{\alpha} = y_{c0} \left( 1 - \frac{k_B \theta}{A} \textrm{ln} \left\lvert \frac{\dot{\gamma_\textrm{p}}^{\alpha}}{\dot{\gamma}_0}\right\rvert \right),
\label{eqn:capture_radius}
\end{equation}
where $y_{c0}$ is the reference annihilation capture radius following \cite{BEYERLEIN2008867}.

Both of Models 1 and 2 were implemented in the finite element program Abaqus. The kinematic variables (i.e., $\mathbf{L}^{{\textrm{p}}}$, $\mathbf{F}^{{\textrm{p}}}$ and $\mathbf{F}^{{\textrm{e}}}$), stresses (i.e., $\mathbf{T}^{{\textrm{e}}}$, $\mathbf{T}$) and state variables are updated via appropriate numerical integration algorithms (\cite{LUSCHER201763,kalidindi1992crystallographic}). Detailed descriptions of the numerical procedures can be found in \cite{LUSCHER201763} and \cite{LEE2023103529} for Models 1 and 2, respectively.

\section{Bayesian parameter inference and global sensitivity analysis}
\label{Sec:Bayesian}

Bayesian model calibration (BMC) introduced in \cite{kennedy2001bayesian} and \cite{doi:10.1198/016214507000000888} is a statistical framework that has been proven useful in quantifying uncertainty of complex physical models (\cite{ 10.1063/1.5051442,bernhard_bayesian_2019,NGUYEN2021104284,pogorelko_dynamic_2023}). We here utilize the BMC procedure to quantify uncertainties in material parameters of the crystal plasticity Models 1 and 2 using experimental data at various strain rates and temperatures. Subsequently, a variance-based global sensitivity analysis is conducted to further quantify the importance of each input material parameter in the model responses, thereby improving our understanding of the underlying deformation mechanisms in bcc molybdenum.

\subsection{Slip systems and fixed material parameters}
\label{Sec:para}

As noted in Section \ref{sec:intro}, active slip systems for plastic deformation in bcc refractory metals remain elusive (\cite{Seeger2002AnomalousMetals,lim_investigating_2020,dequiedt_slip_2023}). Slip in bcc crystals is known to occur in close-packed directions of the $\langle 111 \rangle$ on several possible normal planes ($\{ 110 \}$, $\{ 112 \}$ or $\{ 123 \}$; \cite{Seeger2002AnomalousMetals,weinberger_slip_2013}) or via pencil glides on the maximum resolved shear stress planes (\cite{lim_investigating_2020,weinberger_slip_2013}). Based on the experimental observations summarized in \cite{weinberger_slip_2013}, we considered the commonly accepted $\{ 110 \}\langle 111 \rangle$ and $\{ 112 \}\langle 111 \rangle$ slip systems in this study, which are listed in Table \ref{table:slip_system}.

\begin{table}
\centering
\caption{Slip systems for $\left\{110\right\}\langle 111 \rangle$ and $\left\{112\right\}\langle 111 \rangle$.}
\renewcommand{\arraystretch}{1.4}
\begin{tabular}{l@{\quad}l@{\quad}l @{\qquad \qquad} l@{\quad}l@{\quad}l}
\hline
\multicolumn{3}{c}{$\{ 110 \}$ slip systems} & \multicolumn{3}{c}{$\{ 112 \}$ slip systems}\\
\hline
$\alpha$ & $\mathbf{s}^{\alpha}_0$ & $\mathbf{m}^{\alpha}_0$ & $\alpha$& $\mathbf{s}^{\alpha}_0$ & $\mathbf{m}^{\alpha}_0$ \\
\hline
1 & [1$\overline{1}$1] & (110) & 13 & [$\overline{1}\overline{1}\overline{1}$] & (11$\overline{2}$)\\
2 & [$\overline{1}$11] & (110) & 14 & [$\overline{1}\overline{1}\overline{1}$] & (1$\overline{2}$1) \\
3 & [111] & (1$\overline{1}$0) & 15 & [$\overline{1}\overline{1}\overline{1}$] & ($\overline{2}$11) \\
4 & [11$\overline{1}$] & (1$\overline{1}$0) & 16 & [$\overline{1}$11] & (1$\overline{1}$2) \\
5 & [11$\overline{1}$] & (011) & 17 & [$\overline{1}$11] & (12$\overline{1}$) \\
6 & [1$\overline{1}$1] & (011) & 18 & [1$\overline{1}\overline{1}$] & (211) \\
7 & [$\overline{1}$11] & (01$\overline{1}$) & 19 & [1$\overline{1}$1] & ($\overline{1}$12) \\
8 & [111] & (01$\overline{1}$) & 20 & [$\overline{1}$1$\overline{1}$] & (121) \\
9 & [$\overline{1}$11] & (101) & 21 & [1$\overline{1}$1] & (21$\overline{1}$) \\
10 & [11$\overline{1}$] & (101) & 22 & [$\overline{1}\overline{1}$1] & (112) \\
11 & [111] & (10$\overline{1}$) & 23 & [11$\overline{1}$] & ($\overline{1}$21) \\
12 & [1$\overline{1}$1] & (10$\overline{1}$) & 24 & [11$\overline{1}$] & (2$\overline{1}$1) \\
\hline
\label{table:slip_system}
\end{tabular}
\end{table}

\begin{table}[h!]
\centering
\caption{Fixed crystal plasticity model parameters for bcc molybdenum. These parameters are not varied during calibration.}
\renewcommand{\arraystretch}{1.4}
\begin{tabular}{l@{\quad}l @{\qquad \quad} l@{\quad}l}
\hline
\multicolumn{4}{c}{Shared material parameters}\\
\hline
$C_{11, 0}$ [GPa]  & 484.26 & $k_{B}$ [J/K]                                        & 1.38 $\times$ 10$^{\textrm{-} 23}$\\
$C_{12, 0}$ [GPa]  & 165.7  & $\tau_{\textrm{a,0}}$ [MPa]  & 20.0 \\
$C_{44, 0}$ [GPa]  & 109.4  & $\tau^{\alpha}_{\textrm{r,0 for} \,  \{110\}}$ [MPa]    & 870.0  \\  
$C_{11,T}$ [MPa/K] & -50.5  & $\tau^{\alpha}_{\textrm{r,0 for} \, \{112, T\}}$ [MPa]  & 690.0  \\  
$C_{12,T}$ [MPa/K] & 5.71   & $\tau^{\alpha}_{\textrm{r,0 for} \, \{112, AT\}}$ [MPa] & 1184.0  \\  
$C_{44,T}$ [MPa/K] & -9.11  & $c$ [J/kg-K ] & 251.0 \\
$\rho_0$ [kg/m$^3$]  & 10200  & $\Delta G$ [J] & 1.859 $\times$ 10$^{\textrm{-} 19}$\\
$b$ [nm]   & 0.272  \\  
\hline
\multicolumn{4}{c}{Parameters for Mie-Gr{\"u}niesen equation of state}\\
\hline
\multicolumn{2}{l}{$C_0$ [m/s]} & \multicolumn{2}{l}{5140}\\
\multicolumn{2}{l}{$\Gamma_0$} & \multicolumn{2}{l}{1.59}\\
\multicolumn{2}{l}{$s$} & \multicolumn{2}{l}{1.255}\\
\hline
\multicolumn{4}{c}{Fixed parameters for Model 1}\\
\hline
$B_{0}$ [Pa $\cdot$ s] & 10$^{\textrm{-}5}$ &  $C_{for}$ & 0.1 \\
$f_{D}$ [$\mu$s$^{\textrm{-}1}$] & 10$^5$ & $C_{cop}$ & 0.4   \\ 
$C_{T}/C_{M}$ & 0.7 & $C_{11,P}$   & 6.41 \\
$c_s$ [m/s] & 3583.0 & $C_{12,P}$  & 3.454\\
$\mu_{\star}$ [GPa] & 132.0 & $C_{44,P}$         & 1.396 \\
\hline
\multicolumn{4}{c}{Fixed parameters for Model 2}\\
\hline
$a_{\textrm{copl}}$      & 0.06 & $a_{\textrm{J}}$   & 0.063\\
$a_{\textrm{colli}\,60\degree}$  & 0.6241 & $a_{\textrm{XJ}}$    & 0.047\\
$a_{\textrm{colli}\,90\degree}$   & 0.7225 & $k_1$ & 180\\
$a_{\textrm{colli}\,30\degree}$   & 0.3844\\
\hline
\label{table:para1}
\end{tabular}
\end{table}

The fixed material parameters for the crystal plasticity models are given in Table \ref{table:para1}. The mass density $\rho$, Burgers vector $b$, specific heat $c$, reference shear modulus $\mu_{\star}$, shear wave speed $c_s$ and melting temperature $\theta_m$ are well known through widely accepted material characterization procedures. The elastic constants at 0 K ($C_{11, 0}$, $C_{12, 0}$, and $C_{44, 0}$) and their temperature dependences ($C_{11, T}$, $C_{12, T}$, and $C_{44, T}$) were calculated via linear regression of experimentally measured elastic constants taken from \cite{10.1063/1.1728952}, and the pressure dependences of elastic constants ($C_{11, P}$, $C_{12, P}$ and $C_{44, P}$) were taken from \cite{katahara_pressure_1979}. It is noted that although $C_{12}$ increases with temperature in experiments, the rate of increase is negligibly small ($C_{12, T} \simeq 0$). The reference athermal resistance $\tau_{\textrm{a,0}}$ was taken to be the resolved shear stress at high temperatures reported in \cite{https://doi.org/10.1002/pssb.19660150214}. The activation energy $\Delta G$ was determined based on \cite{hollang_flow_1997} and \cite{dezerald_first-principles_2015}. Parameters for the Mie-Gr{\"u}niesen equation of state ($C_0$, $\Gamma_0$, $s$) were taken from \cite{clayton_nonlinear_2019}. The drag coefficient $B_0$ and attempt frequency $f_D$ for the dislocation velocity of Model 1 were taken from \cite{NGUYEN2021104284}. Although the parameters $C_{for}$, $C_{cop}$ and $C_{T}/C_{M}$ are not well known \textit{a priori}, they were held fixed to reduce the dimensionality of calibration parameter space. Furthermore, as noted in Section \ref{section: umat model}, the dislocation interaction coefficients in Model 2 ($a_{\textrm{copl}}$, $a_{\textrm{colli}\,60\degree}$, $a_{\textrm{colli}\,90\degree}$, $a_{\textrm{colli}\,30\degree}$, $a_{\textrm{J}}$, $a_{\textrm{XJ}}$) were informed by the recent dislocation dynamics simulations of \cite{MADEC2017166}. Here, the interaction coefficients are defined as $a_{\textrm{copl}}$: self- and coplanar interactions, $a_{\textrm{colli}\,60\degree}$: collinear interaction with $\theta = \arccos \lvert \mathbf{n}^{\alpha}_0 \cdot \mathbf{n}^{\beta}_0 \rvert = 60 \degree$, $a_{\textrm{colli}\,90\degree}$: collinear interaction with $\theta = \arccos \lvert \mathbf{n}^{\alpha}_0 \cdot \mathbf{n}^{\beta}_0 \rvert = 90 \degree$, $a_{\textrm{colli}\,30\degree}$: collinear interaction with $\theta = \arccos \lvert \mathbf{n}^{\alpha}_0 \cdot \mathbf{n}^{\beta}_0 \rvert = 30 \degree$, $a_{\textrm{J}}$: junctions between $\{$110$\}$ systems or $\{$112$\}$ systems, and $a_{\textrm{XJ}}$: junctions between $\{$110$\}$ and $\{$112$\}$ systems. Furthermore, the mean free path coefficient $k_1$ was taken from \cite{dequiedt_heterogeneous_2015}.

In this work, the twinning/anti-twinning (T/AT) asymmetry in the Peierls stress is also taken into account, following the recent approach in \cite{nguyen2021dynamic}. In \cite{nguyen2021dynamic}, the T/AT asymmetry was represented by the direction-dependent Peierls stress. The Peierls stresses on the $\{ 110 \}$ planes were assumed to be independent of the shearing direction, and those on the $\{ 112 \}$ plane have been known to be lower for slips in the twinning sense than for the anti-twinning sense. The Peierls stresses on the $\{110\}$ planes ($\tau^{\alpha}_{\textrm{r,0 for} \, \{110\}}$) and on the $\{ 112 \}$ planes in the twinning directions ($\tau^{\alpha}_{\textrm{r,0 for} \, \{112, T\}}$) were identified from the experimental work of \cite{hollang_work_2001}. Since experimental data for the Peierls stress on the $\{ 112 \}$ planes in the anti-twinning directions ($\tau^{\alpha}_{\textrm{r,0 for} \, \{112, AT\}}$) is not available in the literature, it was determined by multiplying the experimental measurement of $\tau^{\alpha}_{\textrm{r,0 for} \, \{112, T\}}$ by the ratio of $\tau^{\alpha}_{\textrm{r,0 for} \, \{112, AT\}}$ to $\tau^{\alpha}_{\textrm{r,0 for} \, \{112, T\}}$ from the atomistic simulation results reported in \cite{wang_generalized_2021} ($\tau^{\alpha}_{\textrm{r,0 for} \, \{112, AT\}} / \tau^{\alpha}_{\textrm{r,0 for} \, \{112, T\}} = 1.7$).

\subsection{Bayesian model calibration}
\label{subsection:Bayesian}
We then perform the Bayesian model calibration procedure using experimental data of the stress-strain behavior to obtain probabilistic estimates of the material parameters used in Models 1 and 2. The SEPIA package (Simulation-Enabled Prediction, Inference, and Analysis; \cite{james_gattiker_2020_4048801}), built upon the framework of \cite{doi:10.1198/016214507000000888}, was used for the BMC procedure. In the SEPIA package, the model parameters are calibrated simultaneously with a Gaussian process-based emulator, which is a statistical model that mimics the input-output relationship of a corresponding computational model with significantly faster evaluation speed. This joint calibration framework for the emulator and model parameters enables computationally efficient evaluation on posterior parameter distributions, which represent the belief of the model parameters, conditioned on the observed, experimental data. Three types of inputs are required for our BMC procedure: 
\begin{enumerate}
    \item Experimental observations. Here, we use the stress-strain data from uniaxial stress loading conditions.
    \item Prior samples that cover input parameter space. Here, the samples represent a uniform, non-informative distribution.
    \item Model responses corresponding to the parameter samples (crystal plasticity models in this work) under the same loading conditions of the experimental observations.
\end{enumerate}
First, numerical simulations using the physics-based constitutive models are conducted with parameter samples obtained from Latin hypercube sampling (\cite{Tang1993}). Then, the Gaussian process-based emulator is constructed by projecting the simulation outputs onto a principal-component basis via singular value decomposition of the simulation output matrix. The SEPIA package automatically computes the minimum number of basis vectors required to capture 99.5\% of the variance in the simulation results. The associated principal-component weights are modeled as the Gaussian processes over the calibration parameters and are inferred jointly with the model parameters using the observations through Markov chain Monte Carlo (MCMC) sampling processes (\cite{doi:10.1198/016214507000000888,higdon_simulationaided_2012,james_gattiker_2020_4048801}). By applying the Bayes’ rule via the Metropolis–Hastings acceptance criterion (\cite{10.1063/1.1699114,Chib1995UnderstandingAlgorithm}), the MCMC explores the high-probability regions in the input parameter space that best matches the experimental data. Once trained, the emulator enables probabilistic predictions at input configurations that were not included in the training process (\cite{doi:10.1198/016214507000000888}).

\begin{table}[t]
\centering
\renewcommand{\arraystretch}{1.3}
\caption{Prior parameter ranges for the Bayesian model calibration of the single crystal plasticity models, and the posterior expectations obtained from the calibration.}
\begin{tabular}{l@{\quad}l@{\quad}l@{\quad}l}
\hline
Parameter & Range & \makecell[l]{Expectation \\ for Model 1} & \makecell[l]{Expectation \\ for Model 2}\\
\hline
\multicolumn{4}{c}{Shared bounds}\\
\hline
$p$ & [0.3, 0.4] & 0.31 & 0.32\\
$q$ & [1.3, 1.5] & 1.45 & 1.41\\
$k_B \theta_m / A$ & [0.4, 1.0] & 0.73 & 0.84\\
log $\dot{\gamma}_0$ [s$^{\textrm{-1}}$] & [6.0, 9.0] & 6.88 & 7.05\\
\hline
\multicolumn{4}{c}{Bounds for Model 1}\\
\hline
$C_M$ & [3.0, 7.0] & 4.73 & \\
$\rho_{0, \, sat}$ [cm$^{\textrm{-2}}$] & [3.0, 7.0]$\times$10$^{11}$ & 5.83$\times$10$^{11}$ & \\
\hline
\multicolumn{4}{c}{Bounds for Model 2}\\
\hline
$y_{c0}$ & [15, 25] $b$ & & 21.28 $b$\\
$k_1$ & [1.0, 2.5] &  & 1.11\\
\hline
\label{table:para4}
\end{tabular}
\end{table}

\begin{figure}[b!]
    \centering
    \includegraphics[width=0.99\linewidth]{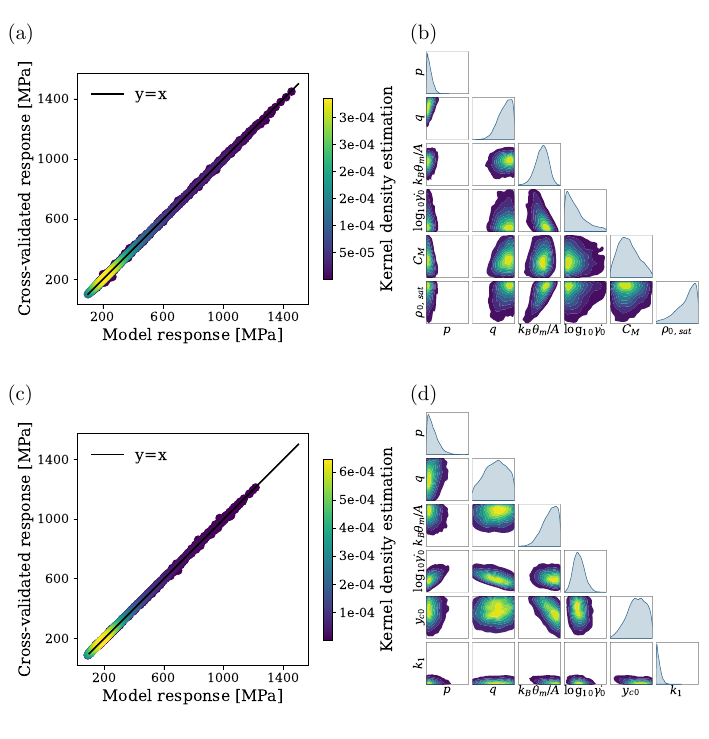}
    \caption{Comparison of cross-validated emulator predictions with the simulated stresses from the crystal plasticity models: (a) Model 1 and (c) Model 2. Each point on the scatter plot is colored according to its Gaussian kernel density estimate. The cross-validation comparisons show that the trained emulators exhibit excellent performance. Pair plots that represent posterior parameter distributions for (b) Model 1 and (d) Model 2. The plots on the diagonal of pair plots show the marginal probability distributions for each parameter, while the plots on the off-diagonal show the bivariate kernel density estimations between pairs of parameters. Overall, the model parameters are well calibrated through the BMC processes.}
    \label{fig1:pair}
\end{figure}

In this work, we calibrated six key parameters associated with the dislocation-mediated crystallographic slips in Models 1 and 2, and the parameter ranges are given in Table \ref{table:para4}. For each of the four parameters shared between the two models (i.e., $p,\,q,\,A$ and $\dot{\gamma}_0$), the bounds were taken to be identical in both models. Upper and lower bounds of all of the six parameters to be calibrated in each model were set by considering acceptable ranges in the literature (\cite{kothari_elasto-viscoplastic_1998,monnet_dislocation-dynamics_2013,LUSCHER201763,nguyen2021dynamic,LEE2023103529,dequiedt_slip_2023}), physical constraints (\cite{KOCKSUF1975,devincre2008}), and the numerical stability during the crystal plasticity simulations. We note that $A$ and $\dot{\gamma}_{0}$ were sampled from bounds over the functional form $k_B \theta_m / A$ and log $\dot{\gamma}_0$ due to the order of magnitude difference between their minimum and maximum values. The total initial dislocation density ($ \sum_{\alpha} \rho^{\alpha} $) was assumed to be $4.8 \times 10^{7}$ cm$^{-2}$ for all simulations. We also assumed an equal mixture of initial mobile and immobile dislocations especially in Model 1. In addition, the loading conditions (strain rate and temperature) of the experimental data used for the model calibration procedure are summarized in Table \ref{tab:loading_conditions}. Experimental data at quasi-static strain rates ($\dot{\epsilon} < 10^1 \, s^{-1} $) and at various temperatures (195 K to 500 K) were taken from \cite{https://doi.org/10.1002/pssb.19660150214}, \cite{https://doi.org/10.1002/pssb.19670190135}, \cite{https://doi.org/10.1002/pssa.2210220236} and \cite{Bulatov2006}\footnote{Since experimental data given in \cite{https://doi.org/10.1002/pssb.19660150214} and \cite{https://doi.org/10.1002/pssb.19670190135} are presented in terms of maximum resolved shear stresses (MRSS), we convert them into axial true stresses. See Ch. 5 of \cite{guiu1965plastic} for more details on the converting procedure.}. Experimental data at high strain rates ($\dot{\epsilon} > 10^1 \, s^{-1} $) and room temperature were taken from \cite{DAVIDSON1966703}. Although the initial dislocation densities may vary across different experimental studies, the same value was used in the simulations presented in this section. The model parameters and the emulator were calibrated via 120,000 MCMC drawings with experimental data and numerical simulation results corresponding to 100 parameter samples taken from uniform distributions over the parameter bounds.

Since the model parameters are jointly calibrated with the emulator, the emulator must achieve sufficient accuracy against the simulation results to yield a valid posterior parameter distribution. To validate the emulator, we performed Leave-One-Out Cross-Validation (\cite{hastie_elements_2009}). For each of the 100 simulation runs corresponding to parameter samples, the emulator is trained to the remaining 99 simulation runs and is used to predict the held-out run as in \cite{10.1063/1.5051442}, \cite{NGUYEN2021104284} and \cite{nguyen2021dynamic}. The emulator predictions and corresponding held-out simulation results are then compared to evaluate accuracy of the trained emulator. Figure \ref{fig1:pair} (a) compares the cross-validated responses predicted by the emulator with the simulation results for Model 1, and Figure \ref{fig1:pair} (c) presents the same comparison for Model 2. In both Figure \ref{fig1:pair} (a) and Figure \ref{fig1:pair} (c), the values aligned on the line of y = x indicate that the trained emulator accurately reproduces the physics model-based simulation results. Here, the coefficients of variation (COV) of the root mean square deviation were found to be less than 2 \% for all cross-validations in both models. This level of emulator accuracy is considered to be sufficient to obtain accurate posterior parameter distributions. Moreover, pair plots, which represent the shape of the posterior parameter distributions, are shown in Figures \ref{fig1:pair} (b) and (d), for Models 1 and 2 respectively. The plots on the diagonal present the marginal distribution of each parameter, and the off-diagonal plots present the bivariate kernel density estimation (KDE) which illustrates a correlation between the pairs of parameters. For Model 1, all parameters are well constrained and calibrated as shown in the diagonal marginal distributions, and $\rho_{0, \, sat}$, $\dot{\gamma}_{0}$ and $k_B \theta_m/A$ are linearly correlated with each other as shown in the bivariate KDEs. Parameters $p$ and $k_B \theta_m / A$ exhibit more constrained marginal distributions than the others, which indicates that the model output is more sensitive to these parameters. For Model 2, most parameters except $q$ are well constrained, and the lack of a strong peak in the marginal distribution of $q$ is attributed to its strong linear correlation with $\dot{\gamma}_{0}$ as shown in the corresponding bivariate KDE. As discussed in \cite{10.1063/1.5051442}, the marginal distributions that are concentrated near the predefined bounds (i.e., $p$, $\dot{\gamma}_{0}$ and $\rho_{0, \, sat}$ in Model 1 and $p$, $k_B \theta_m/A$ and $k_1$ in Model 2) indicate that there is missing physics in the corresponding model, or the predefined bounds in the prior distribution are poorly selected. However, adjusting the bounds for these parameters in the current setting resulted in numerical instability during some sample runs, divergence of the MCMC procedure, or the violation of physical constraints. For example, although the marginal distributions of $p$ in both models (Figures \ref{fig1:pair} (b) and (d)) show a strong preference for the lower bound, decreasing the lower bound of $p$ resulted in numerical issues without yielding meaningful improvements in the model performance. Moreover, further reduction of the lower bounds in $\dot{\gamma_0}$ and $k_1$, as well as increases in the upper bounds of $\rho_{0, \, sat}$ and $k_B \theta_m/A$, were found to be restricted by the acceptable ranges in the literature and physical constraints (\cite{KOCKSUF1975,devincre2008,LUSCHER201763,NGUYEN2021104284,nguyen2021dynamic}).

\begin{figure}[h!]
    \centering
    \includegraphics[width=1\linewidth]{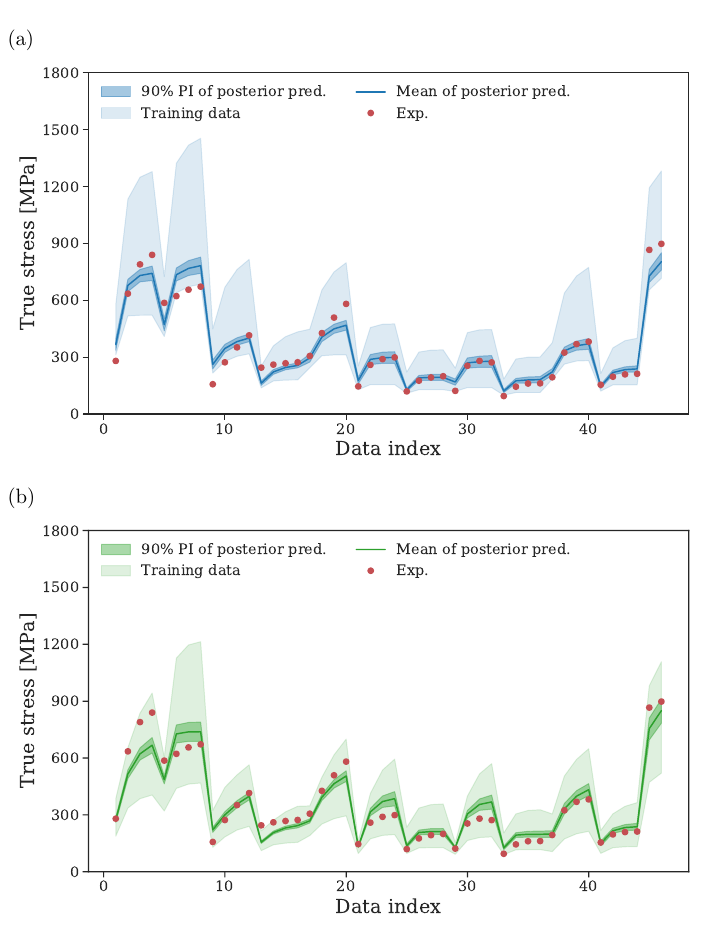}
    \caption{Stress-strain responses of single crystal molybdenum from experiments, 100 training simulations, and 90\% prediction interval (PI) of posterior emulator predictions after Bayesian calibration: (a) Model 1 and (b) Model 2. The solid lines show the mean of posterior emulator predictions. Information on the data is given in Table \ref{tab:loading_conditions}. The calibrated results show good agreement with the experiment data.}
    \label{fig2:emulator_prediction}
\end{figure}

\begin{figure}[h!]
    \centering
    \includegraphics[width=1\linewidth]{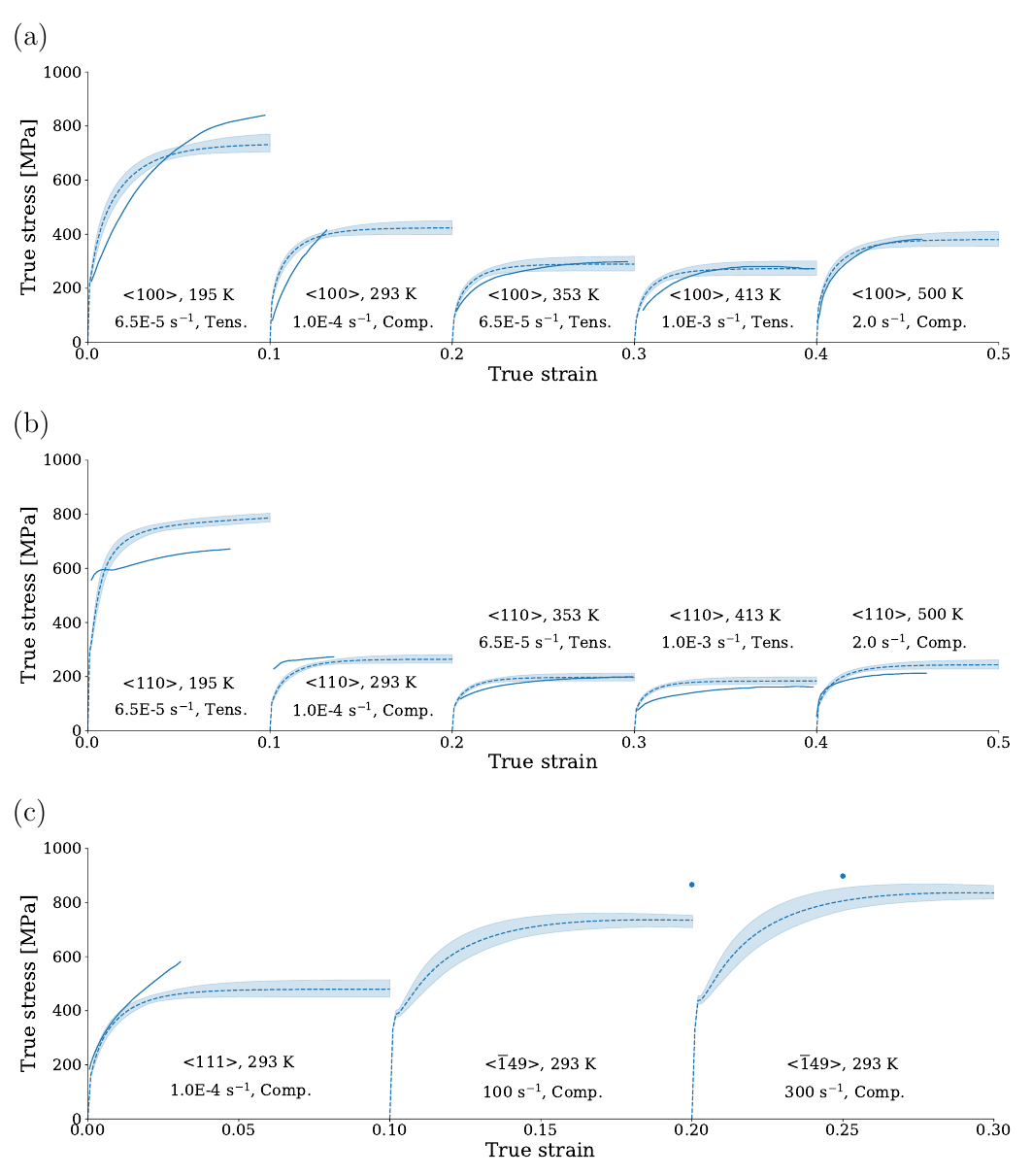}
    \caption{Stress-strain responses of single crystal molybdenum from experiments and Model 1 using 100 parameter samples taken from the posterior distribution (e.g., Figure \ref{fig1:pair} (b)): (a) $\langle100\rangle$, (b) $\langle110\rangle$, (c) $\langle111\rangle$ and $\langle\bar149\rangle$. The solid lines and dots represent the experimental data. The shaded regions show the 90\% credible interval of the simulated responses. The dashed lines show the mean of the simulated responses.}
    \label{fig3:computational_model_prediction}
\end{figure}

\begin{figure}[h!]
    \centering
    \includegraphics[width=1\linewidth]{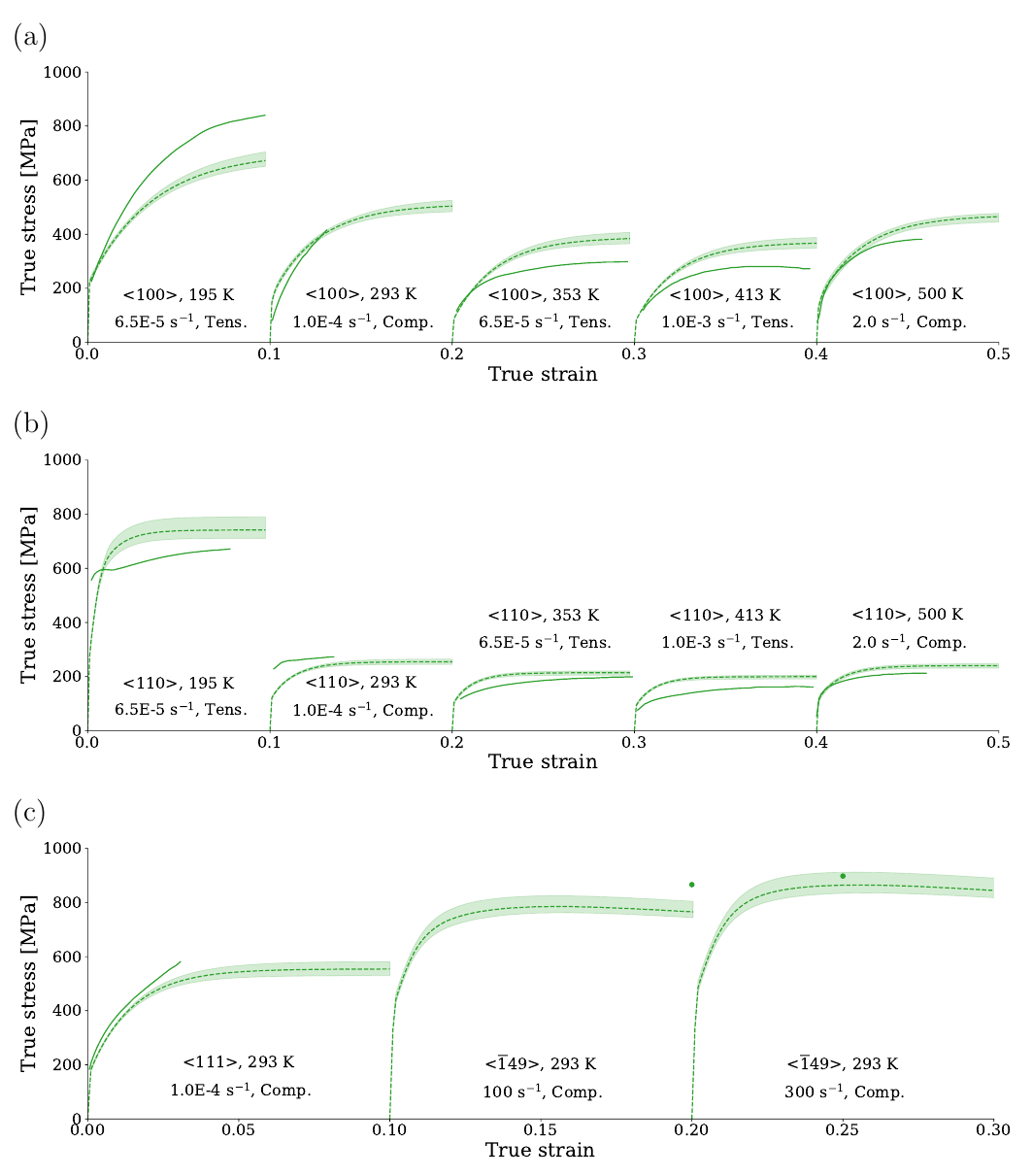}
    \caption{Stress-strain responses of single crystal molybdenum from experiments and Model 2 using 100 parameter samples taken from the posterior distribution (e.g., Figure \ref{fig1:pair} (d)): (a) $\langle100\rangle$, (b) $\langle110\rangle$, (c) $\langle111\rangle$ and $\langle\bar149\rangle$. The solid lines and dots represent the experimental data. The shaded regions show the 90\% credible interval of the simulated responses. The dashed lines show the mean of the simulated responses.}
    \label{fig4:computational_model_prediction}
\end{figure}

Figure \ref{fig2:emulator_prediction} shows a comparison between emulator predictions and experimental observations. The x-axis corresponds to the data index (observation number) given in Table \ref{tab:loading_conditions}, and the y-axis represents the axial stress at each data index. In the SEPIA package, the 90 \% prediction interval of stress data corresponding to 2000 posterior parameter samples and its mean are calculated in a few seconds using the emulator. Both calibrated models reproduce the experimental data reasonably well with RMSDs of 54.51 MPa (COV = 15.7 \%) in Model 1 and 62.30 MPa (COV = 17.8 \%) in Model 2, showing no significant difference in their performance. Further, we compare experimental data with the full crystal plasticity model-based simulations via finite elements. We calculated the responses for 100 parameter sets sampled from the Markov chain that represents posterior parameter distributions, and the 90 \% credible intervals of model responses are presented together with experimental data in Figures \ref{fig3:computational_model_prediction} and \ref{fig4:computational_model_prediction}. As shown, both models with the calibrated parameters capture the overall features in the stress-strain data well across a wide range of temperatures, strain rates, and crystallographic orientations. However, there are apparent discrepancies between experiments and numerical simulations for both Models 1 and 2, especially at low temperatures (195 K). Importantly, the Bayesian calibration procedure quantifies the parameter uncertainties conditional on the assumed physics-based model forms and available data. However, it should be noted that the BMC procedure cannot resolve the fundamental limitations in the assumed model forms. Furthermore, the Bayesian calibration procedure alone cannot clearly identify the sources of discrepancies between the experimental data and the model predictions using the calibrated parameter set. This highlights the need for a systematic procedure to identify the sources of uncertainties that account for these discrepancies. 

\begin{table}[t]
    \centering
    \renewcommand{\arraystretch}{1.2}
    \begin{tabular}{llll}
        \hline
        \textbf{Index} & \textbf{Orientation} & \textbf{Range of strain} & \textbf{Loading condition} \\
        \hline
        1$\sim$4   & $\langle$100$\rangle$ tension & 0.005$\sim$0.1 & \multirow{2}{*}{195 K, $6.5\times10^{-5}$ s$^{-1}$} \\
        5$\sim$8   & $\langle$110$\rangle$ tension & 0.005$\sim$0.08 & \\
        \hline
        9$\sim$12  & $\langle$100$\rangle$ compression & 0.005$\sim$0.03 & \multirow{3}{*}{293 K, $1.0\times10^{-4}$ s$^{-1}$} \\
        13$\sim$16 & $\langle$110$\rangle$ compression & 0.005$\sim$0.035 & \\
        17$\sim$20 & $\langle$111$\rangle$ compression & 0.005$\sim$0.03 & \\
        \hline
        21$\sim$24 & $\langle$100$\rangle$ tension & 0.005$\sim$0.095 & \multirow{2}{*}{353 K, $6.5\times10^{-5}$ s$^{-1}$} \\
        25$\sim$28 & $\langle$110$\rangle$ tension & 0.005$\sim$0.1 & \\
        \hline
        29$\sim$32 & $\langle$100$\rangle$ tension & 0.005$\sim$0.096 & \multirow{2}{*}{413 K, 0.001 s$^{-1}$} \\
        33$\sim$36 & $\langle$110$\rangle$ tension & 0.005$\sim$0.097 & \\
        \hline
        37$\sim$40 & $\langle$100$\rangle$ compression & 0.005$\sim$0.057 & \multirow{2}{*}{500 K, 2 s$^{-1}$} \\
        41$\sim$44 & $\langle$110$\rangle$ compression & 0.005$\sim$0.060 & \\
        \hline
        45 & \multirow{2}{*}{$\langle \overline{1}49 \rangle$ compression} & 0.1 & 293 K, 100 s$^{-1}$ \\
        46 &  & 0.05 & 293 K, 300 s$^{-1}$ \\
        \hline
    \end{tabular}
    \caption{Crystallographic orientations and loading conditions in the data used for the Bayesian model calibration. The 195 K and 353 K data were taken from \cite{https://doi.org/10.1002/pssb.19660150214}, the 293 K data were taken from \cite{https://doi.org/10.1002/pssa.2210220236}, the 413 K data were taken from \cite{https://doi.org/10.1002/pssb.19670190135}, and the 500 K data were taken from the supplementary information of \cite{Bulatov2006}. The data at high strain rates ($\dot{\epsilon} \geq 10^2$ s$^{-1} $) and room temperature were taken from \cite{DAVIDSON1966703}.}
    \label{tab:loading_conditions}
\end{table}

\clearpage
\subsection{Global sensitivity analysis}
\label{sec:sobol}

Variance-based global sensitivity analysis quantifies the variance proportion in the model output ($Y$) attributed to a specific input parameter ($X_i$) (\cite{sobol1993sensitivity,sobol_global_2001}). Despite numerous modeling efforts to better understand the deformation mechanism of bcc single crystals, a detailed analysis of how each model input (e.g., loading conditions, strain levels, material parameters) affects the output or response has been largely lacking for the complex dislocation density-based bcc single crystal plasticity models. Prior work on constitutive modeling with the BMC procedure has focused on either scaling up to large-scale numerical simulations using a single representative or optimal parameter set taken from the posterior parameter distributions or assessing extrapolative performance via probabilistic predictions from the posterior (e.g., \cite{10.1063/1.5051442,NGUYEN2021104284,nguyen2021dynamic}). However, these BMC procedures do not provide a systematic diagnosis of why model predictions succeed or fail. In particular, quantifying residual sources of predictive uncertainties remains difficult in the extrapolative applications. This motivates us to incorporate the global sensitivity analysis to further quantify the contributions of these uncertainties to the model prediction performance and to evaluate the validity of model-specific, governing deformation mechanisms for explaining the deformation behavior at various strain rates and temperatures. The global sensitivity analysis enables a detailed examination of how each model reproduces the experimental data by quantifying the relationships between input parameter uncertainty and model outputs. This procedure allows us to infer the roles of individual model components and to better diagnose fundamental differences between the two models and the associated deformation mechanisms beyond simple comparisons of experimental data and simulation results with calibrated model parameters. 

The first-order sensitivity index, or main-effect Sobol' index (\cite{sobol1993sensitivity}), is defined as,
\begin{equation}
\label{eq:sobol}
    S_i=\frac{V_{X_i}\left(E_{\boldsymbol{X}_{\sim i}}\left[Y \mid X_i\right]\right)}{V(Y)},
\end{equation}
where $E[\cdot]$ is the expectation, $V[\cdot]$ is the variance, and $E_{\boldsymbol{X}_{\sim i}}[\cdot]$ denotes the expectation taken over all input except $X_i$. Here, $E_{\boldsymbol{X}_{\sim i}}\left[Y \mid X_i\right]$ denotes the conditional expectation for fixed $X_i$, and $V_{X_i}\left(E_{\boldsymbol{X}_{\sim i}}\left[Y \mid X_i\right]\right)$ denotes its variance over $X_{i}$. $V_{X_i}\left(E_{\boldsymbol{X}_{\sim i}}\left[Y \mid X_i\right]\right)$ quantifies how the average model output $Y$ varies with respect to $X_{i}$. $V(Y)$ is the total variance of $Y$. For a given range of input parameters (i.e., Table \ref{table:para4} in this work), the first-order sensitivity index $(S_i)$ quantifies the normalized effect of varying a single input parameter $(X_i)$ over its domain, averaged over the variability of all other input parameters. In the limiting case, i.e., $S_i = 0$ implies that the output is completely insensitive to $X_i$, while $S_i = 1$ indicates that the variance of output depends solely on $X_i$. Furthermore, the sensitivity analysis can be extended to analyze high-order effects due to variable interactions. For instance, the second-order sensitivity index is defined as, 
\begin{equation}
\label{eq:sobol2nd}
    S_{ij} = \frac{ V_{X_{ij}} \left( E_{\boldsymbol{X}_{\sim ij}}[Y \mid X_i, X_j] \right) }{ V(Y) } - S_i - S_j,
\end{equation}
where $E_{\boldsymbol{X}_{\sim ij}}[Y \mid X_i, X_j]$ denotes the conditional expectation for fixed $X_i$ and $X_j$, and $V_{X_{ij}} ( E_{\boldsymbol{X}_{\sim ij}}[Y \mid X_i, X_j] )$ denotes its variance over $X_i$ and $X_j$. Since $V_{X_{ij}} ( E_{\boldsymbol{X}_{\sim ij}}[Y \mid X_i, X_j] )$ quantifies how the expected model response $Y$ varies with respect to the $X_i$ and $X_j$, $S_{ij}$ quantifies the contribution of the interaction between two input variables to the output variance, excluding all first-order sensitivity effects. In a similar way, it can be extended to higher-dimensional indices. According to the Sobol' variance decomposition theorem (\cite{sobol1993sensitivity}), the sum of the first-order sensitivity indices and all higher-order sensitivity indices is equal to one. 

In this work, we utilized a Python implementation of Bayesian Adaptive Spline Surfaces (pyBASS; \cite{JSSv094i08}) for the global sensitivity analysis. The pyBASS framework is based on the Bayesian Multivariate Adaptive Regression Spline (\cite{friedman_multivariate_1991,denison_bayesian_1998}), which represent model responses as a sum of tensor-product polynomial spline bases (i.e., piecewise polynomials) with the variables, knots, and interaction orders learned from data. Accordingly, the reversible-jump MCMC of \cite{green_reversible_1995} is used to determine the variables, knot locations, and interaction orders via birth (add), death (remove), and change (modify) moves on the spline basis functions. In the early studies on the global sensitivity analysis, the calculation of sensitivity indices relied heavily on the Monte Carlo estimation of variance components, which required numerous runs of computationally expensive models for convergence (\cite{sobol1993sensitivity,sobol_global_2001}). In contrast, once the pyBASS surrogate is well constructed with fewer model runs, sensitivity indices can be computed analytically since the spline basis of the pyBASS surrogate admits closed-form integrals. Moreover, the pyBASS supports the functional global sensitivity analysis via the functional Sobol’ decomposition (\cite{francom_sensitivity_2017,JSSv094i08}); instead of a single sensitivity index per each parameter, it returns sensitivity index curves $S_i(\epsilon)$ along the functional argument $\epsilon$ (e.g., strain or time) by conditioning the variance decomposition on $\epsilon$ (i.e., not integrating over $\epsilon$). Thus, we constructed the pyBASS surrogate using the simulation results from the same 100 parameter samples used in our BMC procedures presented in Section \ref{subsection:Bayesian} and calculated the sensitivity indices using the pyBASS surrogate. 

\begin{figure}[t!]
    \centering
    \includegraphics[width=0.998\linewidth]{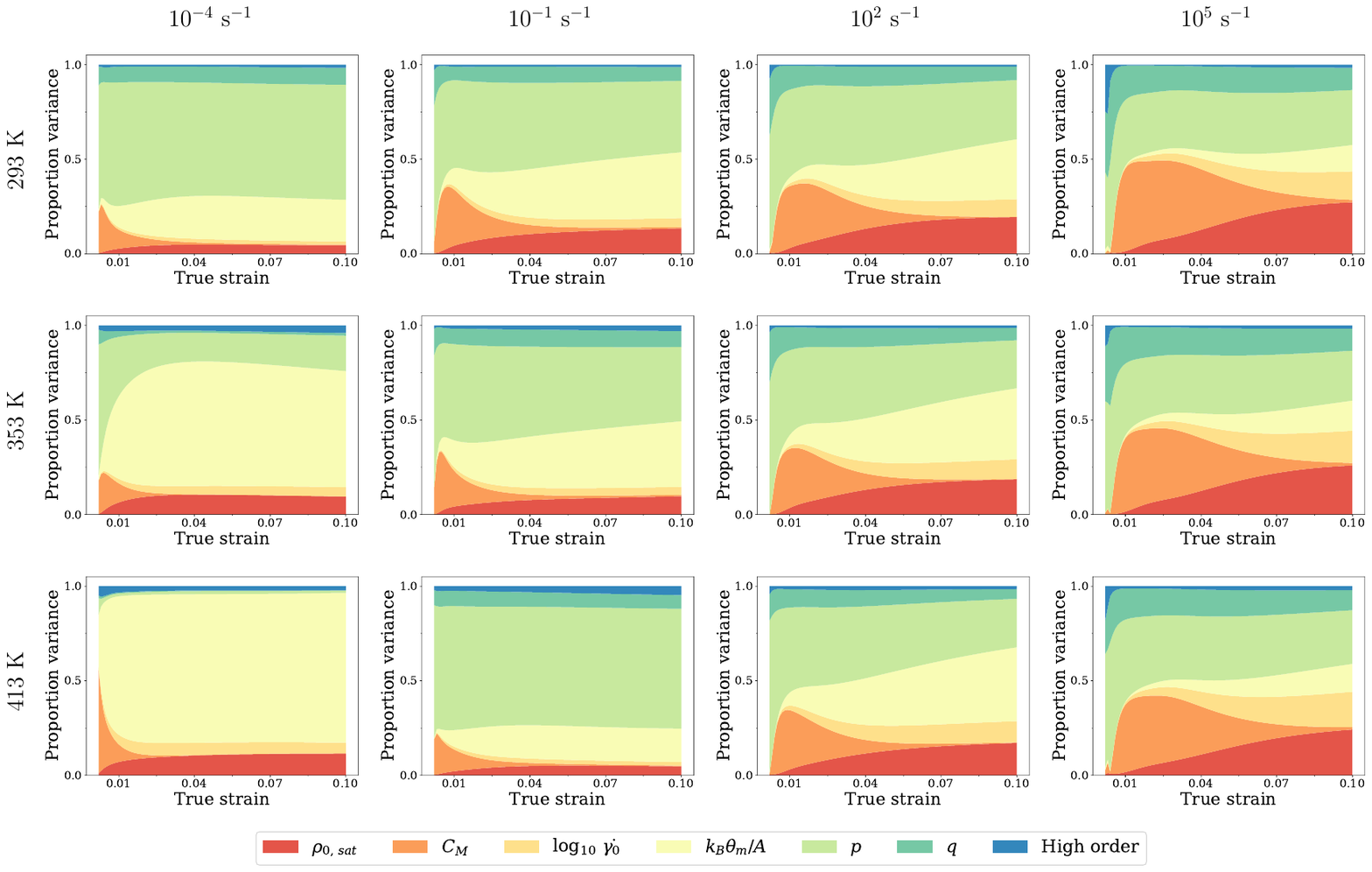}
    \caption{Functional global sensitivity analysis results for [100] single crystal molybdenum using Model 1. Each color band indicates the evolution of the sensitivity index for a given parameter as a function of strain.}
    \label{fig5:GSA}
\end{figure}

\begin{figure}[t!]
    \centering
    \includegraphics[width=0.998\linewidth]{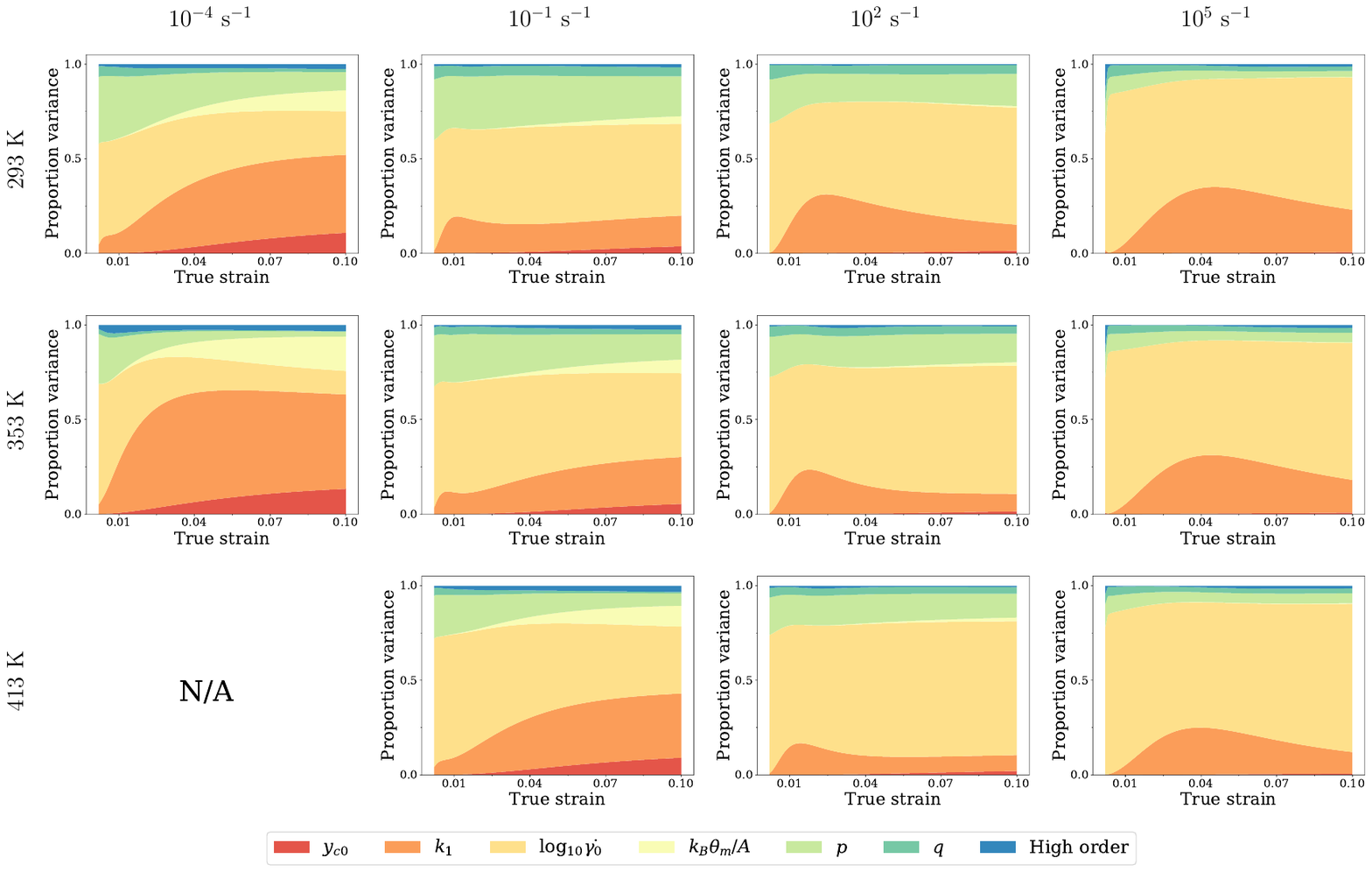}
    \caption{Functional global sensitivity analysis results for [100] single crystal molybdenum using Model 2. Each color band indicates the evolution of the sensitivity index for a given parameter as a function of strain.}
    \label{fig6:GSA}
\end{figure}

We performed the functional global sensitivity analysis on the compressive responses of [100] molybdenum at the 12 loading cases consisting of 4 strain rates (10$^{-4}$, 10$^{-1}$, 10$^{2}$ and 10$^{5}$ s$^{-1}$) and 3 temperatures (293, 353 and 413 K). Figures \ref{fig5:GSA} and \ref{fig6:GSA} show the variations of sensitivity indices as functions of an imposed strain (x-axis) calculated from the functional global sensitivity analysis at each loading condition for Models 1 and 2, respectively. The vertical height of each color band represents the sensitivity index magnitude of the corresponding parameter at the strain level indicated on the x-axis. The higher-order effects were found to have an insignificant impact on the simulation results and are thus incorporated into the blue-colored bands. As shown, all calibration parameters in both models have non-negligible importance, and some of the sensitivity indices (e.g., $C_M$ and $\rho_{0,\mathrm{sat}}$ in Model 1, and $k_1$ and $\dot{\gamma_0}$ in Model 2) are highly correlated with strain rate and temperature. In Model 1 (Figure \ref{fig5:GSA}), the dislocation multiplication coefficient ($C_M$) in Equation \ref{eq:multi} and the parameters ($p$ and $q$) associated with the thermally activated velocity of dislocations in Equation \ref{eq:thermal_act} are shown to be mainly responsible for uncertainty during the initial stage of plastic deformation. As the strain rate increases, $C_M$ has a stronger influence in the early stage of deformation and remains influential to higher strain levels. The strong influence of $C_M$ in the early stage of deformation, especially at high strain rates, indicates that the rate of the mobile dislocation density evolution must be accurately captured to predict the initial yield strength and early stage hardening behavior. The mobile dislocation density significantly influences the dislocation velocity required to accommodate the desired plastic strain rate via Equation \ref{eq_model1:orowan} in Model 1. This relationship between the mobile dislocation density and the dislocation velocity directly impacts the early stage responses in Model 1 as the dislocation velocity is fundamentally stress-dependent. This influence gradually decreases as the dislocation density evolves toward its saturation value of dislocation densities ($\rho_{I, sat}^{\alpha}$ in Equation \ref{eq:sat_dd} and $\rho_{M, sat}^{\alpha} $). Moreover, the model output is more sensitive to the reference saturation dislocation density ($\rho_{0,\mathrm{sat}}$) at high strain rates, indicating that the saturation value of the dislocation, in addition to its evolution rate, must be accurately determined to capture the rate-dependent hardening features observed in the experimental data in Figures \ref{fig3:computational_model_prediction} and \ref{fig4:computational_model_prediction}. We also found that the influence of the parameter $\dot{\gamma_0}$ associated with the rate-dependence of the saturation dislocation density increases with strain rate; further, the parameter $k_B \theta_m/A$ associated with the temperature-dependence of the saturation dislocation density increases with temperature. The negligible influence of $p$ and $q$ and the strong impact of $k_B \theta_m/A$ on the model output at 473 K and 10$^{-4}$ s$^{-1}$ indicate that the thermal energy at these conditions is sufficient to overcome the dislocation barrier through the thermal activation processes without any applied stress. Taken together, Figures \ref{fig1:pair} (b) and \ref{fig5:GSA} show that the parameters which play crucial roles at high strain rates ($C_{\mathrm{M}}$, $\rho_{0,\mathrm{sat}}$, and $\dot{\gamma}_{0}$) exhibit relatively wider posterior marginal (diagonal) distributions, reflecting substantial posterior uncertainty. This is attributed to limited calibration data (only two data points) at high strain rates, which results in less constrained posterior marginal distributions.

As shown in Figure \ref{fig6:GSA}, the influence of parameters associated with dislocation density evolution ($y_{c0}$, $k_1$, and $k_B \theta_m/A$) in Model 2 is highest at the lowest strain rate. At high strain rates, $\dot{\gamma_0}$ primarily contributes to scaling the thermally-activated flow rule (Equation \ref{eqn:slip}) rather than determining the rate dependence of dislocation density annihilation (Equation \ref{eqn:capture_radius}) as the model response becomes insensitive to the other annihilation-related parameters ($y_{c0}$ and $k_B \theta_m/A$) at high strain rates. The diminished influence of $y_{c0}$ and $k_B \theta_m/A$ at high strain rates is attributed to the decrease in the value of $y_{c}^{\alpha}$ as the slip rate increases toward $\dot{\gamma_0}$, as depicted in Equation \ref{eqn:capture_radius} in Model 2. Moreover, the influence of $k_1$ in Model 2 evolves more slowly than that of $C_M$ in Model 1, indicating that the initial yield strength and early stage hardening behavior are not highly sensitive to the dislocation density evolution in Model 2. While the high influence of $C_M$ in Model 1 arises from its treatment of mobile dislocations as active mediators of plastic deformation via Equation \ref{eq_model1:orowan}, the slower increase in the influence of $k_1$ indicates that Model 2 primarily employs the dislocation density for strain hardening rather than for both strain hardening and plastic strain accommodation. Instead, the fixed pre-exponential factor $\dot{\gamma}_0$ controls the stress at the onset of yield in Model 2. Since $\dot{\gamma}_0$ is directly proportional to the slip rate (Equation \ref{eqn:slip} in Model 2), which is analogous to the role of the mobile dislocation density employed in Model 1 (Equation \ref{eq_model1:orowan}), it controls the initial responses. At low strain rates (10$^{-4}$ and 10$^{-1}$ s$^{-1}$), the influence of $p$, $q$, and $\dot{\gamma_0}$ decreases with increasing temperature, and simulations using some parameter samples fail to converge especially at 413 K and 10$^{-4}$ s$^{-1}$. This high temperature and low strain rate condition is close to the limit of applicability of an Arrhenius-type rate equation (e.g., Equation \ref{eqn:slip} in Model 2) with a fixed pre-exponential factor (\cite{KOCKSUF1975, kothari_elasto-viscoplastic_1998}) where the thermal energy is sufficient to overcome the barrier through thermal activation alone without any applied stress. Furthermore, the relatively weak influence of $q$ and $y_{c0}$ is consistent with their broader posterior marginal distributions displayed in Figure \ref{fig1:pair} (d).

In summary, our global sensitivity analysis results have shown apparent mechanistic differences between Model 1 and Model 2. Although the difference in residual errors after the BMC procedure presented in Section \ref{subsection:Bayesian} was found to be negligible (see Figure \ref{fig2:emulator_prediction} on Model 1 and Model 2), sensitivity to dislocation density evolution at the onset of plastic flow differs significantly in the two models. The initial responses of Model 1 are characterized by an evolving mobile dislocation density proportional to the slip rate while those of Model 2 are mainly characterized by a non-evolving pre-exponential factor $\dot{\gamma}_0$. These indicate that the Orowan relation produces the higher sensitivity to the structure of the dislocation density evolution law at the onset of yield, especially in Model 1. Furthermore, under high temperature (413 K) and low strain rate (10$^{-4}$ s$^{-1}$) conditions where the thermal energy alone is sufficient for the barrier crossing without any applied stress, not only does the influence of $p$ and $q$ become negligible in Model 1 but they cause Model 2 to fail to converge. The global sensitivity analysis has identified potential sources of predictive uncertainties arising from data sparsity or the negligible influence of parameters on the model output in the calibration data range. The results presented in Figure \ref{fig5:GSA} show that high strain rate data are currently insufficient to infer the parameters that govern the high strain rate behavior in Model 1 ($C_{\mathrm{M}}$, $\rho_{0,\mathrm{sat}}$, and $\dot{\gamma}_{0}$). In the next section, we further assess the influence of these residual uncertainties on the model predictions for molybdenum single crystals at more extreme loading conditions.

\section{Application: plate impact test}
\label{sec:plate_impact}
\subsection{Experiments and simulation setting}

Here, we further examine the model-specific deformation mechanisms under mechanical extremes by validating the calibrated Models 1 and 2 against the plate impact experiments on [100] single crystal molybdenum by \cite{10.1063/5.0082267}. In their experiments, the single crystal molybdenum specimens of thickness ranging from 0.2 to 8 mm were impacted at 350 ± 10 m/s by the 0.1 to 2.5 mm thick half-hard copper impactors glued to the 5 mm polymethyl methacrylate (PMMA) substrates. To investigate the role of the initial dislocation density on the extreme strain rate behavior ($>$ 10$^{4}$ s$^{-1}$) and spallation, experiments were conducted on three types of molybdenum targets with different levels of prestrains. The measured dislocation densities were $\sum_{\alpha} \rho^{\alpha} = 8 \times 10^4$ cm$^{-2}$ for pristine specimens, $\sum_{\alpha} \rho^{\alpha} = 2.7 \times 10^6$ cm$^{-2}$ for specimens prestrained to 0.6 \%, and $\sum_{\alpha} \rho^{\alpha} = 6.5 \times 10^7$ cm$^{-2}$ for specimens prestrained to 5.4 \%.

The bcc crystal plasticity models (Models 1 and 2) with the Mie-Gr{\"u}niesen equation of state are used for the single crystal molybdenum target, and the Johnson-Cook strength model with the Mie-Gr{\"u}niesen equation of state is employed for the copper impactors in the plate impact simulations. The Johnson-Cook model was calibrated and validated via the proper BMC procedure against experimental data collected at Los Alamos National Laboratory, as summarized in Appendix \ref{appendix:JC}. One-dimensional arrays of three-dimensional hexahedral elements with uniaxial strain constraint are used to model specimens and impactors (\cite{luscher_model_2013,LUSCHER201763}). The number of elements is adjusted depending on the total thickness of the specimen to achieve an element size of approximately 7.7 $\upmu$m for the impactor and 10 $\upmu$m for the target. Frictionless contact between the impactor and the target is assumed to replicate the impact conditions. When employing Model 1, the initial mobile dislocation densities ($\sum_{\alpha} \rho^{\alpha}_M$) are assumed to be $3.2 \times 10^4$ cm$^{-2}$ for pristine, $3.0 \times 10^5$ cm$^{-2}$ for 0.6 \% prestrained and $1.7 \times 10^7$ cm$^{-2}$ for 5.4 \% prestrained case, and initial immobile dislocation densities ($\sum_{\alpha} \rho^{\alpha}_I$) are calculated by simply subtracting the initial mobile dislocation densities from the measured total initial dislocation densities. Since both models used in this work do not account for void growth and its interaction with dislocations (c.f. \cite{wilkerson_dynamic_2014,nguyen_dislocation-based_2017,bronkhorst_local_2021,zhang_data-driven_2023,schmelzer_statistical_2025,schmelzer_thermodynamically_2025}), we simulated the impact behavior up to the point where the shock wave reaches the specimen free surface and begins to be released.

\subsection{Plate impact simulation and global sensitivity analysis}
\label{sec:plate_sim_gsa}

\begin{figure}[b!]
    \centering
    \includegraphics[width=0.65\linewidth]{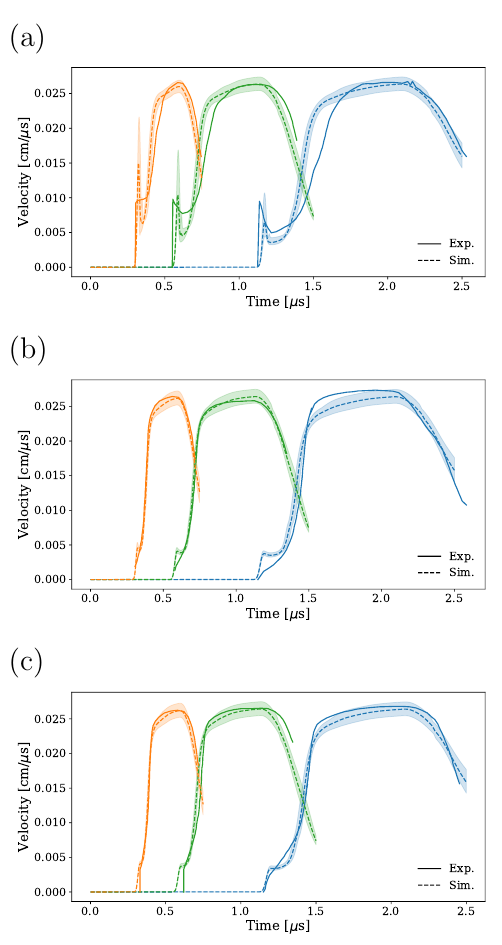}
    \caption{Plate impact responses of [100] single crystal molybdenum predicted using Model 1: (a) pristine, (b) 0.6\% prestrained, and (c) 5.4\% prestrained case. The solid lines represent the experimental data. The shaded regions show the 90\% credible interval of the simulated responses. The dashed lines show the mean of the simulated responses. The orange line corresponds to a copper impactor thickness of 0.78 mm and a molybdenum target thickness of 2.1 mm; the green line to 1.48 mm and 4 mm; and the blue line to 2.5 mm and 8 mm, respectively. The plate impact responses of Model 1 are strongly dependent on the initial dislocation density.}
    \label{fig7:shock_nguyen}
\end{figure}

In addition to the parameter uncertainties of the molybdenum target and copper impactor quantified via the BMC procedures, the uncertainty in the impact velocity (V$_{imp}$) in experiments of \cite{10.1063/5.0082267} was also considered in our plate impact simulations. To this end, we sampled 100 parameter sets from the Markov chains constructed during the BMC procedure and drew impact velocity values from a uniform distribution between 340 and 360 m/s. Then, we compare the experimental and simulated free surface velocities of the molybdenum target. As the free surface velocity is directly proportional to the longitudinal stress, i.e., 
\begin{equation}
\sigma_L = \frac{1}{2}  \rho_0 C_L V_{fs},
\end{equation}
the ability of the model to accurately predict the free surface velocity profile directly validates its predictive capability for yield strength and elastic-plastic transient behavior under these plate impact conditions. Here, $ C_L$ is the longitudinal wave speed, and $V_{fs}$ is the free surface velocity. Figure \ref{fig7:shock_nguyen} shows the free surface velocity profiles obtained from the plate impact simulations using Model 1 with the 100 samples, along with the experimental data from \cite{10.1063/5.0082267}. Figures \ref{fig7:shock_nguyen} (a), (b), and (c) correspond to the pristine, 0.6 \% prestrained, and 5.4 \% prestrained cases, respectively. The thicknesses of the copper impactor and molybdenum target are 0.78 mm and 2.1 mm for the orange line, 1.48 mm and 4 mm for the green line, and 2.5 mm and 8 mm for the blue line, respectively. As the onset of the yield in Model 1 is highly sensitive to the initial dislocation density and its evolution rate, the height of the elastic precursor predicted by Model 1 is also found to be highly dependent on the initial dislocation density. As the initial mobile dislocation density increases, the dislocation velocity to accommodate the imposed strain rate decreases due to Equation \ref{eq_model1:orowan} in Model 1 and the elastic-plastic transition occurs more smoothly. Moreover, both the rise time to the Hugoniot state (the plateau following the elastic precursor) and the peak velocity fall within the predicted range overall. However, the model does not capture the invariance of the elastic precursor amplitude with sample thickness, a distinctive feature of single crystal molybdenum under shock compression reported by \cite{mandal_elastic-plastic_2017} and \cite{10.1063/5.0082267}. Even after further testing with various parameter combinations, the model failed to capture the invariance of the elastic precursor amplitude. Assuming low uncertainty in velocity measurements, the results indicate that this is an inherent limitation of Model 1.

\begin{figure}[t!]
    \centering
    \includegraphics[width=0.65\linewidth]{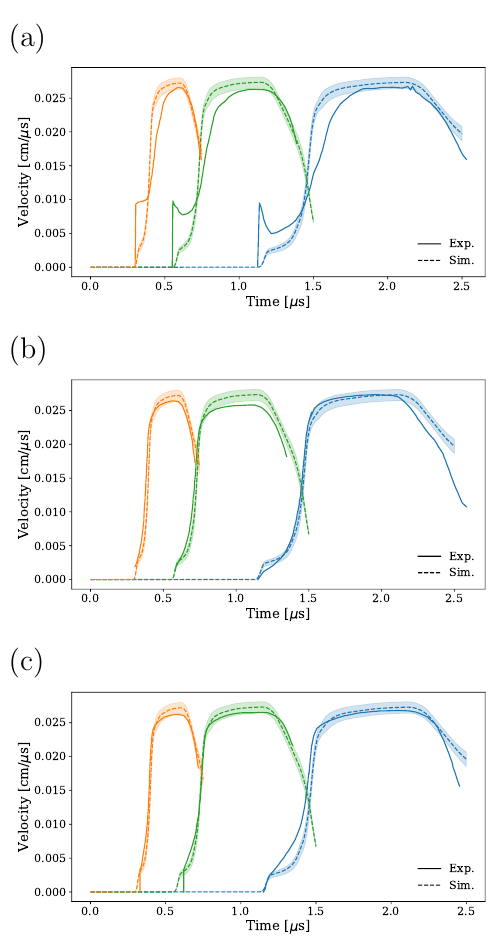}
    \caption{Plate impact responses of [100] single crystal molybdenum predicted using Model 2: (a) pristine, (b) 0.6\% prestrained, and (c) 5.4\% prestrained case. The solid lines represent the experimental data. The shaded regions show the 90\% credible interval of the simulated responses. The dashed lines show the mean of the simulated responses. The orange line corresponds to a copper impactor thickness of 0.78 mm and a molybdenum target thickness of 2.1 mm; the green line to 1.48 mm and 4 mm; and the blue line to 2.5 mm and 8 mm, respectively. The plate impact responses of Model 2 are nearly independent of the initial dislocation density.}
    \label{fig8:shock_lee}
\end{figure}

Figure \ref{fig8:shock_lee} shows the plate impact simulation results with Model 2, using identical layouts to Figure \ref{fig7:shock_nguyen}. Model 2 exhibits no yield drop associated with a stress relaxation behind the elastic precursor especially in the pristine case, and the plate impact response of Model 2 is shown to be nearly insensitive to the initial dislocation density. This observation is consistent with the global sensitivity analysis results under simple stress conditions presented in Figure \ref{fig6:GSA}, which demonstrated that the initial yield behavior of Model 2 was found to be almost insensitive to both the initial dislocation density and its evolution rate. While the simple stress–strain behavior is reasonably captured by the thermally-activated glide mechanism as shown in Figures \ref{fig3:computational_model_prediction} (Model 1) and \ref{fig4:computational_model_prediction} (Model 2), the comparison of plate impact responses between Models 1 and 2 clearly show that two additional mechanisms are essential for capturing the elastic–plastic transition at extreme strain rates: (1) the coupling between the mobile dislocation density and slip rates (Equation \ref{eq_model1:orowan} in Model 1) and (2) a limiting dislocation velocity in the drag-dominated regime (Equation \ref{eqn:running} in Model 1). Furthermore, the comparison of Figures \ref{fig7:shock_nguyen} (c) and \ref{fig8:shock_lee} shows that the plate impact response of Model 1 becomes very close to that of Model 2 with an increasing initial dislocation density (e.g., Figures \ref{fig7:shock_nguyen} (c) and \ref{fig8:shock_lee} (c)). This further demonstrates that Model 2 is valid only when rapid dislocation evolution is not required to accommodate imposed plastic strain rates, i.e., in 0.6 \% and 5.4 \% prestrained specimens in this work.

\begin{figure}[b!]
    \centering
    \includegraphics[width=0.995\linewidth]{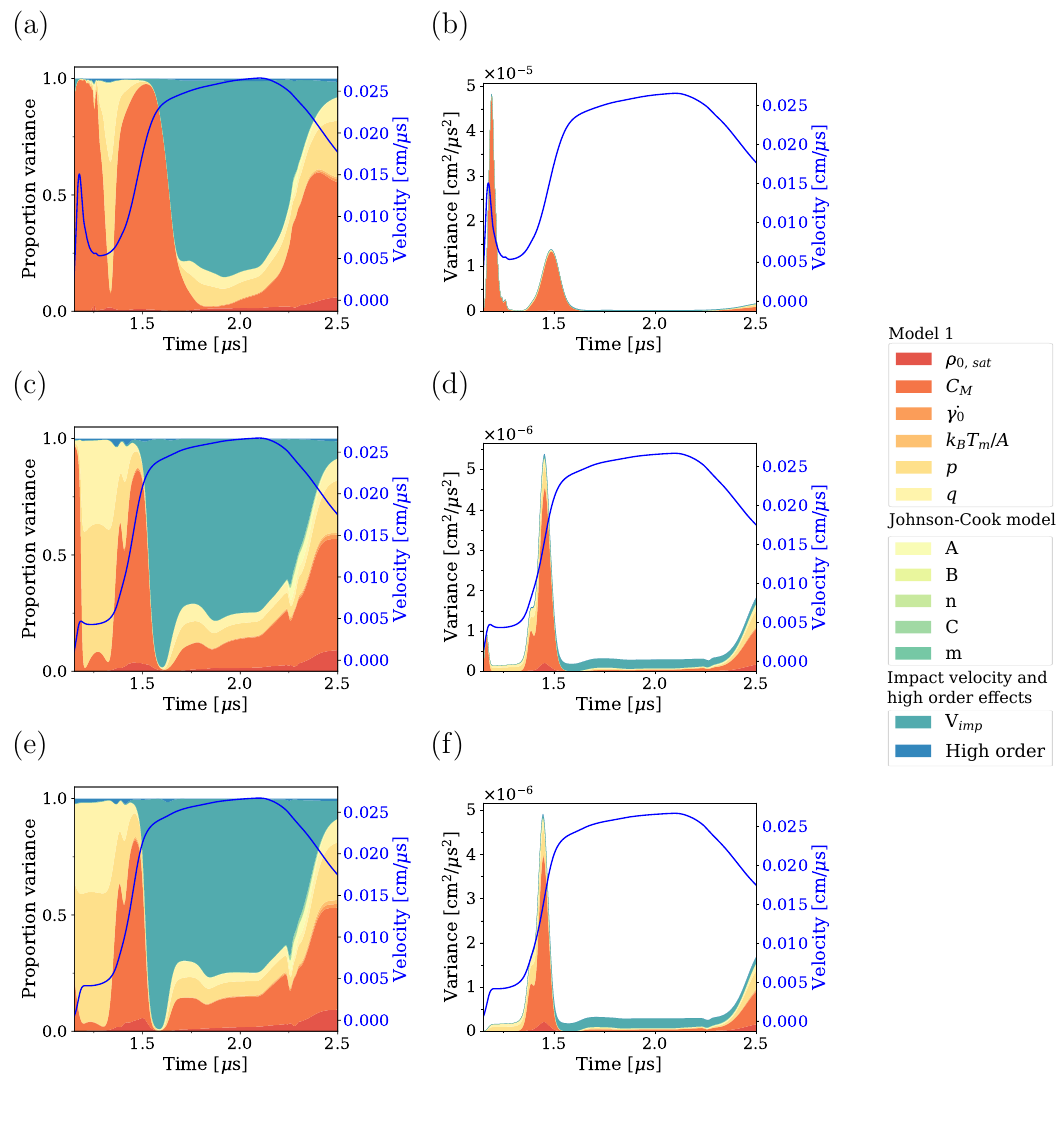}
    \caption{Functional global sensitivity analysis results for the plate impact responses of Model 1 with 2.5 mm copper impactors and 8 mm [100] single crystal molybdenum targets: (a,b) pristine, (c,d) 0.6\% prestrained, and (e,f) 5.4\% prestrained case, with left panels showing the Sobol' indices and right panels showing the total variances. The blue lines represent mean of the free surface velocity profiles used to train the pyBASS emulator.}
    \label{fig9:gsa_nguyen}
\end{figure}

\begin{figure}[t!]
    \centering
    \includegraphics[width=0.995\linewidth]{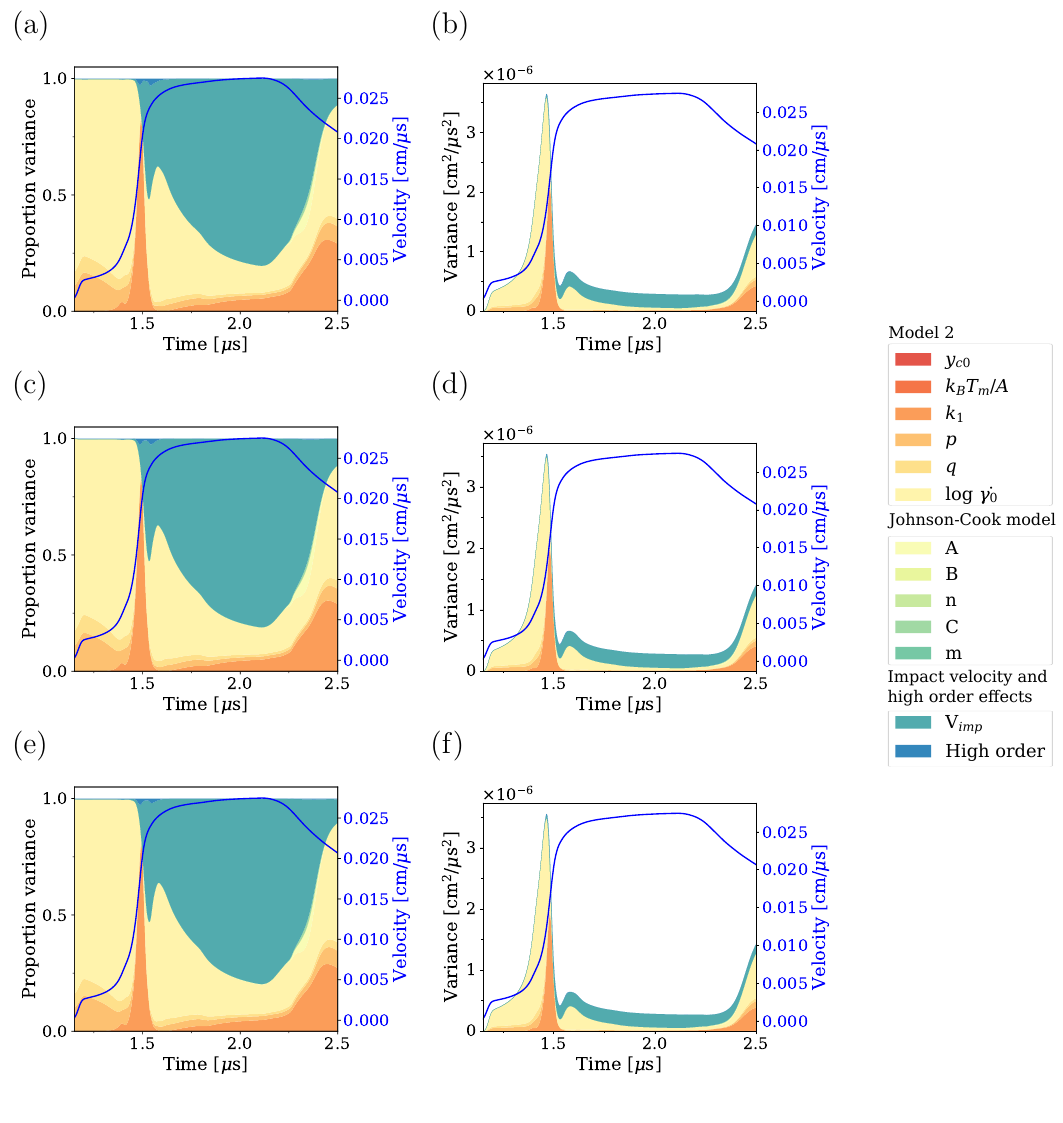}
    \caption{Functional global sensitivity analysis results for the plate impact responses of Model 2 with 2.5 mm copper impactors and 8 mm [100] single crystal molybdenum targets: (a,b) pristine, (c,d) 0.6\% prestrained, and (e,f) 5.4\% prestrained case, with left panels showing the Sobol' indices and right panels showing the total variances. The blue lines represent mean of the free surface velocity profiles used to train the pyBASS emulator.}
    \label{fig10:gsa_lee}
\end{figure}

In order to clearly identify the deformation mechanisms that govern the plate impact responses, additional global sensitivity analyses were performed on the plate impact simulations, which involve multiple coupled sources of uncertainties. The global sensitivity analyses were performed with 12-dimensional (6 crystal plasticity parameters, 5 Johnson-Cook model parameters and V$_{imp}$) uniform distribution over bounds in Tables \ref{table:para4}, \ref{table:jc_para} and [340, 360] m/s for V$_{imp}$. Figures \ref{fig9:gsa_nguyen} and \ref{fig10:gsa_lee} show the global sensitivity analysis results along with the average free surface velocity (blue line), computed from the 100 simulation results used to train the corresponding pyBASS surrogate. The left panels of these figures show stacked plots of the sensitivity indices (and higher-order effects), and the right panels show stacked plots of partitioned variances ($V_{X_i}\left(E_{\boldsymbol{X}_{\sim i}}\left[Y \mid X_i\right]\right)$ in Equation \ref{eq:sobol} and higher-order effects) which measure the contribution of each parameter to the output variance. From the definition of sensitivity indices in Section \ref{sec:sobol}, the sum of partitioned variances equals the total output variance $V(Y)$; thus, the total stacked heights in the right panels directly show how $V(Y)$ changes over time. The Johnson-Cook model parameters for the copper impactors ($A$, $B$, $n$, $C$, $m$)  are not influential for predicting the plate impact responses in the selected parameter range, due to their narrow shear stress difference in these impact tests. 

The parameters that predominantly affect the later stages of the deformation behavior in Figure \ref{fig5:GSA} (i.e., $\rho_{0,\mathrm{sat}}$, $\dot{\gamma}_0$, and $A$) on the uniaxial stress-strain data have a minor influence on the plate-impact responses using Model 1 as shown in Figure \ref{fig9:gsa_nguyen}. Four parameters were identified as having a significant influence on the plate-impact responses as shown in Figures \ref{fig9:gsa_nguyen} (a), (c), and (e): $C_M$, $p$, $q$, and $V_{\mathrm{imp}}$. Here, these model parameters ($C_M$, $p$, and $q$) also dominate the early stage uniaxial stress-strain response, i.e., near the initial yield,  as shown in Figure \ref{fig5:GSA}. As shown in the free surface velocity, the height of the elastic precursor is determined solely by $C_M$ in the pristine case, and the influence of $C_M$ on the elastic precursor gradually decreases as the initial dislocation density increases. Moreover, Figure \ref{fig9:gsa_nguyen} (b), (d), and (f) on stacked plots of partitioned variances clearly show that, under the plate-impact conditions, the output variance is shown to be dominated by $C_M$, and the deformation stages in which $C_M$ has little influence (e.g., at onset of the rise toward the plateau and at the plateau velocity) exhibit relatively small total output variances. Importantly, as the influence of $C_M$ on the Hugoniot elastic limit decreases with an increasing initial dislocation density, the total variance at the Hugoniot elastic limit also decreases. This also suggests that reducing uncertainty in $C_M$ is key to improving the prediction accuracy of Model 1 for the transient, ultrafast strain rate conditions. Furthermore, it shows that the prediction uncertainties in the Hugoniot elastic limit and in the rise time to the Hugoniot state in Figure \ref{fig7:shock_nguyen} reflects residual parameter uncertainties arising from the sparse high-strain-rate data, evidenced by the analysis presented in Sections \ref{subsection:Bayesian} and \ref{sec:sobol}. This further supports that the rapid multiplication of dislocation density in the drag-dominated regime is responsible for the sharp increase and subsequent relaxation of the Hugoniot elastic limit when the initial dislocation density is relatively low. In this regime, the slip rate at the onset of yield is constrained by the initial mobile dislocation density and the shear-wave speed $c_s$, which, via Equation \ref{eqn:running}, induces a rapid increase in the resolved shear stress. When the initial dislocation density is sufficiently high, the elastic–plastic transition at the Hugoniot elastic limit is primarily governed by the thermally activated glide (i.e., the parameters $p$ and $q$); however, the total output variance at the Hugoniot elastic limit is negligible despite the significant contributions of these parameters to the total output variance as shown in Figure \ref{fig9:gsa_nguyen} (b), (d), and (f). This also indicates that precise specification of the thermally activated dislocation velocity is less critical for predicting the plate impact responses with Model 1.

Interestingly, in Model 2 (Figure \ref{fig10:gsa_lee}), the sensitivity indices and output variances under the plate impact conditions were found to be insensitive to the initial dislocation density. Similar to the 5.4 \% prestrained case for Model 1 (Figure \ref{fig9:gsa_nguyen} (e)), the response at the elastic–plastic transition is governed by the parameters associated with the thermally activated glide mechanism (i.e., the parameters, $\dot\gamma_0$, $p$ and $q$). In Model 2, rather than the dislocation density evolution, the fixed pre-exponential factor $\dot{\gamma}_0$, which is directly proportional to the slip rate, mainly controls the initial response of the pristine case. However, unlike $C_M$ in the pristine case of Model 1, uncertainty in parameter $\dot\gamma_0$ does not exhibit large variance at the Hugoniot elastic limit of the pristine case. In conjunction with this observation, the insensitivity to dislocation density evolution parameters (i.e., $y_{c0}$ and $k_1$) at the Hugoniot elastic limit indicates that the dependence of the plate-impact responses on the initial dislocation density cannot be explained without taking the mobile dislocation kinetics for the drag-dominated regimes into account, specifically through the Orowan relation and the limiting dislocation velocity, represented by Equations \ref{eq_model1:orowan} to \ref{eqn:running} in Model 1.; i.e., Model 2 lacks the mobile-dislocation kinetics required to accurately reproduce the responses at the Hugoniot elastic limit. Furthermore, the temporal evolution of the total output variance in the plate impact simulations of Model 2 (Figure \ref{fig10:gsa_lee} (b), (d), and (f)) closely resembles that of Model 1 in the 5.4\% prestrained case (Figure \ref{fig9:gsa_nguyen}(f)). In both Model 1 and Model 2, the total output variances increase as the dislocation multiplication parameters ($C_M$ in Model 1 and $k_1$ in Model 2) become more influential during the rise time to the Hugoniot state. Since the dislocation density contributes primarily to strain hardening rather than plastic strain (and plastic strain rate) accommodation in Model 2, the increase in the influence of $k_1$ during the rise time indicates that the hardening rate in the crystal plasticity models is critical to capture the rise time.

\clearpage

\section{Discussion}
\label{sec:discussion}

The UQ-based model calibration and validation have been employed to identify the key physical assumptions and model components within the physics-based crystal plasticity models that are critical for capturing the deformation behavior of bcc single crystal molybdenum under quasi-static to shock loading conditions. The Bayesian model calibration (BMC) procedure has been demonstrated to be useful for statistically inferring the material parameters in the two representatives, dislocation-mediated crystal plasticity models (labeled Model 1 and Model 2). A common approach in the previous studies (\cite{doi:10.1198/016214507000000888,10.1063/1.5051442,NGUYEN2021104284}) was to identify optimal parameter sets via the BMC procedure for simple loading cases and then use it for predicting and analyzing the material responses upon more complex and extreme loading scenarios (e.g., plate impact or Taylor impact tests). However, due to the coupling effects of multiple uncertainty sources during these complex loading scenarios, it is essential to systematically quantify the contribution of each source to the overall uncertainty in the model predictions. To this end, we have utilized the variance-based functional global sensitivity analysis (\cite{sobol1993sensitivity,francom_sensitivity_2017,JSSv094i08}) in conjunction with the BMC procedure. The functional global sensitivity results presented in Figures \ref{fig5:GSA} and \ref{fig6:GSA} (quasi-static), \ref{fig9:gsa_nguyen}, and \ref{fig10:gsa_lee} (shock loading conditions) have provided key insights into the underlying deformation mechanisms in the representative bcc molybdenum at various loading conditions. The temporal evolution of the sensitivity indices and the partitioned variances calculated during the global sensitivity analysis have clearly shown that dislocation densities play distinct roles in both of Models 1 and 2, especially at the onset of yield. The difference in the responses predictive by Model 1 and Model 2, especially at the Hugoniot elastic limit, has shown that the incorporation of both (1) the Orowan relation and (2) the drag-dominated regime is essential for capturing the significant influence of dislocation density on the elastic-plastic transition behavior. The Hugoniot elastic limit of simulated velocity using Model 2 was found to be primarily governed by the fixed pre-exponential factor $\dot\gamma_0$ rather than dislocation density and its evolution rate. This suggests that treating $\dot\gamma_0$ as a constant is an oversimplification for the predictive modeling of transient, extreme strain rate responses. Furthermore, while many bcc crystal plasticity models have conveniently assumed a fixed $\dot\gamma_0$ (\cite{patra_constitutive_2014,Weinberger2012IncorporatingModels,lim_multi-scale_2015,LEE2023103529}), it should be noted that $\dot\gamma_0$ is fundamentally stress- and microstructure-dependent, as posited in \cite{kothari_elasto-viscoplastic_1998}, \cite{argon_strengthening_2008}, and \cite{monnet_dislocation-dynamics_2013}. Nevertheless, the formulation of Model 2 has been useful for predicting microstructure evolution and hardening behavior from quasi-static to high strain rate regimes ($\approx$ 10$^3$ s$^{-1}$). In particular, Model 2 has been proven to provide significant insight into the fundamental linkages between dislocation mean free path, dislocation interactions, strain hardening, and microstructure evolution in bcc materials at both single and polycrystal levels (\cite{devincre2008,MADEC2017166,bronkhorst_structural_2019,LEE2023103529,dunham_attribution_2025,schmelzer_statistical_2025,schmelzer_thermodynamically_2025}).

The physical insights obtained from our UQ-based model calibration and validation are also shown to guide further model adjustment and development. Although Model 1 can account for the influence of an initial dislocation density on the Hugoniot elastic limit and the rise time to the Hugoniot state, it fails to reproduce the experimentally observed insensitivity of the Hugoniot elastic limit to the specimen thickness, especially for the pristine case (low initial dislocation density). For [100] single-crystal molybdenum, the attenuation of the Hugoniot elastic limit is generally not observed in experiments (\cite{mandal_elastic-plastic_2017,10.1063/5.0082267}), whereas our simulation results using Model 1 have exhibited the significant attenuation. As revealed by the global sensitivity analysis for the pristine case (Figure \ref{fig9:gsa_nguyen} (a)), the magnitude of the elastic precursor was found to depend solely on the rate of mobile dislocation evolution. Therefore, the inability of Model 1 to capture the experimentally observed thickness insensitivity of the Hugoniot elastic limit suggests that an additional dislocation evolution mechanism must be incorporated.

\begin{figure}[b!]
    \centering
    \includegraphics[width=0.5\linewidth]{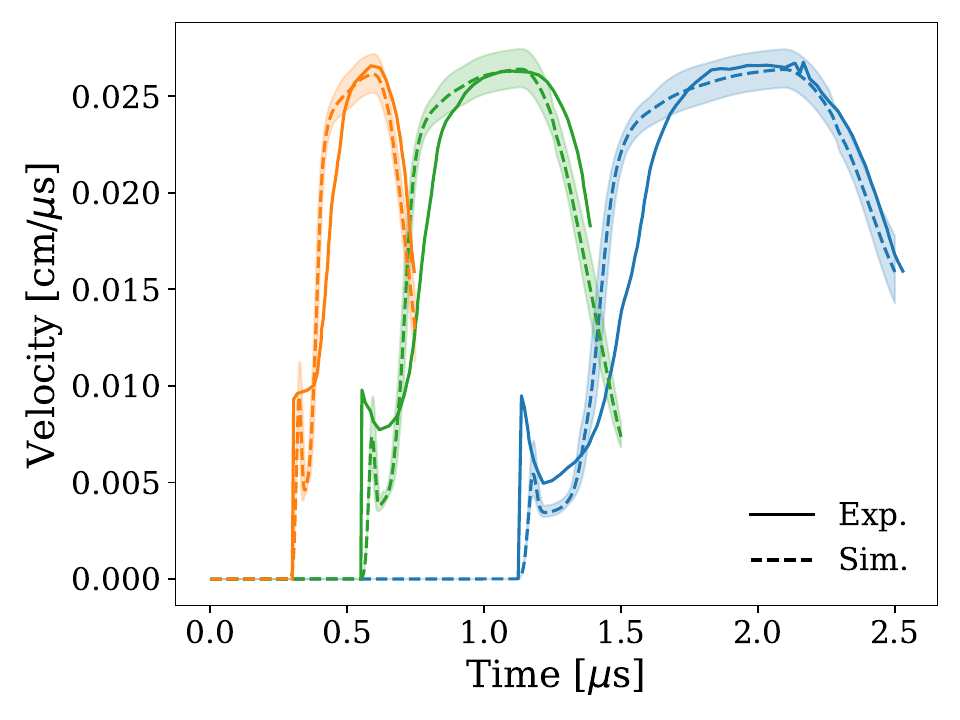}
    \caption{Plate impact response of [100] single crystal molybdenum predicted using Model 1 with the additional dislocation nucleation term. The solid lines represent the experimental data. The shaded regions show the 90\% credible interval of the simulated responses. The dashed lines show the mean of the simulated responses. The orange line corresponds to a copper impactor thickness of 0.78 mm and a molybdenum target thickness of 2.1 mm; the green line to 1.48 mm and 4 mm; and the blue line to 2.5 mm and 8 mm, respectively.}
    \label{fig11:shock_nuc}
\end{figure}

To verify the need for the additional dislocation evolution mechanism, we simply added a stress-dependent dislocation nucleation term (\cite{austin2011dislocation}) to the mobile dislocation density evolution, which is expressed by,
\begin{equation}
\label{eqn:dis_nuc}
    \dot\rho^{\alpha}_{\textrm{nuc}} = \dot\rho_{n0} \textrm{exp} \left( -\frac{\Delta G_{n0}}{k_B \theta} \left\langle 1 - \left(\frac{|\tau^{\alpha}| - \tau^{\alpha}_{\textrm{a}}}{\tau_{n0}} \right)^{p_n} \right\rangle^{q_n} \right),
\end{equation}
where $\dot\rho_{n0}$ is the reference dislocation nucleation rate, $\Delta G_{n0}$ is the activation energy for dislocation nucleation, $\tau_{n0}$ is the resistance to dislocation nucleation, and $p_n$ and $q_n$ are the material parameters that control the stress-dependence. The values of the parameters related to dislocation nucleation are:
\begin{equation*}
    \dot\rho_{n0} = 1 \times 10^{11} \, \textrm{cm}^{-2}, \; \Delta G_{n0} = 5 \times 10^{-19} \, \textrm{J}, \; \tau_{n0} = 3 \, \textrm{GPa}, \; p_n = 0.3, \; q_n=1.5.
\end{equation*}
While $\tau_{n0}$ is commonly taken as the ideal shear strength ($\approx 0.1\,\mu$) (e.g., \cite{lloyd_plane_2014}), here we calibrate $\tau_{n0}$ to plate-impact data and therefore use lower values that capture the elastic-precursor invariance in the [100] molybdenum single crystal under plate impact conditions. The nucleation term in Equation \ref{eqn:dis_nuc} has been incorporated into the mobile dislocation density evolution equation (Equation \ref{eq:mobile_evo}) in Model 1. Through the addition of the dislocation nucleation term, the Hugoniot elastic limits for the shorter propagation distance cases (the orange and green lines) decrease significantly, as shown in Figure \ref{fig11:shock_nuc}, and that of the blue line case shows a marginal change. The simulation results show that the dislocation nucleation mechanism accounts for the rapid stress relaxation leading to attenuation of the elastic precursor near the impact surface. It becomes inactive once a sufficient mobile dislocation density is generated in this region, thereby preventing a significant reduction in the elastic precursor magnitude with an increasing propagation distance or sample thickness. The simulation results for 0.6 \% and 5.4 \% prestrained cases remain almost unaffected by the addition of the nucleation term, as the stress levels involved are insufficient to activate the dislocation nucleation term.

The integrated approach combining the Bayesian calibration, global sensitivity analysis, and uncertainty propagation analysis has been found to facilitate the identification of major sources of discrepancies between experiments and model predictions for bcc molybdenum under quasi-static to extreme loading conditions. Furthermore, it enables the identification of specific model components that need to be further refined. Such systematic diagnostics are not readily achieved using the Bayesian parameter estimation procedures alone. The global sensitivity analysis (Figure \ref{fig9:gsa_nguyen}) and the model response with dislocation nucleation (Figure \ref{fig11:shock_nuc}) suggest that additional dislocation evolution mechanisms must be incorporated for applications in the transient, shock loading conditions. However, we further note that the underlying physical basis for the dislocation nucleation remains unclear, highlighting the need for further investigation. According to the previous studies (\cite{xu_homogeneous_2000,austin2011dislocation}), dislocations are not nucleated even at the stress levels present in the pristine case. Instead, \cite{denoual2024dislocation} considered the depinning of immobile dislocations at high stress levels to be a more important mechanism to accurately capture the dislocation density evolution and plate impact behavior at the polycrystal level. In their work, the stress-induced mobilization of immobile dislocations due to depinning also accounted for the rapid decay of the Hugoniot elastic limit near the impact surface. However, this has not yet been validated and implemented at the single crystal level and involves more complex mathematical and numerical formulations that remain insufficiently examined in the literature; therefore, it should be revisited and precisely compared with the dislocation nucleation mechanisms in future work.

\section{Conclusion}
\label{sec:conclusion}

The deformation behavior of bcc refractory metals remains remarkably challenging to capture within continuum crystal plasticity models. In this work, through uncertainty quantification (UQ), we have identified the key physical assumptions critical for modeling the deformations of bcc single crystal molybdenum under quasi-static to shock loading conditions. Two physics-based bcc crystal plasticity models, each emphasizing distinct key physical assumptions of the structures and kinetics of dislocation evolution, have been validated and analyzed. First, uncertainties in the model input parameters were constrained through the Bayesian calibration procedure against uniaxial stress-strain data spanning a wide range of temperatures (195 to 500 K), strain rates (10$^{-5}$ to 300 s$^{-1}$), and crystallographic orientations. Conditioned on the Bayesian model calibration, predictive capabilities of both models have been found to be very similar for the uniaxial stress-strain responses. The contribution of each material parameter to the model predictions has also been quantified through the global sensitivity analysis. Our analysis has clearly revealed key mechanistic differences between the two models. Model 1, which incorporates the Orowan relation with the kinetics of dislocation motion in both thermally-activated and drag-dominated regimes, has shown high sensitivity to the dislocation multiplication processes near the onset of yield at quasi-static to high strain rates. By contrast, Model 2, employing the simple, thermally-activated glide kinetics with the fixed pre-exponential factor $\dot\gamma_0$, has been found to be much less sensitive to the evolution of dislocation densities at the onset of yield, especially at high strain rates.

We have further examined the two bcc models against the plate impact tests, where multiple sources of uncertainties are coupled. The uncertainty propagation and global sensitivity analysis during the plate impact simulations have suggested that the dislocation density-based crystal plasticity models must incorporate the Orowan relation (Equation \ref{eq_model1:orowan}) with the mobile dislocation kinetics (e.g., through Equations \ref{eqn:mean_velo} to \ref{eqn:running}) to capture the influence of initial dislocation densities at such a transient, extreme loading condition as shock loading. The global sensitivity analysis has further identified that discrepancies between experimental observations and model predictions are primarily attributed to the incomplete representations of the dislocation density evolution in these two bcc crystal plasticity models. In this context, we have demonstrated that incorporating the additional nucleation term into the dislocation evolution law especially in Model 1 can improve its predictive capability for the thickness dependence of the elastic precursor. However, as discussed in Section \ref{sec:discussion}, the physical basis of this dislocation nucleation mechanism remains unclear and further investigation is required in future work.

Significant challenges in identifying and modeling the complex deformation processes in bcc materials at extreme loading scenarios remain beyond the scope of this work. The UQ framework employed here provides a foundation for future investigations. The UQ procedure should be extended to forecasting the extreme events associated with ductile damage and fracture in bcc materials (\cite{pogorelko_dynamic_2023,bronkhorst_local_2021,zhang_data-driven_2023,dunham_attribution_2025,schmelzer_statistical_2025,schmelzer_thermodynamically_2025,saha_numerical_2026}). In addition, the uncertainty quantification procedure can provide a more systematic assessment of the non-Schmid effects due to the non-planar features of dislocation cores in these materials (\cite{GROGER20085401,groger_multiscale_2008b,patra_constitutive_2014,Weygand2015,cho_anomalous_2018}). The proposed UQ procedure can also complement recent efforts in machine learning-based approaches, which increasingly serve as surrogates for computationally expensive large-scale crystal plasticity simulations (\cite{bonatti_cp-fft_2022,liu_learning_2023,eghtesad_machine_2023,schmidt_texture-dependent_2025}). Furthermore, expanding the framework demonstrated in this work to amorphous soft materials will be of critical interest in future work. In particular, the UQ procedure can be used for identifying the key physical mechanisms underlying deformation-fracture processes in polymeric materials in and above their glass transitions, where high-order theories with significant uncertainties have been widely employed (\cite{srivastava_thermo-mechanically-coupled_2010,lee_polyurethane-urea_2023,lee_size-dependent_2024,Lee2024ExtremeCrystals,wagner_foundational_2025,talamini_progressive_2018,jordan_training_2025}).

\section*{Acknowledgement}
This work was supported by the National Research Foundation of Korea (RS-2023-00279843) and Korea Advanced Institute of Science and Technology (N11250018, N11240082). This work was also performed as a component of LDRD project (LDRD-20230128DR) supported by the US Department of Energy through the Los Alamos National Laboratory. Los Alamos National Laboratory is operated by Triad National Security, LLC, for the National Nuclear Security Administration of U.S. Department of Energy (Contract No. 89233218CNA000001).

\appendix
\section{Calibration of Johnson-Cook model for half hard copper}
\label{appendix:JC}
For the plate impact simulations presented in Section \ref{sec:plate_impact}, the Johnson-Cook model (\cite{johnson_constitutive_1983}), a particular type of the Mises plasticity model with analytical forms of the hardening law and rate dependence, have been used for the copper impactor. The flow stress for the Johnson-Cook model is expressed by,
\begin{equation}
    \bar{\sigma} = \left[A + B ({\bar \epsilon}^{pl})^n \right] \left[ 1+C \,\textrm{ln}\left( \frac{\dot {\bar \epsilon}^{pl}}{\dot{\epsilon_0}}\right) \right] \left[ 1 - {\hat \theta}^m \right]
\end{equation}
with
\begin{equation}
     {\hat \theta} \equiv \begin{cases}
0 & \text{if } \theta < \theta_{t}, \\
\frac{\theta-\theta_{t}}{\theta_{m}-\theta_{t}} & \text{if } \theta_{t} \leq \theta \leq \theta_{m}, \\
1 & \text{if } \theta > \theta_{m}, \\
\end{cases}
\end{equation}
where $A$, $B$, and $n$ are the parameters associated with the static yield stress, $C$ controls the rate dependence, and $m$ controls the temperature dependence. $\dot{\epsilon_0}$ ($= 1.0 \, \textrm{s}^{-1}$) is the reference strain rate, $\theta_{m}$ ($= 1358$ K) is the melting temperature, and $\theta_{t}$ ($= 298$ K) is the transition temperature of copper. In this work, we have used the built-in Johnson-Cook model in Abaqus, and the parameter bounds for the Bayesian calibration procedure are given in Table \ref{table:jc_para}.

\begin{figure}[h]
    \centering
    \includegraphics[width=1\linewidth]{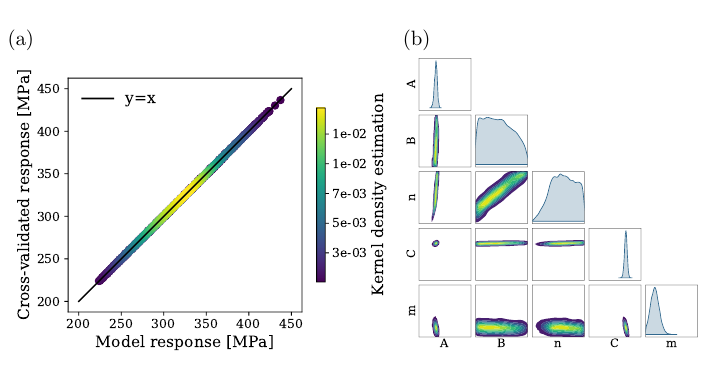}
    \caption{(a) Comparison of cross-validated predictions of emulator with the simulated stresses from the Johnson-Cook model. Each point on the scatter plot is colored according to its Gaussian kernel density estimate. The cross-validation comparisons show that the trained emulators exhibit excellent performance. (b) Pair plots that represent posterior parameter distributions. The plots on the diagonal of pair plots show the marginal probability distributions for each parameter, while the plots on the off-diagonal show the bivariate kernel density estimations between pairs of parameters. Overall, the model parameters are well calibrated through the BMC processes.}
    \label{figa1:jc_calibration}
\end{figure}

\begin{figure}[h]
    \centering
    \includegraphics[width=1\linewidth]{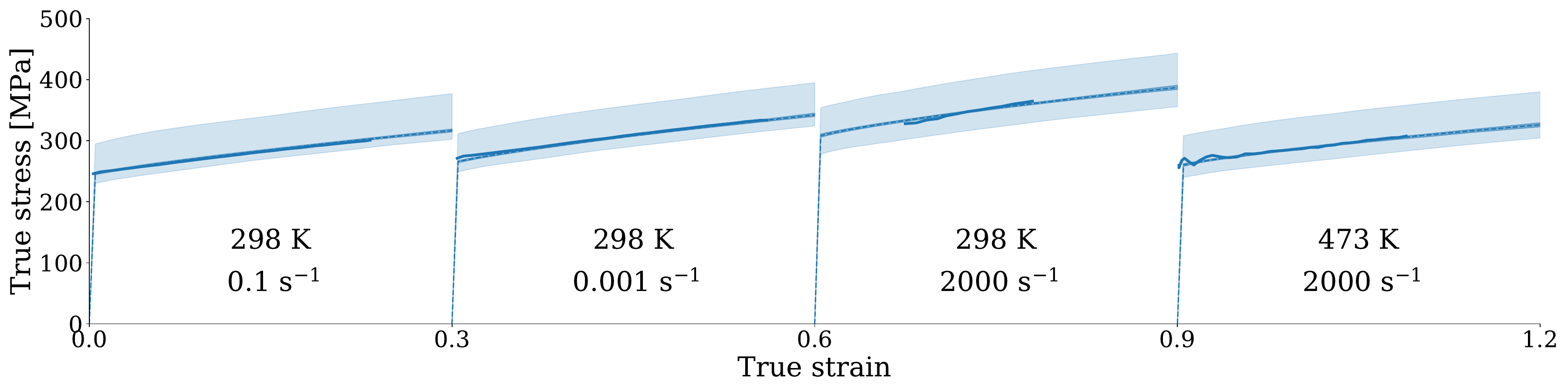}
    \caption{Stress-strain responses of the half-hard polycrystalline copper from experiments and the Johnson-Cook model. The solid lines represent the experimental data. The light blue shaded regions represent the training simulation data corresponding to the prior parameter ranges (e.g., Table \ref{table:jc_para}). The dark blue shaded regions, which are nearly indistinguishable from the dashed lines, show the 90\% credible interval of the simulated responses corresponding to the posterior parameter distribution (e.g., Figure \ref{figa1:jc_calibration} (b)). The dashed lines show the mean of these simulated responses. The calibrated results show an excellent agreement with the experiment data.}
    \label{figa2:jc_emul}
\end{figure}

The Bayesian model calibration was performed with 200 sample model runs against experimental data collected at Los Alamos National Laboratory. The posterior distributions after the Bayesian model calibration are shown in Figure \ref{figa1:jc_calibration}. The parameters $A$, $C$, and $m$ are well constrained with narrow marginal distributions, while a strong positive correlation is observed between the parameters $B$ and $n$. Figure \ref{figa2:jc_emul} shows the calibrated model responses along with the experimental data. The calibrated model accurately captures both rate and temperature dependence. The narrow 90\% prediction interval, nearly indistinguishable from the posterior mean, indicates a high confidence in the model predictions. For the plate impact simulations presented in Section \ref{sec:plate_impact}, the Johnson-Cook model is further modified with the Mie-Gr{\"u}niesen equation of state (Equation \ref{eqn:eos}), and the parameters for the Mie-Gr{\"u}niesen equation of state are:
\begin{equation*}
    \rho_0=8960 \,\textrm{kg/m}^3,  \quad C_0 = 3933 \, \textrm{m/s}, \quad \Gamma_0=1.99, \quad s=1.5.
\end{equation*}
These parameters were taken from \cite{macdonald_thermodynamic_1981} and \cite{mitchell_shock_1981}.

\begin{table}[t]
\centering
\renewcommand{\arraystretch}{1.3}
\caption{Prior parameter ranges for the Bayesian model calibration of the Johnson-Cook model for the half-hard polycrystalline copper, and the posterior expectations obtained from the calibration.}
\begin{tabular}{l@{\quad}l@{\quad}l}
\hline
Parameter & Range & Expectation \\
\hline
$A$ [MPa] & [250.0,320.0] & 272.8 \\
$B$ [MPa] & [200.0,250.0] & 221.8 \\
$n$ & [0.7,0.9] & 0.82\\
$C$ & [0.005,0.02] & 0.0156\\
$m$ & [1.0,1.2] & 1.04\\
\hline
\label{table:jc_para}
\end{tabular}
\end{table}

\clearpage

\printbibliography

@article{LEE2023103529,
title = {Deformation, dislocation evolution and the non-Schmid effect in body-centered-cubic single- and polycrystal tantalum},
journal = {International Journal of Plasticity},
volume = {163},
pages = {103529},
year = {2023},
doi = {https://doi.org/10.1016/j.ijplas.2023.103529},
url = {https://www.sciencedirect.com/science/article/pii/S0749641923000153},
author = {Seunghyeon Lee and Hansohl Cho and Curt A. Bronkhorst and Reeju Pokharel and Donald W. Brown and Bjørn Clausen and Sven C. Vogel and Veronica Anghel and George T. Gray and Jason R. Mayeur}
}

@article{doi:10.1198/016214507000000888,
author = {Dave Higdon and James Gattiker and Brian Williams and Maria Rightley},
title = {Computer Model Calibration Using High-Dimensional Output},
journal = {Journal of the American Statistical Association},
volume = {103},
number = {482},
pages = {570-583},
year  = {2008},
publisher = {Taylor & Francis},
doi = {10.1198/016214507000000888},
}

@article{NGUYEN2021104284,
title = {Bayesian calibration of a physics-based crystal plasticity and damage model},
journal = {Journal of the Mechanics and Physics of Solids},
volume = {149},
pages = {104284},
year = {2021},
doi = {https://doi.org/10.1016/j.jmps.2020.104284},
url = {https://www.sciencedirect.com/science/article/pii/S0022509620304889},
author = {Thao Nguyen and Devin C. Francom and D.J. Luscher and J.W. Wilkerson},
}

@article{MADEC2017166,
title = {Dislocation strengthening in FCC metals and in BCC metals at high temperatures},
journal = {Acta Materialia},
volume = {126},
pages = {166-173},
year = {2017},
doi = {https://doi.org/10.1016/j.actamat.2016.12.040},
url = {https://www.sciencedirect.com/science/article/pii/S1359645416309764},
author = {Ronan Madec and Ladislas P. Kubin},
}

@article{BEYERLEIN2008867,
title = {A dislocation-based constitutive law for pure Zr including temperature effects},
journal = {International Journal of Plasticity},
volume = {24},
number = {5},
pages = {867-895},
year = {2008},
doi = {https://doi.org/10.1016/j.ijplas.2007.07.017},
url = {https://www.sciencedirect.com/science/article/pii/S074964190700109X},
author = {I.J. Beyerlein and C.N. Tomé},
}

@article{10.1063/1.1728952,
    author = {Bolef, D. I. and De Klerk, J.},
    title = "{Elastic Constants of Single‐Crystal Mo and W between 77° and 500°K}",
    journal = {Journal of Applied Physics},
    volume = {33},
    number = {7},
    pages = {2311-2314},
    year = {2004},
    doi = {10.1063/1.1728952},
    url = {https://doi.org/10.1063/1.1728952},
}

@article{hollang_flow_1997,
	title = {The Flow Stress of Ultra-High-Purity Molybdenum Single Crystals},
	volume = {160},
	language = {en},
	number = {2},
	journal = {physica status solidi (a)},
	author = {Hollang, L. and Hommel, M. and Seeger, A.},
	year = {1997},
	pages = {329--354},
}

@article{hollang_work_2001,
	title = {Work hardening and flow stress of ultrapure molybdenum single crystals},
	volume = {319-321},
	journal = {Materials Science and Engineering: A},
	author = {Hollang, L and Brunner, D and Seeger, A},
	year = {2001},
	pages = {233--236},
}

@article{dezerald_first-principles_2015,
	title = {First-principles prediction of kink-pair activation enthalpy on screw dislocations in bcc transition metals: V, Nb, Ta, Mo, W, and Fe},
	volume = {91},
	language = {en},
	number = {9},
	journal = {Physical Review B},
	author = {Dezerald, L. and Proville, L. and Ventelon, Lisa and Willaime, F. and Rodney, D.},
	year = {2015},
	pages = {094105}
}

@article{lim_investigating_2020,
	title = {Investigating active slip planes in tantalum under compressive load: {Crystal} plasticity and slip trace analyses of single crystals},
	volume = {185},
	journal = {Acta Materialia},
	author = {Lim, Hojun and Carroll, Jay D. and Michael, Joseph R. and Battaile, Corbett C. and Chen, Shuh Rong and D. Lane, J. Matthew},
	year = {2020},
	pages = {1--12}
}

@book{clayton_nonlinear_2019,
	series = {Shock Wave and High Pressure Phenomena},
	title = {Nonlinear Elastic and Inelastic Models for Shock Compression of Crystalline Solids},
	isbn = {978-3-030-15329-8 978-3-030-15330-4},
	publisher = {Springer International Publishing},
	author = {Clayton, John D.},
	year = {2019}
}

@article{katahara_pressure_1979,
	title = {Pressure derivatives of the elastic moduli of {BCC} {Ti}-{V}-{Cr}, {Nb}-{Mo} and {Ta}-{W} alloys},
	volume = {9},
	number = {5},
	urldate = {2024-11-08},
	journal = {Journal of Physics F: Metal Physics},
	author = {Katahara, K W and Manghnani, M H and Fisher, E S},
	year = {1979},
	pages = {773--790}
}

@article{dezerald_plastic_2016,
	title = {Plastic anisotropy and dislocation trajectory in BCC metals},
	volume = {7},
	url = {https://www.nature.com/articles/ncomms11695},
	doi = {10.1038/ncomms11695},
	journal = {Nature Communications},
	author = {Dezerald, Lucile and Rodney, David and Clouet, Emmanuel and Ventelon, Lisa and Willaime, François},
	year = {2016},
	pages = {11695},
}

@article{10.1063/1.5051442,
    author = {Walters, David J. and Biswas, Ayan and Lawrence, Earl C. and Francom, Devin C. and Luscher, Darby J. and Fredenburg, D. Anthony and Moran, Kelly R. and Sweeney, Christine M. and Sandberg, Richard L. and Ahrens, James P. and Bolme, C. A.},
    title = "{Bayesian calibration of strength parameters using hydrocode simulations of symmetric impact shock experiments of Al-5083}",
    journal = {Journal of Applied Physics},
    volume = {124},
    number = {20},
    pages = {205105},
    year = {2018},
    doi = {10.1063/1.5051442},
}

@article{https://doi.org/10.1002/pssb.19660150214,
author = {Guiu, F. and Pratt, P. L.},
title = {The Effect of Orientation on the Yielding and Flow of Molybdenum Single Crystals},
journal = {physica status solidi (b)},
volume = {15},
number = {2},
pages = {539-552},
doi = {https://doi.org/10.1002/pssb.19660150214},
url = {https://onlinelibrary.wiley.com/doi/abs/10.1002/pssb.19660150214},
year = {1966}
}

@phdthesis{guiu1965plastic,
  title={Plastic deformation of molybdenum single crystals},
  author={Guiu, Francisco},
  year={1965},
  school={University of London}
}

@article{https://doi.org/10.1002/pssb.19670190135,
author = {Guiu, F.},
title = {Temperature and Strain Rate Dependence of the Flow Stress in Molybdenum},
journal = {physica status solidi (b)},
volume = {19},
number = {1},
pages = {339-351},
doi = {https://doi.org/10.1002/pssb.19670190135},
year = {1967}
}

@article{wang_generalized_2021,
	title = {Generalized stacking fault energies and Peierls stresses in refractory body-centered cubic metals from machine learning-based interatomic potentials},
	volume = {192},
	journal = {Computational Materials Science},
	author = {Wang, Xiaowang and Xu, Shuozhi and Jian, Wu-Rong and Li, Xiang-Guo and Su, Yanqing and Beyerlein, Irene J.},
	year = {2021},
	pages = {110364},
}

@article{https://doi.org/10.1002/pssa.2210220236,
author = {Irwin, G. J. and Guiu, F. and Pratt, P. L.},
title = {The influence of orientation on slip and strain hardening of molybdenum single crystals},
journal = {physica status solidi (a)},
volume = {22},
number = {2},
pages = {685-698},
doi = {https://doi.org/10.1002/pssa.2210220236},
year = {1974}
}

@Article{Bulatov2006,
author={Bulatov, Vasily V.
and Hsiung, Luke L.
and Tang, Meijie
and Arsenlis, Athanasios
and Bartelt, Maria C.
and Cai, Wei
and Florando, Jeff N.
and Hiratani, Masato
and Rhee, Moon
and Hommes, Gregg
and Pierce, Tim G.
and de la Rubia, Tomas Diaz},
title={Dislocation multi-junctions and strain hardening},
journal={Nature},
year={2006},
volume={440},
number={7088},
pages={1174-1178},
}

@article{weinberger_slip_2013,
	title = {Slip planes in bcc transition metals},
	volume = {58},
	number = {5},
	journal = {International Materials Reviews},
	author = {Weinberger, C R and Boyce, B L and Battaile, C C},
	year = {2013},
	pages = {296--314}
}

@article{GROGER20085401,
title = {Multiscale modeling of plastic deformation of molybdenum and tungsten: I. Atomistic studies of the core structure and glide of 1/2〈111〉 screw dislocations at 0K},
journal = {Acta Materialia},
volume = {56},
number = {19},
pages = {5401-5411},
year = {2008},
doi = {https://doi.org/10.1016/j.actamat.2008.07.018},
author = {R. Gröger and A.G. Bailey and V. Vitek},
}

@article{groger_multiscale_2008b,
	title = {Multiscale modeling of plastic deformation of molybdenum and tungsten: II. Yield criterion for single crystals based on atomistic studies of glide of 1/2〈111〉 screw dislocations},
	volume = {56},
	doi = {10.1016/j.actamat.2008.07.037},
	number = {19},
	journal = {Acta Materialia},
	author = {Gröger, R. and Racherla, V. and Bassani, J.L. and Vitek, V.},
	year = {2008},
	pages = {5412--5425}
}

@article{10.1063/5.0082267,
    author = {Kanel, G. I. and Garkushin, G. V. and Savinykh, A. S. and Razorenov, S. V. and Paramonova, I. V. and Zaretsky, E. B.},
    title = "{Effect of small pre-strain on the resistance of molybdenum [100] single crystal to high strain rate deformation and fracture}",
    journal = {Journal of Applied Physics},
    volume = {131},
    number = {9},
    pages = {095903},
    year = {2022}
}

@article{denison_bayesian_1998,
	title = {Bayesian {MARS}},
	volume = {8},
	doi = {10.1023/A:1008824606259},
	number = {4},
	urldate = {2025-02-10},
	journal = {Statistics and Computing},
	author = {Denison, D. G. T. and Mallick, B. K. and Smith, A. F. M.},
	year = {1998},
	pages = {337--346}
}

@article{mandal_elastic-plastic_2017,
	title = {Elastic-plastic deformation of molybdenum single crystals shocked along [100]},
	volume = {121},
	doi = {10.1063/1.4974475},
	number = {4},
	journal = {Journal of Applied Physics},
	author = {Mandal, A. and Gupta, Y. M.},
	year = {2017},
	pages = {045903}
}

@article{cho_anomalous_2018,
	title = {Anomalous plasticity of body-centered-cubic crystals with non-{Schmid} effect},
	volume = {139-140},
	doi = {10.1016/j.ijsolstr.2018.01.029},
	journal = {International Journal of Solids and Structures},
	author = {Cho, Hansohl and Bronkhorst, Curt A. and Mourad, Hashem M. and Mayeur, Jason R. and Luscher, D.J.},
	year = {2018},
	pages = {138--149}
}

@article{asaro_overview_1985,
	title = {Overview no. 42 {Texture} development and strain hardening in rate dependent polycrystals},
	volume = {33},
	issn = {00016160},
	number = {6},
	journal = {Acta Metallurgica},
	author = {Asaro, R.J. and Needleman, A.},
	year = {1985},
	pages = {923--953},
}

@article{stainier_micromechanical_2002,
	title = {A micromechanical model of hardening, rate sensitivity and thermal softening in {BCC} single crystals},
	volume = {50},
	number = {7},
	journal = {Journal of the Mechanics and Physics of Solids},
	author = {L. Stainier and A.M. Cuitiño and M. Ortiz},
	year = {2002},
	pages = {1511--1545}
}

@article{dequiedt_slip_2023,
    title = {Slip system interactions in {BCC} single crystals: {System} deactivation and segregation},
    volume = {184},
    journal = {Mechanics of Materials},
    author = {Dequiedt, J.L.},
    year = {2023},
    pages = {104730},
}

@article{JSSv094i08,
 title={BASS: An R Package for Fitting and Performing Sensitivity Analysis of Bayesian Adaptive Spline Surfaces},
 volume={94},
 url={https://www.jstatsoft.org/index.php/jss/article/view/v094i08},
 doi={10.18637/jss.v094.i08},
 number={8},
 journal={Journal of Statistical Software},
 author={Francom, Devin and Sansó, Bruno},
 year={2020},
 pages={1–36}
}

@software{james_gattiker_2020_4048801,
  author       = {James Gattiker and
                  Natalie Klein and
                  Grant Hutchings and
                  Earl Lawrence},
  title        = {lanl/SEPIA: v1.1},
  year         = 2020,
  publisher    = {Zenodo},
  version      = {v1.1},
  doi          = {10.5281/zenodo.4048801},
  url          = {https://doi.org/10.5281/zenodo.4048801}
}

@article{kalidindi1992crystallographic,
  title={Crystallographic texture evolution in bulk deformation processing of FCC metals},
  author={Kalidindi, Surya R and Bronkhorst, Curt A and Anand, Lallit},
  journal={Journal of the Mechanics and Physics of Solids},
  volume={40},
  number={3},
  pages={537--569},
  year={1992},
  publisher={Elsevier}
}

@article{denoual2024dislocation,
  title={Dislocation storage-release-recovery model for metals under strain rates from 10- 3 to 107 s- 1, and application to tantalum},
  author={Denoual, Christophe and Pellegrini, Yves-Patrick and Lafourcade, Paul and Madec, Ronan},
  journal={Journal of Applied Physics},
  volume={135},
  number={4},
  year={2024},
  pages={045101},
  publisher={AIP Publishing}
}

@article{LUSCHER201763,
title = {A dislocation density-based continuum model of the anisotropic shock response of single crystal α-cyclotrimethylene trinitramine},
journal = {Journal of the Mechanics and Physics of Solids},
volume = {98},
pages = {63-86},
year = {2017},
doi = {https://doi.org/10.1016/j.jmps.2016.09.005},
author = {D.J. Luscher and F.L. Addessio and M.J. Cawkwell and K.J. Ramos}
}

@article{EOrowan_1940,
doi = {10.1088/0959-5309/52/1/303},
url = {https://dx.doi.org/10.1088/0959-5309/52/1/303},
year = {1940},
publisher = {},
volume = {52},
number = {1},
pages = {8},
author = {E Orowan},
title = {Problems of plastic gliding},
journal = {Proceedings of the Physical Society}
}

@book{KOCKSUF1975,
	address = {Oxford ; New York},
	edition = {1st ed},
	series = {Progress in materials science},
	title = {Thermodynamics and kinetics of slip},
	isbn = {978-0-08-017964-3},
	number = {v. 19},
	publisher = {Pergamon Press},
	author = {Kocks, U. F. and Argon, Ali S. and Ashby, M. F.},
	year = {1975},
}

@article{austin2011dislocation,
  title={A dislocation-based constitutive model for viscoplastic deformation of fcc metals at very high strain rates},
  author={Austin, Ryan A and McDowell, David L},
  journal={International Journal of Plasticity},
  volume={27},
  number={1},
  pages={1--24},
  year={2011},
  publisher={Elsevier}
}

@article{friedman_multivariate_1991,
	title = {Multivariate {Adaptive} {Regression} {Splines}},
	volume = {19},
	doi = {10.1214/aos/1176347963},
	number = {1},
	urldate = {2025-02-10},
	journal = {The Annals of Statistics},
	author = {Friedman, Jerome H.},
	year = {1991}
}

@article{devincre2008,
author = {B. Devincre  and T. Hoc  and L. Kubin },
title = {Dislocation Mean Free Paths and Strain Hardening of Crystals},
journal = {Science},
volume = {320},
number = {5884},
pages = {1745-1748},
year = {2008},
doi = {10.1126/science.1156101},
URL = {https://www.science.org/doi/abs/10.1126/science.1156101}}

@article{DAVIDSON1966703,
title = {The deformation behavior of high purity polycrystalline iron and single crystal molybdenum as a function of strain rate at 300°k},
journal = {Acta Metallurgica},
volume = {14},
number = {6},
pages = {703-710},
year = {1966},
doi = {https://doi.org/10.1016/0001-6160(66)90117-9},
url = {https://www.sciencedirect.com/science/article/pii/0001616066901179},
author = {D.L Davidson and U.S Lindholm and L.M Yeakley},
}

@article{bertin_crystal_2023,
	title = {Crystal plasticity model of {BCC} metals from large-scale {MD} simulations},
	volume = {260},
	issn = {13596454},
	doi = {10.1016/j.actamat.2023.119336},
	journal = {Acta Materialia},
	author = {Bertin, Nicolas and Carson, Robert and Bulatov, Vasily V. and Lind, Jonathan and Nelms, Matthew},
	year = {2023},
	pages = {119336},
}

@article{kennedy2001bayesian,
  title={Bayesian calibration of computer models},
  author={Kennedy, Marc C and O'Hagan, Anthony},
  journal={Journal of the Royal Statistical Society: Series B (Statistical Methodology)},
  volume={63},
  number={3},
  pages={425--464},
  year={2001},
  publisher={Wiley Online Library}
}

@article{nguyen2021dynamic,
  title={Dynamic crystal plasticity modeling of single crystal tantalum and validation using Taylor cylinder impact tests},
  author={Nguyen, Thao and Fensin, Saryu J and Luscher, Darby J},
  journal={International Journal of Plasticity},
  volume={139},
  pages={102940},
  year={2021},
  publisher={Elsevier}
}

@article{nguyen2024calibration,
	title = {Calibration and validation of the foundation for a multiphase strength model for tin},
	volume = {135},
	issn = {0021-8979, 1089-7550},
        doi = {10.1063/5.0207405},
        language = {en},
	number = {22},
	journal = {Journal of Applied Physics},
	author = {Nguyen, Thao and Burakovsky, Leonid and Fensin, Saryu J. and Luscher, Darby J. and Prime, Michael B. and Cady, Carl and Gray, George T. and Jones, David R. and Martinez, Daniel T. and Rowland, Richard L. and Sjue, Sky and Sturtevant, Blake T. and Valdez, James A.},
	month = jun,
	year = {2024},
	pages = {225105}
}

@article{sargsyan2014dimensionality,
  title={Dimensionality reduction for complex models via Bayesian compressive sensing},
  author={Sargsyan, Khachik and Safta, Cosmin and Najm, Habib N and Debusschere, Bert J and Ricciuto, Daniel and Thornton, Peter},
  journal={International Journal for Uncertainty Quantification},
  volume={4},
  number={1},
  year={2014},
  publisher={Begel House Inc.}
}

@article{huan2018global,
  title={Global sensitivity analysis and estimation of model error, toward uncertainty quantification in scramjet computations},
  author={Huan, Xun and Safta, Cosmin and Sargsyan, Khachik and Geraci, Gianluca and Eldred, Michael S and Vane, Zachary P and Lacaze, Guilhem and Oefelein, Joseph C and Najm, Habib N},
  journal={AIAA Journal},
  volume={56},
  number={3},
  pages={1170--1184},
  year={2018},
  publisher={American Institute of Aeronautics and Astronautics}
}

@article{sobol1993sensitivity,
  title={Sensitivity estimates for nonlinear mathematical models},
  author={Sobol', IM},
  journal={Math. Model. Comput. Exp.},
  volume={1},
  pages={407},
  year={1993}
}

@article{sobol_global_2001,
	title = {Global sensitivity indices for nonlinear mathematical models and their {Monte} {Carlo} estimates},
	volume = {55},
	language = {en},
	number = {1-3},
	urldate = {2025-02-15},
	journal = {Mathematics and Computers in Simulation},
	author = {Sobol', I.M},
	year = {2001},
	pages = {271--280}
}

@article{dequiedt_heterogeneous_2015,
	title = {Heterogeneous deformation in ductile {FCC} single crystals in biaxial stretching: the influence of slip system interactions},
	volume = {83},
	doi = {10.1016/j.jmps.2015.05.020},
	journal = {Journal of the Mechanics and Physics of Solids},
	author = {Dequiedt, J.L. and Denoual, C. and Madec, R.},
	year = {2015},
	pages = {301--318},
}

@article{francom_sensitivity_2017,
	title = {Sensitivity Analysis and Emulation for Functional Data using Bayesian Adaptive Splines},
	doi = {10.5705/ss.202016.0130},
	urldate = {2025-02-11},
	journal = {Statistica Sinica},
	author = {Francom,, Devin and Sanso, Bruno and Kupresanin, Ana and Johannesson, Gardar},
	year = {2017},
}

@article{sedighiani_efficient_2020,
	title = {An efficient and robust approach to determine material parameters of crystal plasticity constitutive laws from macro-scale stress–strain curves},
	volume = {134},
	journal = {International Journal of Plasticity},
	author = {Sedighiani, K. and Diehl, M. and Traka, K. and Roters, F. and Sietsma, J. and Raabe, D.},
	year = {2020},
	pages = {102779}
}

@article{nguyen_dislocation-based_2017,
	title = {A dislocation-based crystal plasticity framework for dynamic ductile failure of single crystals},
	volume = {108},
	journal = {Journal of the Mechanics and Physics of Solids},
	author = {Nguyen, Thao and Luscher, D.J. and Wilkerson, J.W.},
	year = {2017},
	pages = {1--29}
}

@article{wilkerson_dynamic_2014,
	title = {A dynamic void growth model governed by dislocation kinetics},
	volume = {70},
	journal = {Journal of the Mechanics and Physics of Solids},
	author = {Wilkerson, J.W. and Ramesh, K.T.},
	year = {2014},
	pages = {262--280}
}

@article{regazzoni_dislocation_1987,
    title = {Dislocation kinetics at high strain rates},
    volume = {35},
    number = {12},
    journal = {Acta Metallurgica},
    author = {Regazzoni, G. and Kocks, U.F. and Follansbee, P.S.},
    year = {1987},
    pages = {2865--2875},
}

@article{estrin_local_1986,
    title = {Local strain hardening and nonuniformity of plastic deformation},
    volume = {34},
    number = {12},
    journal = {Acta Metallurgica},
    author = {Estrin, Y. and Kubin, L.P.},
    year = {1986},
    pages = {2455--2464},
}

@article{lloyd_plane_2014,
    title = {Plane wave simulation of elastic-viscoplastic single crystals},
    volume = {69},
    issn = {00225096},
    url = {https://linkinghub.elsevier.com/retrieve/pii/S0022509614000775},
    doi = {10.1016/j.jmps.2014.04.009},
    language = {en},
    urldate = {2025-03-25},
    journal = {Journal of the Mechanics and Physics of Solids},
    author = {Lloyd, J.T. and Clayton, J.D. and Austin, R.A. and McDowell, D.L.},
    month = sep,
    year = {2014},
    pages = {14--32},
}

@article{zhang_data-driven_2023,
    title = {Data-driven statistical reduced-order modeling and quantification of polycrystal mechanics leading to porosity-based ductile damage},
    volume = {179},
    issn = {00225096},
    url = {https://linkinghub.elsevier.com/retrieve/pii/S0022509623001904},
    doi = {10.1016/j.jmps.2023.105386},
    language = {en},
    urldate = {2025-02-18},
    journal = {Journal of the Mechanics and Physics of Solids},
    author = {Zhang, Yinling and Chen, Nan and Bronkhorst, Curt A. and Cho, Hansohl and Argus, Robert},
    month = oct,
    year = {2023},
    pages = {105386},
}

@article{dunham_attribution_2025,
    title = {Attribution of heterogeneous stress distributions in low-grain polycrystals under conditions leading to damage},
    volume = {186},
    issn = {07496419},
    url = {https://linkinghub.elsevier.com/retrieve/pii/S0749641925000191},
    doi = {10.1016/j.ijplas.2025.104258},
    language = {en},
    urldate = {2025-02-18},
    journal = {International Journal of Plasticity},
    author = {Dunham, Samuel D. and Zhang, Yinling and Chen, Nan and Alleman, Coleman and Bronkhorst, Curt A.},
    month = mar,
    year = {2025},
    pages = {104258},
}

@article{robbe_global_2023,
    title = {Global sensitivity analysis of a coupled multiphysics model to predict surface evolution in fusion plasma–surface interactions},
    volume = {226},
    issn = {09270256},
    url = {https://linkinghub.elsevier.com/retrieve/pii/S0927025623002239},
    doi = {10.1016/j.commatsci.2023.112229},
    language = {en},
    urldate = {2025-01-07},
    journal = {Computational Materials Science},
    author = {Robbe, Pieterjan and Blondel, Sophie and Casey, Tiernan A. and Lasa, Ane and Sargsyan, Khachik and Wirth, Brian D. and Najm, Habib N.},
    month = jun,
    year = {2023},
    pages = {112229},
}

@article{kothari_elasto-viscoplastic_1998,
    title = {Elasto-viscoplastic constitutive equations for polycrystalline metals: {Application} to tantalum},
    volume = {46},
    copyright = {https://www.elsevier.com/tdm/userlicense/1.0/},
    issn = {00225096},
    shorttitle = {Elasto-viscoplastic constitutive equations for polycrystalline metals},
    url = {https://linkinghub.elsevier.com/retrieve/pii/S0022509697000379},
    doi = {10.1016/S0022-5096(97)00037-9},
    language = {en},
    number = {1},
    urldate = {2025-02-25},
    journal = {Journal of the Mechanics and Physics of Solids},
    author = {Kothari, M. and Anand, L.},
    month = jan,
    year = {1998},
    pages = {51--83},
}

@article{nelms_uncertainty_2024,
    title = {Uncertainty quantification of material parameters in modeling coupled metal and high explosive experiments},
    volume = {136},
    issn = {0021-8979, 1089-7550},
    url = {https://pubs.aip.org/jap/article/136/19/195101/3320673/Uncertainty-quantification-of-material-parameters},
    doi = {10.1063/5.0226642},
    language = {en},
    number = {19},
    urldate = {2025-02-19},
    journal = {Journal of Applied Physics},
    author = {Nelms, Matthew and Schill, William and William Kuo, I.-F. and Barton, Nathan and Schmidt, Kathleen},
    month = nov,
    year = {2024},
    pages = {195101},
}

@article{xu_homogeneous_2000,
    title = {Homogeneous nucleation of dislocation loops under stress in perfect crystals},
    volume = {80},
    issn = {0950-0839, 1362-3036},
    url = {http://www.tandfonline.com/doi/abs/10.1080/09500830050134318},
    doi = {10.1080/09500830050134318},
    language = {en},
    number = {9},
    urldate = {2025-04-17},
    journal = {Philosophical Magazine Letters},
    author = {Xu, Guanshui and Argon, Ali S.},
    month = aug,
    year = {2000},
    pages = {605--611},
}

@article{mitchell_shock_1981,
    title = {Shock compression of aluminum, copper, and tantalum},
    volume = {52},
    issn = {0021-8979},
    url = {https://doi.org/10.1063/1.329160},
    doi = {10.1063/1.329160},
    number = {5},
    urldate = {2025-05-16},
    journal = {Journal of Applied Physics},
    author = {Mitchell, A. C. and Nellis, W. J.},
    month = may,
    year = {1981},
    pages = {3363--3374},
}

@article{macdonald_thermodynamic_1981,
    title = {Thermodynamic properties of fcc metals at high temperatures},
    volume = {24},
    copyright = {http://link.aps.org/licenses/aps-default-license},
    issn = {0163-1829},
    url = {https://link.aps.org/doi/10.1103/PhysRevB.24.1715},
    doi = {10.1103/PhysRevB.24.1715},
    language = {en},
    number = {4},
    urldate = {2025-05-16},
    journal = {Physical Review B},
    author = {MacDonald, Rosemary A. and MacDonald, William M.},
    month = aug,
    year = {1981},
    pages = {1715--1724},
}

@article{bronkhorst_local_2021,
    title = {Local micro-mechanical stress conditions leading to pore nucleation during dynamic loading},
    volume = {137},
    issn = {07496419},
    url = {https://linkinghub.elsevier.com/retrieve/pii/S0749641920307488},
    doi = {10.1016/j.ijplas.2020.102903},
    language = {en},
    urldate = {2025-06-04},
    journal = {International Journal of Plasticity},
    author = {Bronkhorst, C.A. and Cho, H. and Marcy, P.W. and Vander Wiel, S.A. and Gupta, S. and Versino, D. and Anghel, V. and Gray, G.T.},
    month = feb,
    year = {2021},
    pages = {102903},
}

@article{depriester_crystal_2023,
    title = {Crystal {Plasticity} simulations of in situ tensile tests: {A} two-step inverse method for identification of {CP} parameters, and assessment of {CPFEM} capabilities},
    volume = {168},
    issn = {07496419},
    shorttitle = {Crystal {Plasticity} simulations of in situ tensile tests},
    url = {https://linkinghub.elsevier.com/retrieve/pii/S074964192300181X},
    doi = {10.1016/j.ijplas.2023.103695},
    language = {en},
    urldate = {2025-06-04},
    journal = {International Journal of Plasticity},
    author = {Depriester, D. and Goulmy, J.P. and Barrallier, L.},
    month = sep,
    year = {2023},
    pages = {103695},
}

@article{monnet_dislocation-dynamics_2013,
    title = {Dislocation-dynamics based crystal plasticity law for the low- and high-temperature deformation regimes of bcc crystal},
    volume = {61},
    copyright = {https://www.elsevier.com/tdm/userlicense/1.0/},
    issn = {13596454},
    url = {https://linkinghub.elsevier.com/retrieve/pii/S135964541300503X},
    doi = {10.1016/j.actamat.2013.07.002},
    language = {en},
    number = {16},
    urldate = {2025-02-25},
    journal = {Acta Materialia},
    author = {Monnet, Ghiath and Vincent, Ludovic and Devincre, Benoit},
    month = sep,
    year = {2013},
    pages = {6178--6190},
}

@article{venkatraman_bayesian_2022,
    title = {Bayesian analysis of parametric uncertainties and model form probabilities for two different crystal plasticity models of lamellar grains in α+β {Titanium} alloys},
    volume = {154},
    copyright = {https://www.elsevier.com/tdm/userlicense/1.0/},
    issn = {0749-6419},
    url = {https://linkinghub.elsevier.com/retrieve/pii/S0749641922000717},
    doi = {10.1016/j.ijplas.2022.103289},
    language = {en},
    urldate = {2025-07-11},
    journal = {International Journal of Plasticity},
    author = {Venkatraman, Aditya and McDowell, David L. and Kalidindi, Surya R.},
    month = jul,
    year = {2022},
    note = {Publisher: Elsevier BV},
    pages = {103289},
}

@inproceedings{higdon_simulationaided_2012,
    address = {New York, NY},
    title = {Simulation-{Aided} {Inference} in {Cosmology}},
    isbn = {978-1-4614-3520-4},
    booktitle = {Statistical {Challenges} in {Modern} {Astronomy} {V}},
    publisher = {Springer New York},
    author = {Higdon, David and Lawrence, Earl and Heitmann, Katrin and Habib, Salman},
    editor = {Feigelson, Eric D. and Babu, G. Jogesh},
    year = {2012},
    pages = {41--57},
}

@article{sills_dislocation_2018,
    title = {Dislocation {Networks} and the {Microstructural} {Origin} of {Strain} {Hardening}},
    volume = {121},
    issn = {0031-9007, 1079-7114},
    url = {https://link.aps.org/doi/10.1103/PhysRevLett.121.085501},
    doi = {10.1103/PhysRevLett.121.085501},
    language = {en},
    number = {8},
    urldate = {2024-10-31},
    journal = {Physical Review Letters},
    author = {Sills, Ryan B. and Bertin, Nicolas and Aghaei, Amin and Cai, Wei},
    month = aug,
    year = {2018},
    pages = {085501},
}

@article{ma_dislocation_2006,
    title = {A dislocation density based constitutive model for crystal plasticity {FEM} including geometrically necessary dislocations},
    volume = {54},
    copyright = {https://www.elsevier.com/tdm/userlicense/1.0/},
    issn = {13596454},
    url = {https://linkinghub.elsevier.com/retrieve/pii/S1359645406000747},
    doi = {10.1016/j.actamat.2006.01.005},
    language = {en},
    number = {8},
    urldate = {2025-09-25},
    journal = {Acta Materialia},
    author = {Ma, A. and Roters, F. and Raabe, D.},
    month = may,
    year = {2006},
    pages = {2169--2179},
}

@article{madec_role_2003,
    title = {The {Role} of {Collinear} {Interaction} in {Dislocation}-{Induced} {Hardening}},
    volume = {301},
    issn = {0036-8075, 1095-9203},
    url = {https://www.science.org/doi/10.1126/science.1085477},
    doi = {10.1126/science.1085477},
    language = {en},
    number = {5641},
    urldate = {2025-09-25},
    journal = {Science},
    author = {Madec, R. and Devincre, B. and Kubin, L. and Hoc, T. and Rodney, D.},
    month = sep,
    year = {2003},
    pages = {1879--1882},
}

@article{bernhard_bayesian_2019,
    title = {Bayesian estimation of the specific shear and bulk viscosity of quark–gluon plasma},
    volume = {15},
    issn = {1745-2473, 1745-2481},
    url = {https://www.nature.com/articles/s41567-019-0611-8},
    doi = {10.1038/s41567-019-0611-8},
    language = {en},
    number = {11},
    urldate = {2025-10-10},
    journal = {Nature Physics},
    author = {Bernhard, Jonah E. and Moreland, J. Scott and Bass, Steffen A.},
    month = nov,
    year = {2019},
    pages = {1113--1117},
}

@article{Seeger2001,
    title = {{Why anomalous slip in body-centred cubic metals?}},
    year = {2001},
    journal = {Materials Science and Engineering A},
    author = {Seeger, Alfred},
    pages = {254--260},
    volume = {319-321},
    doi = {10.1016/S0921-5093(01)00958-3},
    issn = {09215093}
}

@article{Seeger2002AnomalousMetals,
    title = {{Anomalous Slip - A Feature of High-Purity Body-Centred Cubic Metals}},
    year = {2002},
    journal = {physica status solidi (a)},
    author = {Seeger, A. and Wasserbäch, W.},
    number = {1},
    month = {1},
    pages = {27--50},
    volume = {189},
    url = {https://onlinelibrary.wiley.com/doi/10.1002/1521-396X(200201)189:1%3C27::AID-PSSA27%3E3.0.CO;2-T},
    doi = {10.1002/1521-396X(200201)189:1<27::AID-PSSA27>3.0.CO;2-T},
    issn = {0031-8965}
}

@article{Tang1993,
    title = {{Orthogonal Array-Based Latin Hypercubes}},
    year = {1993},
    journal = {Journal of the American Statistical Association},
    author = {Tang, Boxin},
    number = {424},
    month = {12},
    pages = {1392--1397},
    volume = {88},
    url = {http://www.tandfonline.com/doi/abs/10.1080/01621459.1993.10476423},
    doi = {10.1080/01621459.1993.10476423},
    issn = {0162-1459}
}

@article{10.1063/1.1699114,
    title = {{Equation of State Calculations by Fast Computing Machines}},
    year = {1953},
    journal = {The Journal of Chemical Physics},
    author = {Metropolis, Nicholas and Rosenbluth, Arianna W. and Rosenbluth, Marshall N and Teller, Augusta H. and Teller, Edward},
    number = {6},
    month = {6},
    pages = {1087--1092},
    volume = {21},
    url = {https://pubs.aip.org/jcp/article/21/6/1087/202680/Equation-of-State-Calculations-by-Fast-Computing},
    doi = {10.1063/1.1699114},
    issn = {0021-9606}
}

@article{Chib1995UnderstandingAlgorithm,
    title = {{Understanding the Metropolis-Hastings Algorithm}},
    year = {1995},
    journal = {The American Statistician},
    author = {Chib, Siddhartha and Greenberg, Edward},
    number = {4},
    month = {11},
    pages = {327--335},
    volume = {49},
    url = {https://www.tandfonline.com/doi/abs/10.1080/00031305.1995.10476177#:~:text=We provide a detailed%2C introductory exposition of the,method is given along with guidance on implementation. http://www.tandfonline.com/doi/abs/10.1080/00031305.1995.10476177},
    doi = {10.1080/00031305.1995.10476177},
    issn = {0003-1305}
}

@article{green_reversible_1995,
    title = {Reversible jump {Markov} chain {Monte} {Carlo} computation and {Bayesian} model determination},
    volume = {82},
    issn = {0006-3444, 1464-3510},
    url = {https://academic.oup.com/biomet/article-lookup/doi/10.1093/biomet/82.4.711},
    doi = {10.1093/biomet/82.4.711},
    language = {en},
    number = {4},
    urldate = {2025-10-12},
    journal = {Biometrika},
    author = {Green, Peter J.},
    year = {1995},
    pages = {711--732},
}

@article{luscher_model_2013,
    title = {A model for finite-deformation nonlinear thermomechanical response of single crystal copper under shock conditions},
    volume = {61},
    issn = {00225096},
    url = {https://linkinghub.elsevier.com/retrieve/pii/S0022509613000914},
    doi = {10.1016/j.jmps.2013.05.002},
    language = {en},
    number = {9},
    urldate = {2024-10-15},
    journal = {Journal of the Mechanics and Physics of Solids},
    author = {Luscher, Darby J. and Bronkhorst, Curt A. and Alleman, Coleman N. and Addessio, Francis L.},
    month = sep,
    year = {2013},
    pages = {1877--1894},
}

@article{schmelzer_thermodynamically_2025,
    title = {Thermodynamically consistent and microstructure informed porosity-based dynamic ductile damage model},
    issn = {00225096},
    url = {https://linkinghub.elsevier.com/retrieve/pii/S0022509625003114},
    doi = {10.1016/j.jmps.2025.106336},
    language = {en},
    urldate = {2025-08-25},
    journal = {Journal of the Mechanics and Physics of Solids},
    author = {Schmelzer, Noah J. and Lieberman, Evan J. and Gray, George T. and Bronkhorst, Curt A.},
    month = aug,
    year = {2025},
    pages = {106336},
}

@article{schmelzer_statistical_2025,
    title = {Statistical evaluation of microscale stress conditions leading to void nucleation in the weak shock regime},
    volume = {188},
    issn = {07496419},
    url = {https://linkinghub.elsevier.com/retrieve/pii/S0749641925000774},
    doi = {10.1016/j.ijplas.2025.104318},
    language = {en},
    urldate = {2025-04-03},
    journal = {International Journal of Plasticity},
    author = {Schmelzer, Noah J. and Lieberman, Evan J. and Chen, Nan and Dunham, Samuel D. and Anghel, Veronica and Gray, George T. and Bronkhorst, Curt A.},
    month = may,
    year = {2025},
    pages = {104318},
}

@article{Cereceda2016,
    title = {{Unraveling the temperature dependence of the yield strength in single-crystal tungsten using atomistically-informed crystal plasticity calculations}},
    year = {2016},
    journal = {International Journal of Plasticity},
    author = {Cereceda, David and Diehl, Martin and Roters, Franz and Raabe, Dierk and Perlado, J. Manuel and Marian, Jaime},
    pages = {242--265},
    volume = {78},
    publisher = {Elsevier Ltd},
    url = {http://dx.doi.org/10.1016/j.ijplas.2015.09.002},
    doi = {10.1016/j.ijplas.2015.09.002},
    issn = {07496419},
    arxivId = {1506.02224}
}

@article{lim_multi-scale_2015,
    title = {A multi-scale model of dislocation plasticity in α-{Fe}: {Incorporating} temperature, strain rate and non-{Schmid} effects},
    volume = {73},
    issn = {07496419},
    shorttitle = {A multi-scale model of dislocation plasticity in α-{Fe}},
    url = {https://linkinghub.elsevier.com/retrieve/pii/S0749641914002356},
    doi = {10.1016/j.ijplas.2014.12.005},
    language = {en},
    urldate = {2025-10-18},
    journal = {International Journal of Plasticity},
    author = {Lim, H. and Hale, L.M. and Zimmerman, J.A. and Battaile, C.C. and Weinberger, C.R.},
    month = oct,
    year = {2015},
    pages = {100--118},
}

@article{schill_simultaneous_2021,
    title = {Simultaneous inference of the compressibility and inelastic response of tantalum under extreme loading},
    volume = {130},
    issn = {0021-8979, 1089-7550},
    url = {https://pubs.aip.org/jap/article/130/5/055901/158716/Simultaneous-inference-of-the-compressibility-and},
    doi = {10.1063/5.0056437},
    language = {en},
    number = {5},
    urldate = {2025-10-19},
    journal = {Journal of Applied Physics},
    author = {Schill, W. J. and Austin, R. A. and Schimdt, K. L. and Brown, J. L. and Barton, N. R.},
    month = aug,
    year = {2021},
    pages = {055901},
}

@article{mie_zur_1903,
    title = {Zur kinetischen {Theorie} der einatomigen {Körper}},
    volume = {316},
    issn = {0003-3804, 1521-3889},
    url = {https://onlinelibrary.wiley.com/doi/10.1002/andp.19033160802},
    doi = {10.1002/andp.19033160802},
    language = {en},
    number = {8},
    urldate = {2025-10-19},
    journal = {Annalen der Physik},
    author = {Mie, Gustav},
    month = jan,
    year = {1903},
    pages = {657--697},
}

@article{gruneisen_theorie_1912,
    title = {Theorie des festen {Zustandes} einatomiger {Elemente}},
    volume = {344},
    issn = {0003-3804, 1521-3889},
    url = {https://onlinelibrary.wiley.com/doi/10.1002/andp.19123441202},
    doi = {10.1002/andp.19123441202},
    language = {en},
    number = {12},
    urldate = {2025-10-19},
    journal = {Annalen der Physik},
    author = {Grüneisen, E.},
    month = jan,
    year = {1912},
    pages = {257--306},
}

@article{Queyreau2009,
    title = {{Slip systems interactions in {$\alpha$}-iron determined by dislocation dynamics simulations}},
    year = {2009},
    journal = {International Journal of Plasticity},
    author = {Queyreau, Sylvain and Monnet, Ghiath and Devincre, Benoît},
    number = {2},
    pages = {361--377},
    volume = {25},
    doi = {10.1016/j.ijplas.2007.12.009},
    issn = {07496419}
}

@incollection{Madec2004,
    title = {{Dislocation Interactions and Symmetries in BCC Crystals}},
    year = {2004},
    author = {Madec, R. and Kubin, L. P.},
    pages = {69--78},
    url = {http://link.springer.com/10.1007/978-1-4020-2111-4_7},
    doi = {10.1007/978-1-4020-2111-4{\_}7}
}

@article{Weygand2015,
    title = {{Multiscale Simulation of Plasticity in bcc Metals}},
    year = {2015},
    journal = {Annual Review of Materials Research},
    author = {Weygand, Daniel and Mrovec, Matous and Hochrainer, Thomas and Gumbsch, Peter},
    pages = {369--390},
    volume = {45},
    doi = {10.1146/annurev-matsci-070214-020852},
    issn = {15317331}
}

@book{argon_strengthening_2008,
    address = {Oxford},
    series = {Oxford series on materials modelling},
    title = {Strengthening mechanisms in crystal plasticity},
    isbn = {978-0-19-170571-7},
    language = {eng},
    number = {4},
    publisher = {Oxford University Press},
    author = {Argon, Ali S.},
    year = {2008},
    doi = {10.1093/acprof:oso/9780198516002.001.0001},
}

@article{Weinberger2012IncorporatingModels,
    title = {{Incorporating atomistic data of lattice friction into BCC crystal plasticity models}},
    year = {2012},
    journal = {International Journal of Plasticity},
    author = {Weinberger, Christopher R. and Battaile, Corbett C. and Buchheit, Thomas E. and Holm, Elizabeth A.},
    pages = {16--30},
    volume = {37},
    publisher = {Elsevier Ltd},
    url = {http://dx.doi.org/10.1016/j.ijplas.2012.03.012},
    doi = {10.1016/j.ijplas.2012.03.012},
    issn = {07496419}
}

@book{hastie_elements_2009,
    address = {New York, NY},
    edition = {2nd ed},
    series = {Springer series in statistics},
    title = {The elements of statistical learning: data mining, inference, and prediction},
    isbn = {978-0-387-84857-0 978-0-387-84858-7},
    shorttitle = {The elements of statistical learning},
    publisher = {Springer},
    author = {Hastie, Trevor and Tibshirani, Robert and Friedman, J. H.},
    year = {2009},
}

@article{bronkhorst_structural_2019,
    title = {Structural representation of additively manufactured {316L} austenitic stainless steel},
    volume = {118},
    issn = {07496419},
    url = {https://linkinghub.elsevier.com/retrieve/pii/S0749641918307654},
    doi = {10.1016/j.ijplas.2019.01.012},
    language = {en},
    urldate = {2025-06-13},
    journal = {International Journal of Plasticity},
    author = {Bronkhorst, C.A. and Mayeur, J.R. and Livescu, V. and Pokharel, R. and Brown, D.W. and Gray, G.T.},
    month = jul,
    year = {2019},
    pages = {70--86},
}

@article{srivastava_thermo-mechanically-coupled_2010,
    title = {A thermo-mechanically-coupled large-deformation theory for amorphous polymers in a temperature range which spans their glass transition},
    volume = {26},
    copyright = {https://www.elsevier.com/tdm/userlicense/1.0/},
    issn = {07496419},
    url = {https://linkinghub.elsevier.com/retrieve/pii/S0749641910000057},
    doi = {10.1016/j.ijplas.2010.01.004},
    language = {en},
    number = {8},
    urldate = {2024-10-24},
    journal = {International Journal of Plasticity},
    author = {Srivastava, Vikas and Chester, Shawn A. and Ames, Nicoli M. and Anand, Lallit},
    month = aug,
    year = {2010},
    pages = {1138--1182},
}

@article{jordan_training_2025,
    title = {Training of a physics-based thermo-viscoplasticity model on big data for polypropylene},
    volume = {184},
    issn = {07496419},
    url = {https://linkinghub.elsevier.com/retrieve/pii/S0749641924003061},
    doi = {10.1016/j.ijplas.2024.104179},
    language = {en},
    urldate = {2024-12-18},
    journal = {International Journal of Plasticity},
    author = {Jordan, Benoit and Mohr, Dirk},
    month = jan,
    year = {2025},
    pages = {104179},
}

@article{lee_size-dependent_2024,
    title = {Size-dependent fracture in elastomers: {Experiments} and continuum modeling},
    volume = {8},
    issn = {2475-9953},
    shorttitle = {Size-dependent fracture in elastomers},
    url = {https://link.aps.org/doi/10.1103/PhysRevMaterials.8.115602},
    doi = {10.1103/PhysRevMaterials.8.115602},
    language = {en},
    number = {11},
    urldate = {2025-02-03},
    journal = {Physical Review Materials},
    author = {Lee, Jaehee and Lee, Jeongun and Yun, Seounghee and Kim, Sanha and Lee, Howon and Chester, Shawn A. and Cho, Hansohl},
    month = nov,
    year = {2024},
    pages = {115602},
}

@article{talamini_progressive_2018,
    title = {Progressive damage and rupture in polymers},
    volume = {111},
    issn = {00225096},
    url = {https://linkinghub.elsevier.com/retrieve/pii/S0022509617303939},
    doi = {10.1016/j.jmps.2017.11.013},
    language = {en},
    urldate = {2025-10-24},
    journal = {Journal of the Mechanics and Physics of Solids},
    author = {Talamini, Brandon and Mao, Yunwei and Anand, Lallit},
    month = feb,
    year = {2018},
    pages = {434--457},
}

@inproceedings{johnson_constitutive_1983,
    address = {Hague, Netherlands},
    title = {A constitutive model and data for metals subjected to large strains, high strain rates and high temperatures},
    booktitle = {Proceedings of the 7th {International} {Symposium} on {Ballistics}},
    author = {Johnson, Gordon R and Cook, William H.},
    year = {1983},
}

@article{lee_polyurethane-urea_2023,
    title = {A polyurethane-urea elastomer at low to extreme strain rates},
    volume = {280},
    issn = {00207683},
    url = {https://linkinghub.elsevier.com/retrieve/pii/S0020768323002573},
    doi = {10.1016/j.ijsolstr.2023.112360},
    language = {en},
    urldate = {2025-10-25},
    journal = {International Journal of Solids and Structures},
    author = {Lee, Jaehee and Veysset, David and Hsieh, Alex J. and Rutledge, Gregory C. and Cho, Hansohl},
    month = sep,
    year = {2023},
    pages = {112360},
}

@article{wagner_foundational_2025,
    title = {A foundational framework for the mesoscale modeling of dynamic elastomers and gels},
    volume = {194},
    issn = {00225096},
    url = {https://linkinghub.elsevier.com/retrieve/pii/S0022509624003806},
    doi = {10.1016/j.jmps.2024.105914},
    language = {en},
    urldate = {2025-10-25},
    journal = {Journal of the Mechanics and Physics of Solids},
    author = {Wagner, Robert J. and Silberstein, Meredith N.},
    month = jan,
    year = {2025},
    pages = {105914},
}

@article{saha_numerical_2026,
    title = {Numerical and data-driven modeling of spall failure in polycrystalline ductile materials},
    volume = {448},
    issn = {00457825},
    url = {https://linkinghub.elsevier.com/retrieve/pii/S0045782525007650},
    doi = {10.1016/j.cma.2025.118493},
    language = {en},
    urldate = {2025-10-31},
    journal = {Computer Methods in Applied Mechanics and Engineering},
    author = {Saha, Indrashish and Graham-Brady, Lori},
    month = jan,
    year = {2026},
    pages = {118493},
}

@article{pogorelko_dynamic_2023,
    title = {Dynamic tensile fracture of iron: {Molecular} dynamics simulations and micromechanical model based on dislocation plasticity},
    volume = {167},
    issn = {07496419},
    shorttitle = {Dynamic tensile fracture of iron},
    url = {https://linkinghub.elsevier.com/retrieve/pii/S074964192300164X},
    doi = {10.1016/j.ijplas.2023.103678},
    language = {en},
    urldate = {2025-11-27},
    journal = {International Journal of Plasticity},
    author = {Pogorelko, Viсtor V. and Mayer, Alexander E.},
    month = aug,
    year = {2023},
    pages = {103678},
}

@article{groger_symmetry-adapted_2021,
    title = {Symmetry-adapted single crystal yield criterion for non-{Schmid} materials},
    volume = {146},
    url = {https://doi.org/10.1016/j.ijplas.2021.103101},
    doi = {10.1016/j.ijplas.2021.103101},
    journal = {International Journal of Plasticity},
    author = {Gröger, Roman},
    year = {2021},
    note = {Publisher: Elsevier Ltd},
    pages = {103101},
}

@article{dindarlou_optimization_2024,
    title = {Optimization of crystal plasticity parameters with proxy materials data for alloy single crystals},
    volume = {174},
    issn = {07496419},
    url = {https://linkinghub.elsevier.com/retrieve/pii/S0749641924000214},
    doi = {10.1016/j.ijplas.2024.103894},
    language = {en},
    urldate = {2025-11-28},
    journal = {International Journal of Plasticity},
    author = {Dindarlou, Shahram and Castelluccio, Gustavo M.},
    month = mar,
    year = {2024},
    pages = {103894},
}

@article{rousselier_macroscopic_2012,
    title = {Macroscopic plasticity modeling of anisotropic aluminum extrusions using a {Reduced} {Texture} {Methodology}},
    volume = {30-31},
    copyright = {https://www.elsevier.com/tdm/userlicense/1.0/},
    issn = {07496419},
    url = {https://linkinghub.elsevier.com/retrieve/pii/S0749641911001689},
    doi = {10.1016/j.ijplas.2011.10.004},
    language = {en},
    urldate = {2025-11-28},
    journal = {International Journal of Plasticity},
    author = {Rousselier, Gilles and Luo, Meng and Mohr, Dirk},
    month = mar,
    year = {2012},
    pages = {144--165},
}

@article{tsekpuia_microstructure-based_2023,
    title = {A microstructure-based three-scale homogenization model for predicting the elasto-viscoplastic behavior of duplex stainless steels},
    volume = {164},
    issn = {07496419},
    url = {https://linkinghub.elsevier.com/retrieve/pii/S074964192300061X},
    doi = {10.1016/j.ijplas.2023.103575},
    language = {en},
    urldate = {2025-11-28},
    journal = {International Journal of Plasticity},
    author = {Tsekpuia, Eyram and Guery, Adrien and Gey, Nathalie and Berbenni, Stéphane},
    month = may,
    year = {2023},
    pages = {103575},
}

@article{lhadi_micromechanical_2018,
    title = {Micromechanical modeling of the effect of elastic and plastic anisotropies on the mechanical behavior of β-{Ti} alloys},
    volume = {109},
    issn = {07496419},
    url = {https://linkinghub.elsevier.com/retrieve/pii/S0749641918301839},
    doi = {10.1016/j.ijplas.2018.05.010},
    language = {en},
    urldate = {2025-11-28},
    journal = {International Journal of Plasticity},
    author = {Lhadi, Safaa and Berbenni, Stéphane and Gey, Nathalie and Richeton, Thiebaud and Germain, Lionel},
    month = oct,
    year = {2018},
    pages = {88--107},
}

@article{patra_constitutive_2014,
    title = {Constitutive equations for modeling non-{Schmid} effects in single crystal bcc-{Fe} at low and ambient temperatures},
    volume = {59},
    url = {http://dx.doi.org/10.1016/j.ijplas.2014.03.016},
    doi = {10.1016/j.ijplas.2014.03.016},
    journal = {International Journal of Plasticity},
    author = {Patra, Anirban and Zhu, Ting and McDowell, David L.},
    year = {2014},
    note = {Publisher: Elsevier Ltd},
    pages = {1--14},
}

@article{patra__2023,
    title = {ρ -{CP}: {Open} source dislocation density based crystal plasticity framework for simulating temperature- and strain rate-dependent deformation},
    volume = {224},
    issn = {09270256},
    shorttitle = {ρ -{CP}},
    url = {https://linkinghub.elsevier.com/retrieve/pii/S0927025623001763},
    doi = {10.1016/j.commatsci.2023.112182},
    language = {en},
    urldate = {2025-11-28},
    journal = {Computational Materials Science},
    author = {Patra, Anirban and Chaudhary, Suketa and Pai, Namit and Ramgopal, Tarakram and Khandelwal, Sarthak and Rao, Adwitiya and McDowell, David L.},
    month = may,
    year = {2023},
    pages = {112182},
}

@article{bonatti_cp-fft_2022,
    title = {From {CP}-{FFT} to {CP}-{RNN}: {Recurrent} neural network surrogate model of crystal plasticity},
    volume = {158},
    issn = {07496419},
    shorttitle = {From {CP}-{FFT} to {CP}-{RNN}},
    url = {https://linkinghub.elsevier.com/retrieve/pii/S074964192200208X},
    doi = {10.1016/j.ijplas.2022.103430},
    language = {en},
    urldate = {2025-11-28},
    journal = {International Journal of Plasticity},
    author = {Bonatti, Colin and Berisha, Bekim and Mohr, Dirk},
    month = nov,
    year = {2022},
    pages = {103430},
}

@article{liu_learning_2023,
    title = {Learning macroscopic internal variables and history dependence from microscopic models},
    volume = {178},
    issn = {00225096},
    url = {https://linkinghub.elsevier.com/retrieve/pii/S0022509623001333},
    doi = {10.1016/j.jmps.2023.105329},
    language = {en},
    urldate = {2025-11-29},
    journal = {Journal of the Mechanics and Physics of Solids},
    author = {Liu, Burigede and Ocegueda, Eric and Trautner, Margaret and Stuart, Andrew M. and Bhattacharya, Kaushik},
    month = sep,
    year = {2023},
    pages = {105329},
}

@article{eghtesad_machine_2023,
    title = {Machine learning-enabled identification of micromechanical stress and strain hotspots predicted via dislocation density-based crystal plasticity simulations},
    volume = {166},
    issn = {07496419},
    url = {https://linkinghub.elsevier.com/retrieve/pii/S0749641923001328},
    doi = {10.1016/j.ijplas.2023.103646},
    language = {en},
    urldate = {2025-11-29},
    journal = {International Journal of Plasticity},
    author = {Eghtesad, Adnan and Luo, Qixiang and Shang, Shun-Li and Lebensohn, Ricardo A. and Knezevic, Marko and Liu, Zi-Kui and Beese, Allison M.},
    month = jul,
    year = {2023},
    pages = {103646},
}

@article{schmidt_texture-dependent_2025,
    title = {A texture-dependent yield criterion based on {Support} {Vector} {Classification}},
    volume = {188},
    issn = {07496419},
    url = {https://linkinghub.elsevier.com/retrieve/pii/S0749641925000701},
    doi = {10.1016/j.ijplas.2025.104311},
    language = {en},
    urldate = {2025-11-29},
    journal = {International Journal of Plasticity},
    author = {Schmidt, Jan and Kalidindi, Surya R. and Hartmaier, Alexander},
    month = may,
    year = {2025},
    pages = {104311},
}

@article{Lee2024ExtremeCrystals,
    title = {{Extreme resilience and dissipation in heterogeneous elasto-plastomeric crystals}},
    year = {2024},
    journal = {Soft Matter},
    author = {Lee, Gisoo and Lee, Jaehee and Lee, Seunghyeon and Rudykh, Stephan and Cho, Hansohl},
    publisher = {Royal Society of Chemistry},
    url = {http://xlink.rsc.org/?DOI=D3SM01076G},
    doi = {10.1039/D3SM01076G},
    issn = {1744-683X}
}

@article{noii_bayesian_2021,
    title = {Bayesian inversion for unified ductile phase-field fracture},
    volume = {68},
    issn = {0178-7675, 1432-0924},
    url = {https://link.springer.com/10.1007/s00466-021-02054-w},
    doi = {10.1007/s00466-021-02054-w},
    language = {en},
    number = {4},
    urldate = {2025-11-30},
    journal = {Computational Mechanics},
    author = {Noii, Nima and Khodadadian, Amirreza and Ulloa, Jacinto and Aldakheel, Fadi and Wick, Thomas and François, Stijn and Wriggers, Peter},
    month = oct,
    year = {2021},
    pages = {943--980},
}

@article{noii_bayesian_2022,
    title = {Bayesian {Inversion} with {Open}-{Source} {Codes} for {Various} {One}-{Dimensional} {Model} {Problems} in {Computational} {Mechanics}},
    volume = {29},
    issn = {1134-3060, 1886-1784},
    url = {https://link.springer.com/10.1007/s11831-022-09751-6},
    doi = {10.1007/s11831-022-09751-6},
    language = {en},
    number = {6},
    urldate = {2025-11-30},
    journal = {Archives of Computational Methods in Engineering},
    author = {Noii, Nima and Khodadadian, Amirreza and Ulloa, Jacinto and Aldakheel, Fadi and Wick, Thomas and François, Stijn and Wriggers, Peter},
    month = oct,
    year = {2022},
    pages = {4285--4318},
}

\end{document}